\documentclass[lettersize,journal]{IEEEtran}
\usepackage{makecell}
\usepackage{amsmath,amsfonts}
\usepackage{algorithmic}
\usepackage{algorithm}
\usepackage{array}
\usepackage[caption=false,font=normalsize,labelfont=sf,textfont=sf]{subfig}
\usepackage{textcomp}
\usepackage{stfloats}
\usepackage{url,hyperref}
\usepackage{verbatim}
\usepackage{graphicx}
\usepackage{bm}
\usepackage{cite}
\usepackage{xcolor}
\usepackage{tabularx}
\begin{document}

\title{A Unifying View of OTFS and Its Many Variants}

\author{Qinwen~Deng,~\IEEEmembership{Member,~IEEE, }Yao~Ge,~\IEEEmembership{Member,~IEEE, }and Zhi Ding,~\IEEEmembership{Fellow,~IEEE}
\thanks{This work was supported by the National Science Foundation under Grant 2332760. The work of Yao Ge was supported by the RIE2020 Industry Alignment Fund—Industry Collaboration Projects (IAF-ICP) Funding Initiative, as well as cash and in-kind contribution from the industry partner(s). \textit{(Corresponding author: Yao~Ge.)}}
\thanks{Qinwen~Deng was with Department of Electrical and Computer Engineering, University of California at Davis, Davis, CA 95616 USA, he is now with the School of Electrical and Electronic Engineering, Nanyang Technological University, Singapore 639798 (e-mail: qinwen.deng@ntu.edu.sg).}
\thanks{Yao~Ge is with the Continental-NTU Corporate Lab, Nanyang Technological University, Singapore 639798 (e-mail: yao.ge@ntu.edu.sg).}
\thanks{Zhi~Ding is with Department of Electrical and Computer Engineering, University of California at Davis, Davis, CA 95616 USA (e-mail: zding@ucdavis.edu).}
\\
\vspace*{-.3cm}}



\maketitle

\IEEEpubid{This paper has been accepted for publication in IEEE Communications Surveys \& Tutorials}

\begin{abstract}
High mobility environment leads to severe Doppler effects and poses serious challenges to the conventional physical layer based on the widely popular orthogonal frequency division multiplexing (OFDM). The recent emergence of orthogonal time frequency space (OTFS) modulation, along with its many related variants, presents a promising solution to overcome such channel Doppler effects. This paper aims to clearly establish the relationships among the various manifestations of OTFS. Among these related modulations, we identify their connections, common features, and distinctions. Building on existing works, this work provides a general overview of various OTFS-related detection schemes and performance comparisons. We first provide an overview of OFDM and filter bank multi-carrier (FBMC) by demonstrating OTFS as a precoded FBMC through the introduction of inverse symplectic finite Fourier transform (ISFFT). We explore the relationship between OTFS and related modulation schemes with similar characteristics. We provide an effective channel model for high-mobility channels and offer a unified detection representation. We provide numerical comparisons of power spectrum density (PSD) and bit error rate (BER) to underscore the benefit of these modulation schemes in high-mobility scenarios. We also evaluate various detection schemes, revealing insights into their efficacies. We discuss opportunities and challenges for OTFS in high mobility, setting the stage for future research and development in this field.
\end{abstract}

\begin{IEEEkeywords}
    OTFS, doubly selective channels, Doppler effect, OFDM, high-mobility scenarios.
\end{IEEEkeywords}
\IEEEpubidadjcol

\section{Introduction}
Orthogonal frequency division multiplexing (OFDM) has dominated the physical layer of modern wireless communication, being widely utilized in wireless LAN (e.g. WiFi), and cellular networks (e.g. 4G and 5G) owing to its high spectrum efficiency and robustness to multipath channel distortion \cite{Survey_5G_1,Survey_5G_2,4607239,1266914,6923528}. Despite its advantages and success, OFDM continues to face significant application challenges, including significant out-of-band emissions (OoBE), a substantial cyclic prefix (CP) overhead under large delay spread, and sensitivity to strong Doppler effect in high-mobility scenarios. To tackle spectrum efficiency issues due to OoBE and CP, alternatives to OFDM have been proposed over the years. They include filter bank multi-carrier (FBMC) \cite{FBMC_2}, filtered orthogonal frequency division multiplexing (F-OFDM) \cite{F_OFDM}, generalized frequency division multiplexing (GFDM) \cite{GFDM_1,GFDM_2}, and universal filtered multi-carrier \cite{UFMC_1,UFMC_2}. More recently, wireless communication serving high-mobility scenarios, such as in high-speed railways, vehicular networks, and unmanned aerial or underwater vehicles (UAVs or UUVs), have become a major design consideration \cite{7383229,Survey_5G_4}. To this end, orthogonal time frequency space (OTFS) modulation presents a promising format \cite{OTFS_Intro1, OTFS_1}. {OTFS was first introduced by Monk et al. \cite{monk2016otfsorthogonaltime} in 2016 to address the challenges in high-mobility wireless channels. Unlike traditional schemes such as OFDM and OFDMA, which operate in the time-frequency (TF) domain, OTFS maps data symbols onto the delay-Doppler (DD) domain. By doing so, OTFS exploits the delay-Doppler diversity of high mobility channels to combat Doppler-induced channel fading, as demonstrated in early studies such as \cite{OTFS_Div1, OTFS_Div2,10818747}. Consequently, OTFS is specifically tailored to mitigate high-speed channel fading due to the Doppler effect, and offers a novel modulation for wireless transmission systems encountering high-mobility networking scenarios \cite{OTFS_Comp1}. Over time, further research findings identified additional benefits of OTFS, including a lower peak-to-average power ratio \cite{OTFS_PAPR1,OTFS_PAPR2} compared to traditional OFDM, which simplifies the power amplifier design and improves transmit power efficiency. Efforts to design more practical OTFS implementations have led to significant advances, including the integration of OTFS with multiple-input multiple-output (MIMO) system by Kollengode et al.\cite{ramachandran2018mimo}, effective receiver designs by Raviteja et al.\cite{Detector_11}, channel estimation algorithm proposed by Raviteja et al.\cite{CE_1}, and uplink synchronization method by Alok et al.\cite{9241423}, among others.}

\IEEEpubidadjcol

\begin{figure}[t]
	\centering
	{\includegraphics[width = 0.46\textwidth]{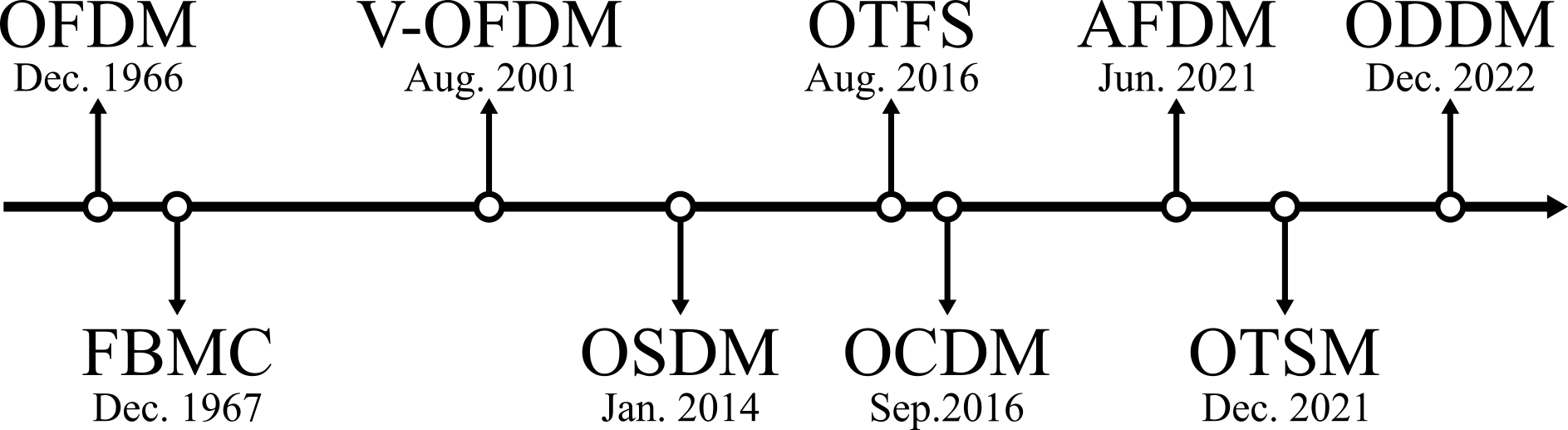}}\\
	\caption{{Timeline of variants modulation schemes related to OTFS.}}
	\label{fig:timeline}
\end{figure}

The past few years have witnessed a number of reported OTFS successes \cite{murali2018otfs,ramachandran2018mimo,CE_1,long2019low,OTFS_2,li2021performance,ge2021receiver,9110823,9181410,9362336,10159363}. Despite many demonstrations of potential OTFS advantages in high-mobility scenarios, multiple proposed variants of OTFS have also demonstrated similar characteristics, often claiming stronger or equivalent benefits in high mobility application scenarios. In fact, the emergence of many OTFS variants, along with their respective claims of performance superiority, tends to create substantial confusion in the literature with respect to their connections, similarities, and differences. For instance, the orthogonal signal-division multiplexing (OSDM) \cite{6495722,7299707,OSDM_1}, initially proposed to address the doubly spread channel in underwater acoustic communications, and Vector OFDM (V-OFDM) \cite{VOFDM_1}, designed to lower CP overhead, share the same expressions with OTFS when using rectangular transceiver pulses and a Nyquist sampling rate. Additionally, recently proposed orthogonal time-sequency multiplexing (OTSM)\cite{OTSM_1} and orthogonal delay-Doppler division multiplexing (ODDM)\cite{ODDM_1}, both for high-mobility communications, follow similar principles as OTFS. These schemes often declare similar or superior performance compared to OTFS in high-mobility scenarios. Meanwhile, another perspective has led to the proposal of chirp-based multi-carrier modulation schemes, such as orthogonal chirp division multiplexing (OCDM)\cite{OCDM_1} and affine frequency division multiplexing (AFDM)\cite{9473655,AFDM_1} for high-mobility communications. These applications of chirp signals to spread data symbols over the entire time-frequency domain are in fact related to the idea of inverse symplectic finite Fourier transform (ISFFT) in OTFS. {A timeline of these modulation schemes is illustrated in Fig.~\ref{fig:timeline}, highlighting their evolution and relationships.}

In view of the multiple variants of OTFS related modulations in the literature and respective assertions with respect to their performance edges, it is vital to present a comprehensive overview to understand their connections and comparisons for the benefit of their effective practical utilization in wireless communication systems. The goal of this work is to clearly analyze and highlight the relationships among the aforementioned OTFS variants. We aim to provide a common thread that unifies these OTFS related modulations and variants, to crystallize their relationship, to highlight their respective strengths and weaknesses, and to elucidate possible confusion with respect to the principles of OTFS. 

\begin{table*}[t!]
	\caption{{Comparison between this paper and Existing Surveys and Tutorials}}
	\begin{tabular}{|p{0.16\textwidth}|p{0.8\textwidth}|}
		\hline
		\multicolumn{1}{|c|}{\textbf{Paper}}                             & \multicolumn{1}{c|}{\textbf{Main Contribution}}                                                                                                                                           \\ \hline
		Wei et al. \cite{OTFS_Survey1}               & OTFS features, challenges, and potentials in next-generation wireless networks.                                                                                           \\
		Yuan et al. \cite{New_survey}                & OTFS fundamentals, transceiver designs, applications, challenges and future directions in next-generation wireless networks.                                                                                       \\
		Hong et al. \cite{OTFS_BOOK}				& Comprehensive explanation of the core concepts, mathematical foundations, and practical applications of OTFS. 																\\
		Zhang et al. \cite{OTFS_Survey4}             & Discusses inter-cell interference and existing anti-jamming solutions for OTFS systems.                                                                                                               \\
		Xiao et al. \cite{OTFS_2}                    & Explores OTFS as a physical layer waveform for IoT applications.                                                                                                                                  \\
		Shi et al. \cite{OTFS_Survey2}               & Advocates OTFS for low Earth orbit (LEO) satellite communications.                                                                                                                       \\
		M. Aldababsa et al. \cite{OTFS_survey_add1} & Focuses on MIMO-OTFS and the integration of OTFS with multiple access techniques (e.g., NOMA).                                                                                           \\
		Cai et al. \cite{8085125}                     & Reviews promising modulation and multiple access techniques for 5G networks.                                                                                                             \\
		Shtaiwei et al. \cite{OTFS_Survey5}          & Highlights the integrated sensing and communication (ISAC) benefits of OTFS.                                                                                                  \\
		Lin et al. \cite{OTFS_Survey3}               & Introduces the properties of ODDM modulation.                                                                                                                                            \\
		Zhou et al. \cite{Survey_5G_3}              & Discusses relationships between different modulation schemes; analyzes BER performance to favor AFDM.                                                             \\
		\textbf{Our Paper}                                             & Provides a unified view of OTFS and related modulation schemes, highlights connections and differences among OTFS variants, addresses confusion in the literature, and provides detailed introduction and analysis of detection schemes and basic performance comparisons.\\
		\hline
	\end{tabular}
	\label{table:Surveys}
\end{table*}

There already exist several survey or tutorial papers on OTFS that contribute to the understanding of OTFS modulation. Nevertheless, they do not address the different OTFS variants and help understand their connections. For example, the survey by Wei et al. \cite{OTFS_Survey1} and by Yuan et al. \cite{New_survey} provide an overview of OTFS, including its features, challenges, and potential applications in next-generation wireless networks. {Hong et al. \cite{OTFS_BOOK} offer a comprehensive explanation of the core concepts, mathematical foundations, and practical applications of OTFS.} Zhang et al. \cite{OTFS_Survey4} discuss the impact of interference and introduce the existing solution to overcome inter-cell interference for OTFS systems. Xiao et al. \cite{OTFS_2} focus on the outlook of OTFS as a physical layer for the Internet of Things (IoT). Shi et al. \cite{OTFS_Survey2} advocate the suitability of OTFS for low earth orbit (LEO) satellite communications. M. Aldababsa et al. \cite{OTFS_survey_add1} explore the multiple-input multiple-output (MIMO)-OTFS systems and the integration of OTFS with multiple access techniques, such as non-orthogonal multiple access (NOMA). In \cite{8085125}, Cai et al. provide an in-depth overview of the most promising modulation techniques and multiple access schemes for 5G networks. Another paper by Shtaiwei et al. \cite{OTFS_Survey5} considers the integrated sensing and communication (ISAC) benefits of OTFS frameworks. On the other hand, Lin et al. \cite{OTFS_Survey3} focus on introducing the properties of ODDM modulation. Zhou et al. \cite{Survey_5G_3} also discuss the relationship of various modulation schemes and attempt to establish a unified framework by introducing the transformation relations between different domains. However, they only analyze the BER performance in one experiment to claim that AFDM is the best candidate waveform for the next-generation wireless networks, which is unilateral. {We summarize and compare the contributions of our work with these existing survey and tutorial papers in Table~\ref{table:Surveys}. }Unlike these and other existing OTFS surveys and tutorials, this paper provides a clear insight into the various OTFS related modulations, by aiming to overcome possible confusion and eliminate potential misunderstanding with respect to these similar modulation schemes. We also provide a more detailed analysis of detection schemes and some basic performance comparisons. 

The following sections delve into these aspects in detail, beginning with conventional modulation models for OFDM and filter bank multi-carrier (FBMC) in Section~\ref{SEC:modulation}. This paves the foundation for OTFS and variants by demonstrating that OTFS can be viewed as FBMC utilizing an ISFFT precoder. Next, we present multiple modulations with similar characteristics as OTFS in Section~\ref{SEC:alternatives}. We demonstrate that both Vector OFDM and OSDM can be seen as a special case of OTFS with rectangular transceiver pulses and Nyquist sampling rate. While not identical to OTFS, OTSM and ODDM follow the same principle of placing data symbols in {non-TF} domains to enhance robustness against channel fading due to high-speed Doppler effects. We present numerical comparisons of these modulations in terms of the power spectrum density (PSD) and receiver bit error rate (BER) in Section~\ref{SEC:PSD}.

We then turn our attention to analyze effects that different channel models have on the OTFS related modulations. Section~\ref{SEC:channelMtx} presents effective system models for the covered modulation schemes under time-selective and frequency-selective fading channels that feature multipath fading models with individual path gain, delay, and Doppler frequency shift. We further illustrate the characteristics of the ``doubly-selective'' fading channels to demonstrate the effects of Doppler shift and multipath delay on OTFS-related signals at wireless receivers. Under such doubly-selective fading channels, we demonstrate a unifying signal detection model to capture these aforementioned modulation schemes under a common receiver architecture\footnote{The Appendix describes another class of chirp-based modulation schemes which also align with OTFS principles by spreading data symbols across the entire time-frequency domain.}.

Upon establishing a common receiver architecture for OTFS and variants, we provide an introduction of several common detection schemes in Section~\ref{SEC:detector}. These OTFS receivers belong to the simpler linear detection methods and the more complex non-linear detection methods. We focus particularly on several practical non-linear detection concepts including decision feedback, message-passing, AI-enhancement, and cross-domain detections. Our experiments further compare their BER performance under typical fading channels. Our presentation shows that designing efficient detectors to balance error performance and computational complexity is important and a challenge for OTFS and related modulations.

\begin{figure}[h!]
	\centering
	{\includegraphics[width = 0.44\textwidth]{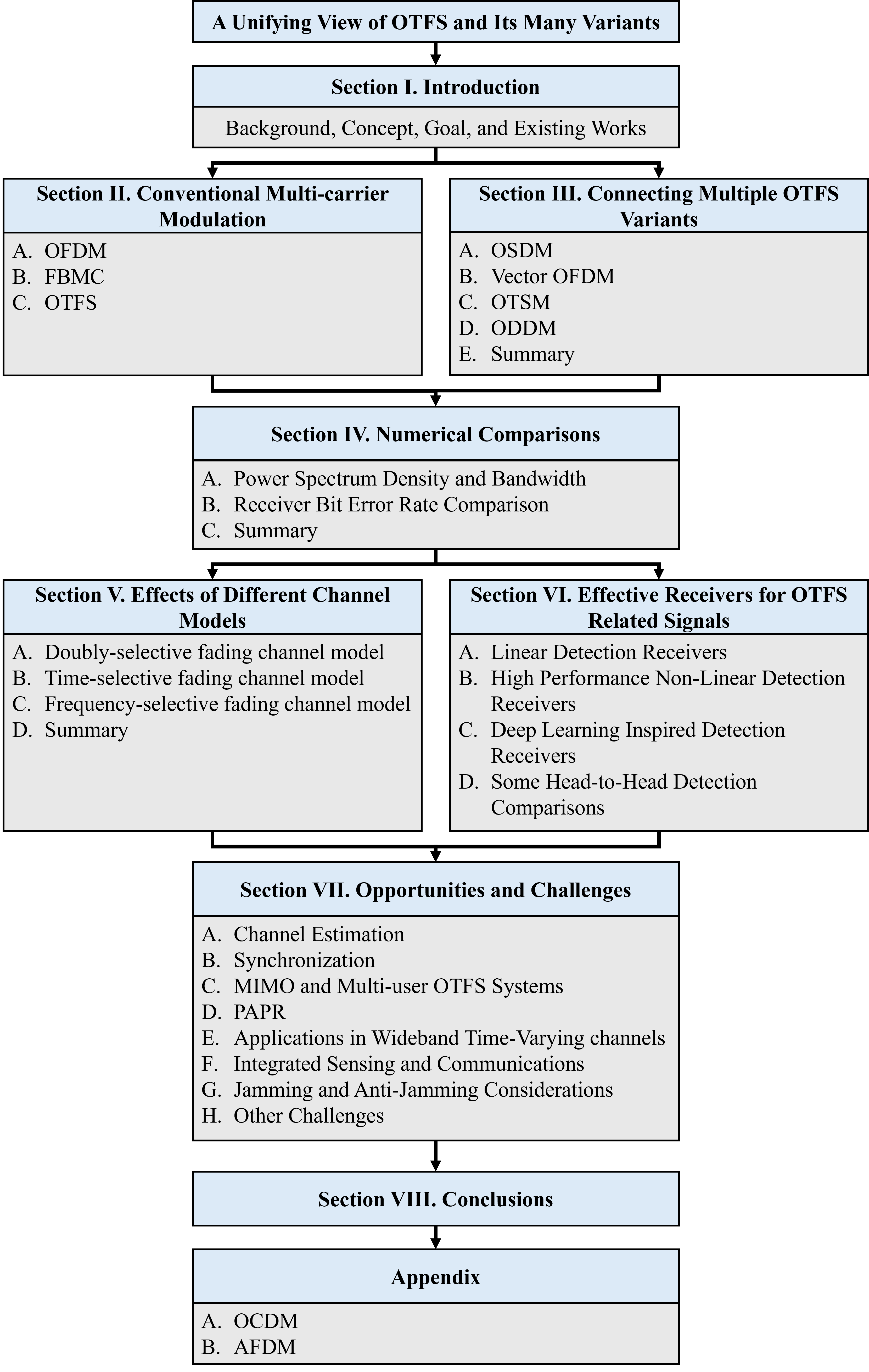}}\\
	\caption{Diagram of overall structure of this paper.}
	\label{fig:structure}
\end{figure}

Finally, we outline the opportunities and challenges associated with OTFS-related modulation schemes designed for high-mobility communications in Section~\ref{SEC:Opportunities}. We identify areas for further research and development to enhance their performance in {high-mobility} environments before presenting concluding remarks. {The overall structure of this paper is illustrated in Fig.~\ref{fig:structure}, and a complete list of acronyms used in this paper is provided in Table~\ref{table:acronyms}.}

\section{Conventional Multi-Carrier Modulation}
\label{SEC:modulation}

Before the introduction of OTFS, this section first describes the widely used multi-carrier modulation schemes, including OFDM and FBMC. OFDM paves the foundation of conventional multi-carrier modulation schemes, whereas FBMC can be seen as a generalization of OFDM. OTFS can be viewed as a precoded FBMC that uses ISFFT as a precoder.

\begin{table*}
	\centering
	\caption{List of Acronyms}
	\label{table:acronyms}
	\begin{tabular}{l l | l l}
		\hline
		\textbf{Acronyms} & \textbf{Definitions} & \textbf{Acronyms} & \textbf{Definitions}\\ \hline
		AEE-OTFS & Autoencoder-based enhanced OTFS & mmWave & Millimeter-wave \\ 
		AFDM & Affine frequency division multiplexing & MP & Message passing \\ 
		AMP & Approximate message passing  & MU-OTFS & Multi-user OTFS \\ 
		AoA & Angle-of-arrival  & NBI & Narrowband interference \\ 
		BP & Belief propagation  & NLE & Nonlinear estimator \\ 
		CAMP & Convolutional AMP  & NOMA & Non-orthogonal multiple access \\ 
		CD-MAMP & Cross-domain MAMP  & OAMP & Orthogonal approximated message passing \\ 
		CD-OAMP & Cross-domain OAMP  & OCDM & Orthogonal chirp division multiplexing \\ 
		CFO & Carrier frequency offset  & ODDM & Orthogonal delay-Doppler division multiplexing \\ 
		CP & Cyclic prefix  & ODSS & Orthogonal delay scale space \\ 
		CPP & Chirp-periodic prefix  & OFDMA & Orthogonal frequency division multiple access \\ 
		CSI & Channel state information  & OOB & Out-of-band \\ 
		DAFT & Discrete affine Fourier transform & OoBE & Out-of-band emissions \\ 
		DD & Delay-Doppler  & OQAM & Offset quadrature amplitude modulation \\ 
		DFE & Decision feedback equalization  & OSDM & Orthogonal signal-division multiplexing \\ 
		DFnT & Discrete Fresnel transform  & OTFS & Orthogonal time frequency space \\ 
		DLID & Deep-learning inspired detection  & OTFS-REC & OTFS with rectangular pulses \\ 
		DSE & Doppler squint effect  & OTFS-SRRC & OTFS with square-root raised cosine pulses \\ 
		EP & Expectation propagation  & OTSM & Orthogonal time-sequency multiplexing \\ 
		FBMC & Filter bank multi-carrier  & PAPR & Peak-to-average power ratio \\ 
		F-OFDM & Filtered orthogonal frequency division multiplexing  & PIN & Periodic impulse noise \\ 
		GFDM & Generalized frequency division multiplexing  & PSD & Power spectrum density \\ 
		GMP & Gaussian message passing  & PS-OFDM & Pulse-shaped OFDM \\ 
		GNN & Graph neural network  & RSMA & Rate splitting multiple access \\ 
		HRIS & Hybrid RIS  & SFFT & Symplectic finite Fourier transform \\ 
		ICF & Iterative clipping and filtering  & SMT & Staggered multitone \\ 
		IID & Independent and identically distributed  & SRRC & Square-root-raised cosine \\ 
		IoT & Internet of Things  & TF & Time-frequency \\ 
		ISAC & Integrated sensing and communication  & TO & Timing offset \\ 
		ISFFT & Inverse symplectic finite Fourier transform  & UAMP & Unitary approximate message passing \\ 
		LE & Linear estimator  & VAMP & Vector approximate message passing \\ 
		LSMR & Least squares minimum residual  & V-OFDM & Vector OFDM \\ 
		MAMP & Memory AMP & WHT & Walsh-Hadamard transform \\ 
		MF & Matched filter & ZP & Zero-padded \\  \hline
	\end{tabular}
\end{table*}

\subsection{OFDM}

\begin{figure}[h!]
    \centering
    \subfloat[]{\includegraphics[width = 0.44\textwidth]{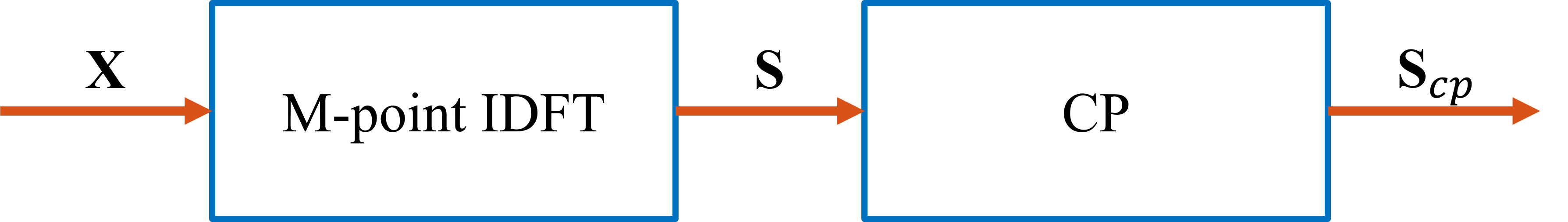}}\\
    \subfloat[]{\includegraphics[width = 0.44\textwidth]{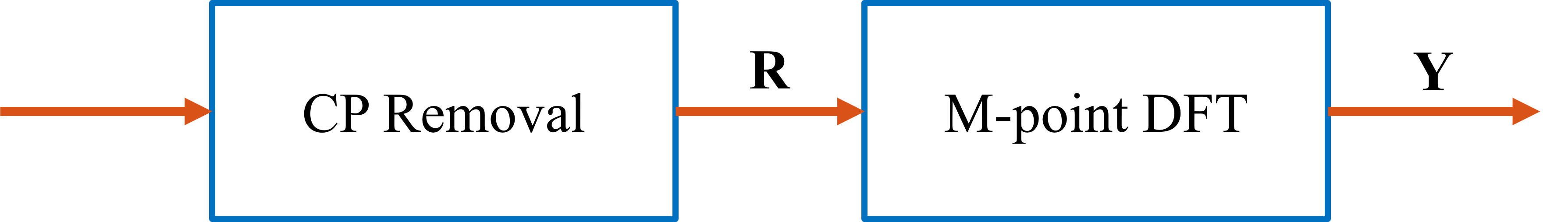}}\\
    \caption{Diagram of OFDM: (a) modulation; (b) demodulation.}
    \label{fig:ofdm}
\end{figure}

As a major legacy modulation technique, OFDM offers several attractive properties such as simple channel estimation, low-complexity equalization, efficient hardware implementation, easy combination with MIMO transmission, and backward compatibility. {As shown in Fig.~\ref{fig:ofdm}}, $\mathbf{X}=\{x_{m,n}\} \in \mathbb{A}^{M\times N}$ is the complex transmitted symbol matrix in the time-frequency (TF) domain, where element $x_{m,n}$ denotes the $m$-th subcarrier complex symbol transmitted at the $n$-th OFDM signal block (often known as the $n$-th OFDM symbol), and $\mathbb{A}$ denotes a finite modulation alphabet (e.g., phase shift keying (PSK) and quadrature amplitude modulation (QAM) symbols). The $n$-th transmitted OFDM signal block containing $M$ data symbols can be written as 
\begin{equation}
    s_n(t) = \sum_{m=0}^{M-1}x_{m,n}e^{j2\pi m \Delta f t},\quad 0\leq t \leq T, 
\end{equation}
where $M$, $T$ and $\Delta f$ denote the number of subcarriers, the symbol duration, and the subcarrier spacing of the OFDM system, respectively. Here, we adapt the orthogonal condition $T\Delta f = 1$. Let $N$ be the number of temporal slots spanned by each OFDM block, then the transmitted signal is written as
\begin{equation}
    s(t) = \sum_{n=0}^{N-1}s_n(t-nT) = \sum_{n=0}^{N-1}\sum_{m=0}^{M-1}x_{m,n}e^{j2\pi m \Delta f (t-nT)}.
\end{equation}
Let $s_{m,n}$ be the sampled transmitted signal associated to the $m$-th subcarrier in $n$-th block with the sampling period of $T/M$. Then, the discrete transmit signal matrix $\mathbf{S}=\{s_{m,n}\} \in \mathbb{C}^{M\times N}$ can be written as
\begin{equation}
    \mathbf{S} = \mathbf{F}_M^H \mathbf{X},
\end{equation}
where $\mathbf{F}_M\in \mathbb{C}^{M\times M}$ is the normalized $M$-point discrete Fourier transform (DFT) matrix. Furthermore, the discrete transmit signal vector $\mathbf{s}=\text{vec}\{\mathbf{S}\}\in \mathbb{C}^{MN\times 1}$ is
\begin{equation}
    \mathbf{s} = (\mathbf{I}_N \otimes \mathbf{F}_M^H) \mathbf{x},
\end{equation}
where $\mathbf{x}=\text{vec}\{\mathbf{X}\}\in \mathbb{A}^{MN\times 1}$ is the transmit symbol vector. {In practical communication systems, the value of $M$ can exceed 2000, resulting in a high PAPR for OFDM signals. Additionally, OFDM is sensitive to channel Doppler and carrier frequency offset (CFO), which cause the inter-carrier interference (ICI) of OFDM signals. Consequently, OFDM may not be the best choice for high-mobility communications where channel Doppler is significant.}

The CP for OFDM is adding to each column of $\mathbf{S}$, resulting in $\mathbf{S}_{cp}\in \mathbb{C}^{(M+N_{cp})\times N}$, where $N_{cp}$ is the length of the CP, which should be sufficient to ensure that the time duration of CP exceeds the maximum channel delay spread to tackle the inter-block interference. {Such a long CP length reduces the spectral efficiency of OFDM compared to other modulation schemes, which will be discussed later. }$\mathbf{S}_{cp}$ is sequentially transmitted over the channel after the vectorization.

The receiver first removes the CP from the received signal. Let $\mathbf{R}\in \mathbb{C}^{M\times N}$ be the reshaped sampled received signal after CP removal. We further transform $\mathbf{R}$ back to the frequency domain signal $\mathbf{Y} \in \mathbb{C}^{M\times N}$ by applying $M$-point DFT  
\begin{equation}
    \mathbf{Y} = \mathbf{F}_M \mathbf{R},
\end{equation} 
where the column vectors of $\mathbf{Y}$ denote the received OFDM symbols. By vectorizing $\mathbf{Y}$, we have an equivalent expression of
\begin{equation}
	\mathbf{y} = \left(\mathbf{I}_N \otimes \mathbf{F}_M \right)\mathbf{x},
\end{equation}
where $\mathbf{y}=\text{vec}({\mathbf{Y}})\in\mathbb{C}^{MN\times 1}$ is the discrete received signal vector. We will discuss the details and performance of different detectors operating on $\mathbf{y}$ in Section~\ref{SEC:detector}.

\subsection{FBMC}

\begin{figure}[h!]
    \centering
    \subfloat[]{\includegraphics[width = 0.48\textwidth]{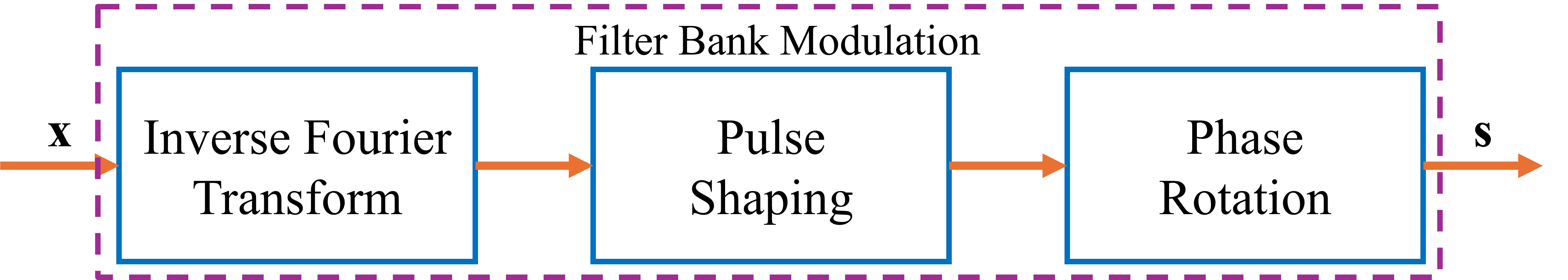}}\\
    \subfloat[]{\includegraphics[width = 0.48\textwidth]{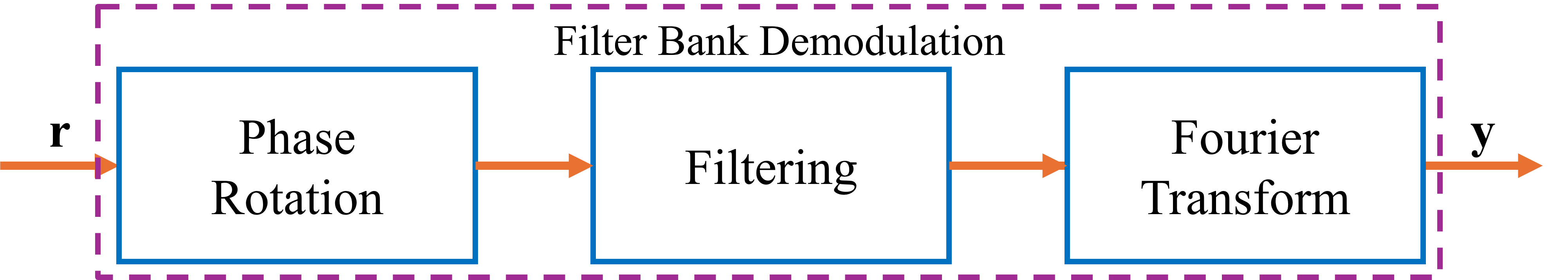}}\\
    \caption{Diagram of FBMC-OQAM: (a) modulation; (b) demodulation.}
    \label{fig:fbmc}
\end{figure}

OFDM uses a rectangular time-domain pulse shaping, which amounts to a frequency domain sinc function with large side lobes, resulting in high out-of-band emission. For this reason, researchers have been looking for waveforms that support variable and controllable pulse shaping to achieve a better trade-off between time-frequency localization. FBMC is a promising solution. Although FBMC has many different variants, we will mainly focus on offset quadrature amplitude modulation (OQAM) as the common FBMC representation. Following Fig.~\ref{fig:fbmc}, the transmit signal of FBMC-OQAM is \cite{FBMC_2,FBMC_6}
\begin{equation}
\label{Eq:FBMC1}
    s(t) = \sum_{n=0}^{N-1}\sum_{m=0}^{M-1}p_{m,n}(t)x_{m,n},
\end{equation}
with the basis pulse
\begin{equation}
\label{Eq:FBMC2}
    p_{m,n}(t) = g(t-nT)e^{j2\pi m\Delta f(t-nT)}e^{j\frac{\pi}{2}(m+n)},
\end{equation}
which combines inverse Fourier transform, pulse shaping, and phase rotation.
Substitute Eq.\eqref{Eq:FBMC2} into Eq.\eqref{Eq:FBMC1}, we have
\begin{equation}
\label{Eq:FBMC3}
    s(t) = \sum_{n=0}^{N-1}\sum_{m=0}^{M-1}g(t-nT)x_{m,n}e^{j2\pi m\Delta f(t-nT)}e^{j\frac{\pi}{2}(m+n)},
\end{equation}
where $M$ is the number of subcarriers, $x_{m,n}$ is the transmitted symbol associated with the $(m,n)$-th resource element in the TF domain. $T$ and $\Delta f$ are two constants that control the actual time and frequency spacing among the transmitted symbols $x_{m,n}$, and $g(t)$ is the underlying shaping pulse equivalent to a prototype filter response. One common choice of $g(t)$ is based on Hermite polynomials $H_n(\cdot)$, as provided in \cite{FBMC_6,FBMC_7}
\begin{equation}
    g(t) = \frac{1}{\sqrt{T_0}}e^{-2\pi (\frac{t}{T_0})^2}\sum_{\substack{i=\{0,4,8,\\12,16,20\}}}a_i H_i(2\sqrt{\pi}\frac{t}{T_0}),
\end{equation}
where $T_0$ is the basic time spacing for the prototype filter, and the value of coefficients $a_i$ are given in \cite{FBMC_6,FBMC_7} as:
\begin{equation}
    \begin{aligned}
        a_0 &= 1.412692577 \quad &a_{12} &=  -2.2611\cdot 10^{-9}\\
        a_4 &= -3.0145\cdot 10^{-3} \quad &a_{16} &=   -4.4570\cdot 10^{-15}\\
        a_8 &= -8.8041 \cdot 10^{-6} \quad &a_{20} &=  1.8633\cdot 10^{-16}
    \end{aligned}
\end{equation}
This prototype filter $g(t)$ guarantees orthogonality of the basis pulses for a time spacing of $T=T_0$ and a frequency spacing of $\Delta f={2}/{T_0}$, so that $T\Delta f=2$. However, a $T\Delta f$ value greater than one indicates a reduced spectral efficiency. On the other hand, according to the Balian-Low theorem \cite{FBMC_BLTheorem}, there exists no possible pulse localized in both time and frequency domains while achieving the maximum spectral efficiency requirement of $T\Delta f=1$. FBMC-OQAM addresses this issue by transmitting only the real-valued symbols, which relaxes the orthogonality condition to a real orthogonality condition. Consequently, both the time spacing and frequency spacing can be reduced by a factor of two, resulting in $T=T_0/2$, $\Delta f={1}/{T_0}$, and $T\Delta f=\frac{1}{2}$. This achieves the same spectrum efficiency as OFDM without CP. {However, this complex filter design also increase the complexity of FBMC in implementation.}

Similarly, by sampling $s(t)$ uniformly at interval $T/M$, the sampled transmitted signal vector $\mathbf{s}\in \mathbb{C}^{N_s\times 1}$ becomes 
\begin{equation}
    \mathbf{s} = \mathbf{P}\mathbf{x},
\end{equation}
where $\mathbf{P}\in \mathbb{C}^{N_s\times MN}$ is the matrix of sampled $p_{m,n}(t)$, whose element is given by \cite{FBMC_6}
\begin{equation}
    P_{i,m+nM} = \sqrt{\frac{T}{M}}p_{m,n}\left(i\frac{T}{M}-3T_0\right),\, i=0,1,\cdots,N_s-1.
\end{equation}
Here, $N_s$ is the number of samples for $s(t)$. For $T=T_0/2$ and the sampling interval of $T/M$, $N_s=M(N+11)$. $s(t)$ is transmitted without adding the CP, which increases the spectral efficiency of FBMC.

At the receiver, the sampled received signal vector $\mathbf{r}\in \mathbb{C}^{N_s\times 1}$ is converted back to the frequency domain by the corresponding receive filter $\mathbf{P}^H$\cite{FBMC_2} via
\begin{equation}
    \mathbf{y} = \mathbf{P}^H\mathbf{r} \in \mathbb{C}^{MN\times 1}
\end{equation}
as the sampled received signal vector in the frequency domain. 

\subsection{OTFS}

\begin{figure*}[t!]
    \centering
    \includegraphics[width = 0.8\textwidth]{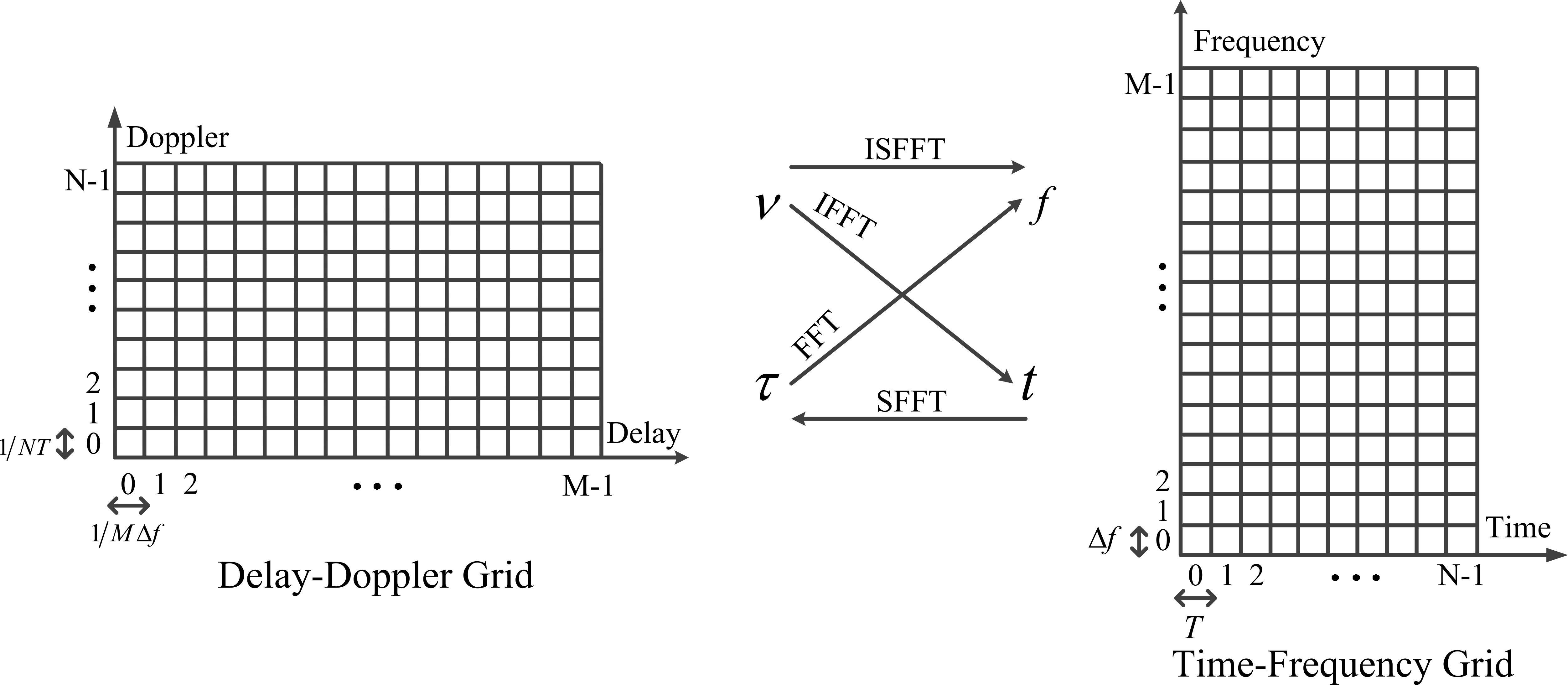}
    \caption{Relationship between delay-Doppler (DD) grid and time-frequency (TF) grid.}
    \label{fig:TFDDgrid}
\end{figure*}

\begin{figure}[h!]
    \centering
    \subfloat[]{\includegraphics[width = 0.48\textwidth]{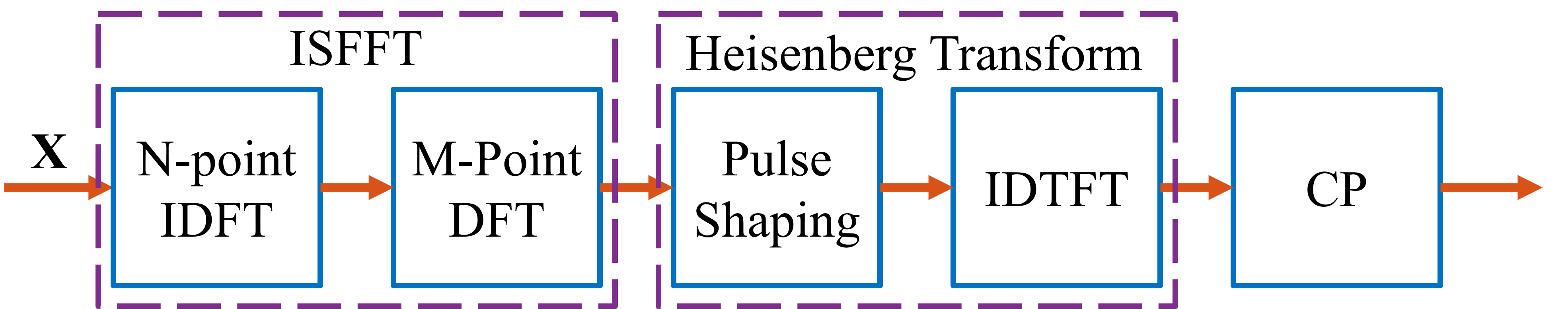}}\\
    \subfloat[]{\includegraphics[width = 0.48\textwidth]{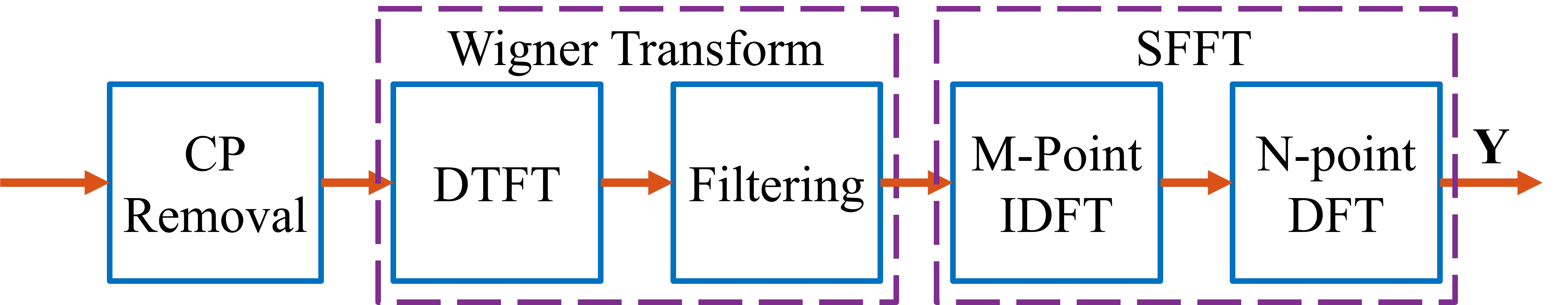}}\\
    \caption{Diagram of OTFS: (a) modulation; (b) demodulation.}
    \label{fig:otfs}
\end{figure}

Recently, OTFS has emerged as a promising modulation for high-mobility scenarios. In order to be robust in doubly selective fading channels, OTFS multiplexes information symbols in the 2-dimensional (2D) delay-Doppler (DD) domain instead of the TF domain in OFDM. As shown in Fig.~\ref{fig:TFDDgrid}, to achieve the maximum spectrum efficiency of $T\Delta f=1$, the lattice in DD domain of OTFS is denoted as
$$\Gamma \hskip -2pt =\hskip -2pt \left\{\hskip -2pt {\left(\frac{m}{{M\Delta f}},\frac{n}{{NT}}\right),m \!=\! 0, \cdots ,M \!-\! 1;n \!=\! 0, \cdots ,N \!-\! 1} \right\},$$
where $M$ and $N$ are the number of subcarriers and time slots; $T$ is the symbol period which is determined to be greater than the maximum multipath delay spread; $\Delta f = \frac{1}{T}$ is the subcarrier spacing which should be larger than the maximum Doppler spread. 

Fig.~\ref{fig:otfs} shows OTFS modulation in two steps. In the first step, OTFS uses ISFFT to transform DD domain symbols $x_{m,n}$ into TF domain symbols $U_{k,\ell},\forall k\in[0,M-1], \ell\in[0,N-1]$ as \cite{OTFS_1}
\begin{equation}
\label{Eq:OTFS1}
    U_{k,\ell} = \frac{1}{\sqrt{MN}}\sum_{m=0}^{M-1}\sum_{n=0}^{N-1}x_{m,n}e^{j2\pi (\frac{n\ell}{N}-\frac{mk}{M})}.
\end{equation}
ISFFT consists of an $N$-point IDFT along the Doppler dimension and a $M$-point DFT along the delay dimension. The $N$-point IDFT transforms Doppler domain signals into the time domain, while the $M$-point DFT transforms delay domain signals into the frequency domain. In the second step, OTFS transform TF domain symbols $U_{k,\ell}$ into time domain transmit signal $s(t)$ via Heisenberg transform:
\begin{equation}
\label{Eq:OTFS2}
    s(t) = \sum_{k=0}^{M-1}\sum_{\ell=0}^{N-1}g(t-\ell T)U_{k,\ell}e^{j2\pi k\Delta f(t-\ell T)},
\end{equation}
where $g(t)$ is the transmit pulse. Unlike the FBMC pulse shape, $g(t)$ in OTFS has the limit time duration of $T$, such that the total duration of $s(t)$ equals to $NT$. One common choice of $g(t)$ in OTFS is the rectangular pulse of duration $T$. 

Compared to Eq.~\eqref{Eq:FBMC3}, Eq.~\eqref{Eq:OTFS2} has the same form of expression if we ignore the last phase shift term in Eq.~\eqref{Eq:FBMC3} introduced by OQAM. Therefore, OTFS can be viewed as a precoded FBMC that uses ISFFT as the precoder. Substituting Eq.~\eqref{Eq:OTFS1} into Eq.~\eqref{Eq:OTFS2}, we can generate the OTFS modulation signal
\begin{multline}\label{Eq:OTFS_s}
s(t) = \frac{1}{\sqrt{MN}}\sum_{k=0}^{M-1}\sum_{\ell=0}^{N-1}\sum_{m=0}^{M-1}\sum_{n=0}^{N-1}g(t-\ell T)x_{m,n}\\ \cdot e^{j2\pi (\frac{n\ell}{N}-\frac{mk}{M})}e^{j2\pi k\Delta f(t-\ell T)}.
\end{multline}

By sampling $s(t)$ with the period of $T/M$, the inverse discrete time fourier transform (IDTFT) in Heisenberg transform Eq.~\eqref{Eq:OTFS2} becomes a $M$-point IDFT in frequency, which matches the $M$-point DFT in ISFFT. As a result, if the time duration of $g(t)$ is no larger than $T$, the discrete transmit signal matrix $\mathbf{S}\in \mathbb{C}^{M\times N}$ can be written as \cite{8516353}
\begin{equation}
\label{Eq:OTFS4}
    \mathbf{S}=\mathbf{G}\mathbf{X}\mathbf{F}_N^H,
\end{equation}
with element $\mathbf{S}_{k,\ell} = g(k\frac{T}{M})\sum_{n=0}^{N-1}x_{k,n}e^{j2\pi \frac{n\ell}{N}}$. $\mathbf{G}\in \mathbb{C}^{M\times M}$ is a diagonal matrix whose diagonal elements are sampled from $g(t)$, i.e., $\mathbf{G}=\text{diag}\left(g(0),g(\frac{T}{M}),\cdots, g(\frac{(M-1)T}{M}\right)$. Furthermore, the discrete transmit signal vector is given by
\begin{equation}
\label{Eq:OTFS5}
\mathbf{s}=\text{vec}\{\mathbf{S}\}= (\mathbf{F}_N^H \otimes \mathbf{G}) \mathbf{x}\, \in\,\mathbb{C}^{MN\times 1}.
\end{equation}

Compared to conventional OFDM, OTFS only needs to add one CP of length $N_{cp}$ for the entire transmission block, resulting in $s_{cp}\in\mathbb{C}^{(MN+N_{cp})\times 1}$, which has a higher spectrum efficiency than conventional OFDM.

At the receiver, the received signal is processed via a CP removal, leading to a time-domain signal $r(t)$. Signal $r(t)$ is then transformed back to the TF domain by using Wigner transform, expressed as
\begin{equation}
    Y(t,f) = \int g_{rx}^*(t'-t)r(t')e^{-j2\pi f(t'-t)}dt',
\end{equation}
where $g_{rx}(t)$ is the receiver pulse. This Wigner transform is the inverse of Heisenberg transform, which can be further decomposed into a DTFT and a pulse shaping filter. The baseband received signal $Y(t,f)$ can be converted into a matrix $\mathbf{Y}\in \mathbb{C}^{M\times N}$ by sampling as
\begin{align*}
     Y_{k,\ell} &= Y(t,f)|_{t=\ell T,f=k\Delta f}, \\
    \forall &\ell =0,\cdots,N-1, k=0,\cdots,M-1.
\end{align*}
The final step transforms $\mathbf{Y}$ into DD domain by symplectic finite Fourier transform (SFFT) as
\begin{align*}
    y_{m,n} = \frac{1}{\sqrt{MN}} \sum_{k=0}^{M-1}\sum_{\ell=0}^{N-1}Y_{k,\ell}e^{-j2\pi (\frac{n\ell}{N}-\frac{mk}{M})}.
\end{align*}
By uniformly sampling the received signal with the sampling interval $\frac{T}{M}$, the transformed receive signal vector in DD domain becomes
\begin{equation}
\label{Eq:OTFS_r}
    \mathbf{y} = (\mathbf{F}_N \otimes \mathbf{G}_{rx}) \mathbf{r},
\end{equation}
where $\mathbf{r}\in \mathbb{C}^{MN\times 1}$ is sampled received signal vector in time domain and $\mathbf{G}_{rx}\in \mathbb{C}^{M\times M}$ is a diagonal matrix whose diagonal elements are sampled from $g_{rx}(t)$, {i.e., $\mathbf{G}_{rx}=\text{diag}\left(g_{rx}(0),g_{rx}(\frac{T}{M}),\cdots, g_{rx}(\frac{(M-1)T}{M}\right)$.} This expression can be rewritten as
\begin{equation}
	\label{Eq:OTFS_r2}
	\mathbf{Y} = \mathbf{G}_{rx}\mathbf{R}\mathbf{F}_N,
\end{equation}
where matrix $\mathbf{R}\in \mathbb{C}^{M\times N}$ is the devectorized form of $\mathbf{r}$.

The choice of pulses $g(t)$ and $g_{rx}(t)$ is important for the performance of OTFS. One common selection is the rectangular pulse of a duration of $T$ for both $g(t)$ and $g_{rx}(t)$. In this case, both $\mathbf{G}$ and $\mathbf{G}_{rx}$ become identity matrices, which simplifies the modulation and demodulation process as well as the analysis. However, rectangular pulses do not perform well in out-of-band (OOB) control. In contrast, a pair of square-root-raised cosine (SRRC) pulses offers better performance. We will discuss the performance analysis of these two cases later in Sections~\ref{SEC:PSD} and \ref{SEC:channelMtx}.

\section{Connecting Multiple OTFS Variants}
\label{SEC:alternatives}

In this section, we will introduce other modulation schemes with similar characteristics to OTFS. We will show that vector OFDM and OSDM have the same expression with OTFS when using rectangular transceiver pulses and a Nyquist sampling rate. Additionally, we show that even schemes like OTSM and ODDM are not identical to OTFS, they follow the same principle of placing data symbols in a specific transform domain other than TF domain to enhance robustness against Doppler effects of high mobility systems.

\subsection{OSDM}

\begin{figure}[h!]
    \centering
    \subfloat[]{\includegraphics[width = 0.48\textwidth]{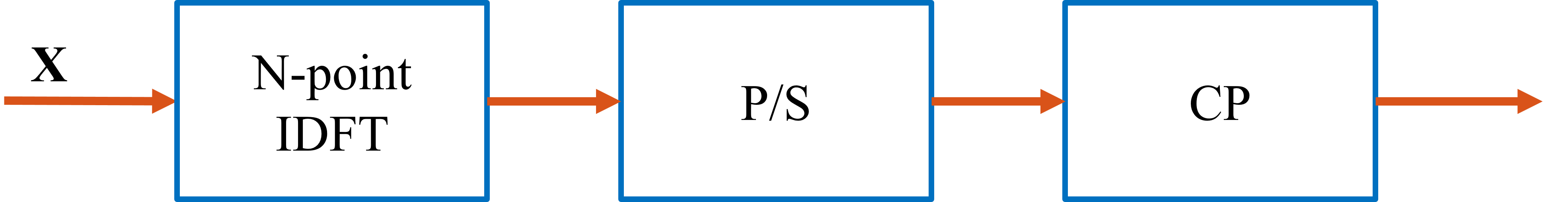}}\\
    \subfloat[]{\includegraphics[width = 0.48\textwidth]{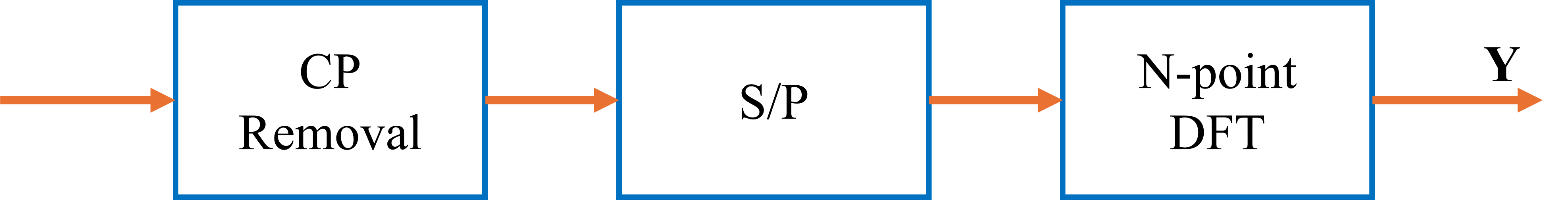}}\\
    \caption{Diagram of OSDM: (a)  modulation; (b)  demodulation.}
    \label{fig:osdm}
\end{figure}

OSDM partitions the transmitted block $\mathbf{X}\in \mathbb{A}^{M\times N}$ into $N$ symbol vectors of length $M$, i.e., 
\begin{equation}
    \mathbf{X}_n \!=\! [\mathbf{X}_{nM},\mathbf{X}_{nM+1},\cdots,\mathbf{X}_{nM+M-1}]^T,
    \; n=0,\cdots, N-1.
\end{equation}
These $N$ symbol vectors would form a $M\times N$ symbol matrix similar to the symbol matrix of OTFS in DD domain. {As shown in Fig.~\ref{fig:osdm},} the OSDM process can be expressed as \cite{6495722,OSDM_1}
\begin{equation}
    \label{Eq:OSDM1}
    \mathbf{s} = (\mathbf{F}_N^H \otimes \mathbf{I}_M)\mathbf{x},
\end{equation}
which has the same expression of OTFS in Eq.~(\ref{Eq:OTFS5}). Then, the signal is transmitted after adding the CP.

At the receiver, $\mathbf{r}\in \mathbb{C}^{MN\times 1}$ is the sampled received signal vector in time domain after CP removal, the block of OSDM received signal vector in time-frequency domain is given by
\begin{equation}
    \label{Eq:OSDM2}
    \mathbf{y} = (\mathbf{F}_N \otimes \mathbf{I}_M)\mathbf{r} \in 
     \mathbb{C}^{MN\times 1}.
\end{equation}
We can observe that Eqs.\eqref{Eq:OSDM1} and \eqref{Eq:OSDM2} are identical to the OTFS modulation and demodulation Eqs.\eqref{Eq:OTFS5} and \eqref{Eq:OTFS_r} when using rectangular pulses. Therefore, OSDM can be viewed as a special case of OTFS from the perspective of mathematical expressions.

\subsection{Vector OFDM}

\begin{figure}[h!]
    \centering
    \subfloat[]{\includegraphics[width = 0.48\textwidth]{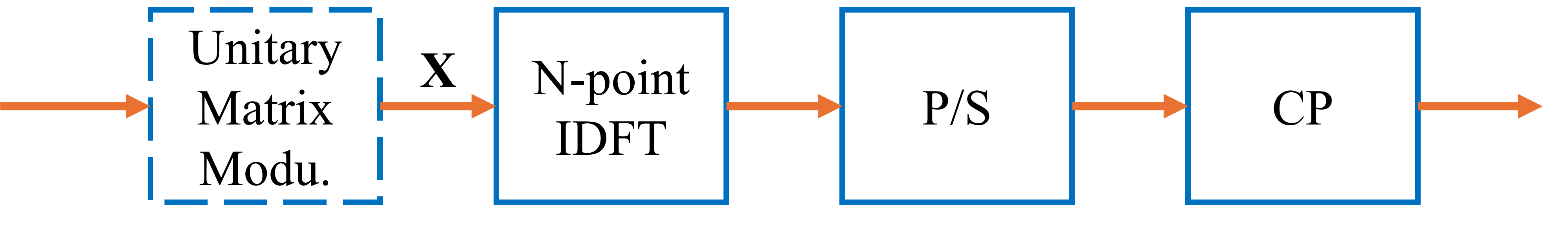}}\\
    \subfloat[]{\includegraphics[width = 0.48\textwidth]{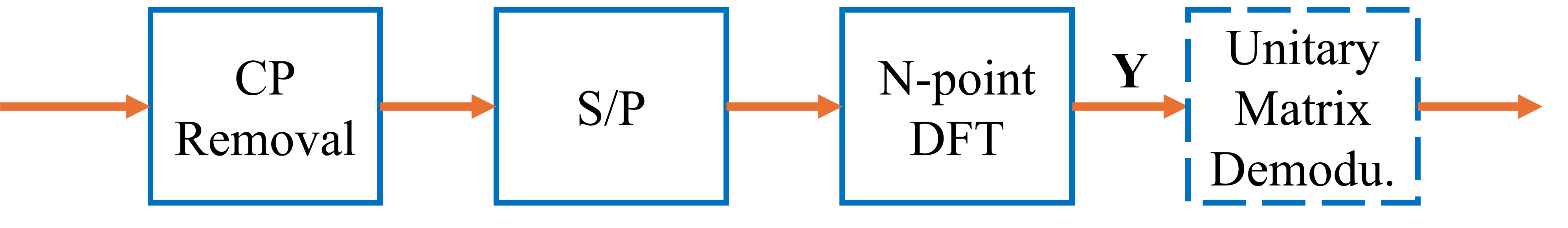}}\\
    \caption{Diagram of V-OFDM: (a) modulation; (b) demodulation.}
    \label{fig:vofdm}
\end{figure}

Unlike the conventional OFDM, Vector OFDM (V-OFDM) modulates symbols blockwise. In the initial design of V-OFDM~\cite{VOFDM_1}, the authors use a unitary matrix modulation to map the data bits into the transmit signal matrix $\mathbf{X}$. {As shown in Fig.~\ref{fig:vofdm},} given a transmit signal matrix of $\mathbf{X}\in \mathbb{A}^{M\times N}$, V-OFDM performs component-wise vector $N$-point IDFT over $\mathbf{X}$ written as \cite{VOFDM_1,VOFDM_2}
\begin{equation}
    \label{Eq:VOFDM1}
    \mathbf{S}=\mathbf{X}\mathbf{F}_N^H,
\end{equation}
with element $\mathbf{S}_{k,\ell} = \frac{1}{N}\sum_{n=0}^{N-1}x_{k,n}e^{j2\pi \frac{n\ell}{N}}$. This expression has the same form with Eq.~(\ref{Eq:OTFS4}) whenever $g(t)$ uses the rectangular pulse. Therefore, V-OFDM can be viewed as a variant of OTFS, despite it does not explicitly present DD domain modulation. 

The transmit signal is sent after adding the CP, whose time duration is longer than the maximum delay of the channel. At the receiver, V-OFDM applies an $N$-point DFT on the received signal matrix $\mathbf{R}\in \mathbb{C}^{M\times N}$ after CP removal, i.e., 
\begin{equation}
    \label{Eq:VOFDM2}
    \mathbf{Y} = \mathbf{R}\mathbf{F}_N.
\end{equation}
We can observe that Eqs.\eqref{Eq:VOFDM1} and \eqref{Eq:VOFDM2} are identical to the OTFS modulation and demodulation Eqs.\eqref{Eq:OTFS4} and \eqref{Eq:OTFS_r2} when using rectangular pulses. Therefore, by only viewing the process between $\mathbf{X}$ and $\mathbf{Y}$, V-OFDM can also be seen as a special case of OTFS from the perspective of mathematical expressions.

\subsection{OTSM}

\begin{figure}[h!]
    \centering
    \subfloat[]{\includegraphics[width = 0.48\textwidth]{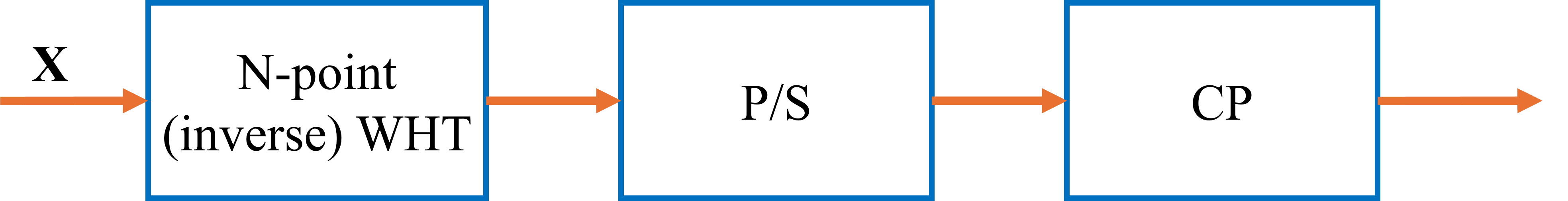}}\\
    \subfloat[]{\includegraphics[width = 0.48\textwidth]{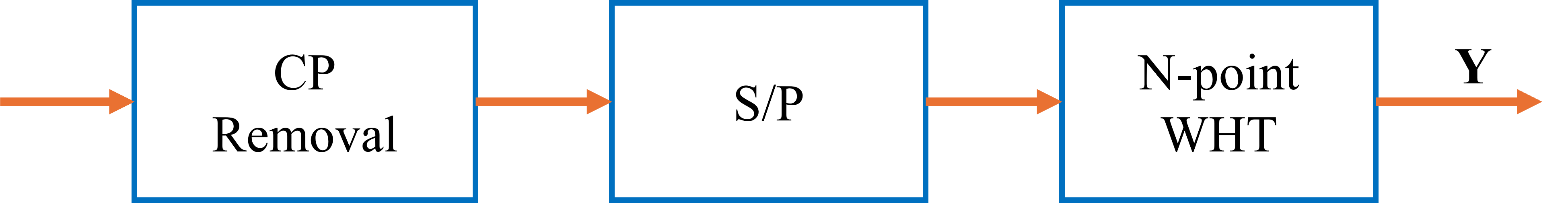}}\\
    \caption{Diagram of OTSM: (a) modulation; (b)  demodulation.}
    \label{fig:otsm}
\end{figure}

Unlike OTFS, OTSM puts the transmit signal matrix $\mathbf{X}\in \mathbb{A}^{M\times N}$ on the delay-sequency domain, where $M$ corresponds to the delay dimension and $N$ corresponds to the sequency dimension. Another difference between OTSM and OTFS is the transmit information symbol vector in OTSM $\mathbf{x}_{rw}\in \mathbb{A}^{MN\times 1}$ is a row-wise vectorization of $\mathbf{X}$. The total time duration and bandwidth of one transmitted signal frame in OTSM are $T_f = NT$ and $B=M\Delta f$, where $\Delta f = 1/T$, which is the same as OTFS. {As shown in Fig.~\ref{fig:otsm},} an $N$-point (inverse) Walsh-Hadamard transform (WHT) is applied on the sequency dimension of $\mathbf{X}$ as
\begin{equation}
	\mathbf{\Bar{X}} = \mathbf{X}\mathbf{W}_{N},
\end{equation}
where $\mathbf{W}_N$ is an $N$-point normalized WHT matrix. Note that the normalized inverse WHT is identical to the normalized WHT. {By using the real-valued WHT instead of complex-valued ISFFT and Heisenberg transform in OTFS, OTSM have a lower complexity.} Then $\mathbf{\Bar{X}}$ is column-wise vectorized into the transmit signal vector $\mathbf{s}\in \mathbb{C}^{MN\times 1}$. These two steps can be expressed as\cite{OTSM_1}
\begin{equation}
	\begin{aligned}
		\mathbf{s} 
		&= \mathbf{\Pi}\left(\mathbf{I}_M \otimes \mathbf{W}_N\right)\mathbf{x}_{rw}\\
		&= \left(\mathbf{W}_N \otimes \mathbf{I}_M\right)\left(\mathbf{\Pi}\,\mathbf{x}_{rw}\right),
	\end{aligned}
\end{equation}
where $\mathbf{\Pi}$ is the row-column interleaver matrix caused by the column-wise vectorization, corresponding to the parallel to serial block in Fig.~\ref{fig:otsm}. A direct comparison with OTFS exemplified by Eq.~(\ref{Eq:OTFS5}) shows a mere replacement of the $N$-point inverse Fourier transform by WHT in addition to a simple row-column interleaver matrix $\mathbf{\Pi}$ applied. A CP of length larger than the maximum delay is added to the time domain signal $\mathbf{s}$ before transmission.

The OTSM receiver also removes the CP from the received signal to yield a time domain received signal vector $\mathbf{r}\in \mathbb{C}^{MN\times1}$, which is column-wise reshaped into a $M$-by-$N$ matrix $\mathbf{\Bar{Y}} = \text{vec}^{-1}(\mathbf{r})$. Next, the delay-time domain matrix $\mathbf{\Bar{Y}}$ is sequentially transformed back into the delay-sequency domain by an $N$-point WHT as
\begin{equation}
    \mathbf{Y} = \left[\mathbf{y}_0,\mathbf{y}_1,\cdots, \mathbf{y}_{M-1}\right]^T = \mathbf{\Bar{Y}}\mathbf{W}_N \, \in \mathbb{C}^{M\times N},
\end{equation}
where $\mathbf{y}_i\in \mathbb{C}^{N\times 1}$ denotes the $i$-th row vector of $\mathbf{Y}$. The row-wise vectorized OTSM signal vector can be rewritten as 
\begin{equation}
\mathbf{y}_{rw}=\left[\mathbf{y}_0^T,\mathbf{y}_1^T,\cdots,\mathbf{y}_{M-1}^T\right]^T = \left(\mathbf{I}_M \otimes \mathbf{W}_N\right)\cdot\left(\mathbf{\Pi}^T\mathbf{r}\right).
\end{equation}

\subsection{ODDM}

\begin{figure}[h!]
	\centering
	\subfloat[]{\includegraphics[width = 0.48\textwidth]{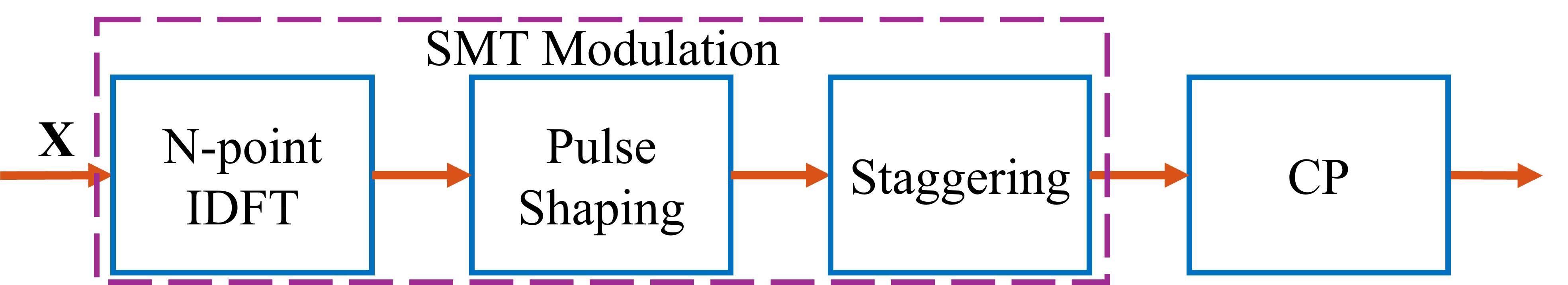}}\\
	\subfloat[]{\includegraphics[width = 0.48\textwidth]{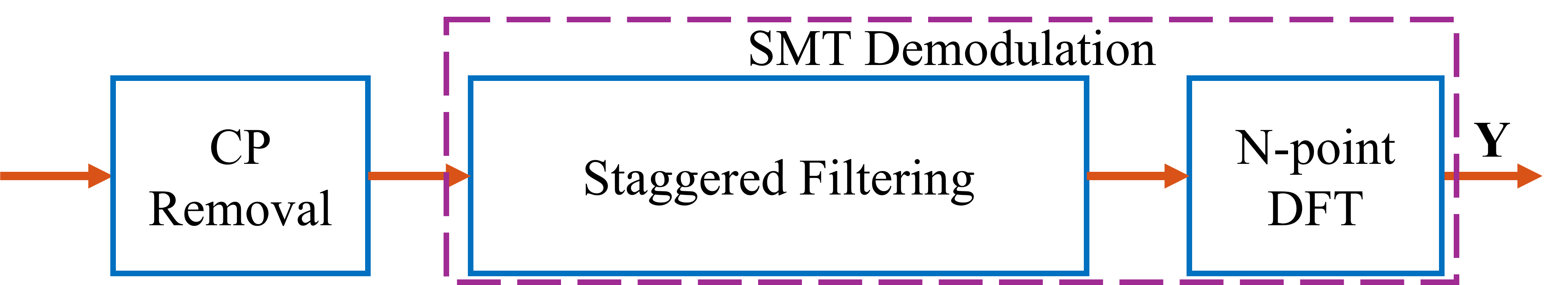}}\\
	\caption{Diagram of ODDM: (a)  modulation; (b) demodulation.}
	\label{fig:oddm}
\end{figure}
ODDM is another competitive waveform and independently developed by Hai et al. \cite{ODDM_1}. Similar to OTFS, ODDM modulation also places the data symbols on the DD domain as $\mathbf{X}\in \mathbb{A}^{M\times N}$. The main difference between ODDM and OTFS lies in how they convert DD domain symbols into the time domain. While OTFS uses the ISFFT and Heisenberg transform, ODDM introduces a ${T}/{M}$-interval stagger in this transform, which can be viewed as a type of staggered multitone (SMT) modulation. Fig.~\ref{fig:oddm} shows that ODDM applies pulse shape $g(t)$, thus, the time domain transmit signal is given by
\begin{equation}\label{Eq:ODDM_s}
	s(t) = \sum_{m=0}^{M-1}\sum_{n=0}^{N-1}\sum_{\ell=0}^{N-1}g\left(t-\frac{mT}{M}-\ell T\right)x_{m,n}e^{j2\pi \frac{n\ell}{N}}.
\end{equation}
One choice of $g(t)$ is a symmetric square-root Nyquist-I pulse with truncated support of $(-Q\frac{T}{M},Q\frac{T}{M})$, where $Q$ is an integer and $2Q\ll M$. This choice of $g(t)$ is ISI-free for the symbol interval of $\frac{T}{M}$ with a corresponding matched filter at the receiver. 

Taking uniform samples of the ODDM transmit signal in Eq.~\eqref{Eq:ODDM_s} with sampling interval $\frac{T}{M}$, the discrete representation of the $(k+M\ell)$-th sample is given by
\begin{equation}
	s[k+M\ell] = g(0)\sum_{n=0}^{N-1}x_{k,n}e^{j2\pi \frac{n\ell}{N}},
\end{equation}
which takes a similar form to the element in Eq.~\eqref{Eq:OTFS4}. 

Despite this similarity, the time domain transmit signal of ODDM is different from OTFS. The time domain transmit signal for OTFS in Eq.~\eqref{Eq:OTFS_s} can be rewritten as
\begin{multline}\label{Eq:OTFS_s_alternative}
	s(t) =\sum_{m=0}^{M-1}\sum_{n=0}^{N-1}\sum_{\ell=0}^{N-1}x_{m,n}e^{j2\pi\frac{n\ell}{N}}\\
	\cdot \left[\frac{1}{\sqrt{MN}}\sum_{k=0}^{M-1}g(t-\ell T) e^{j2\pi k\left(\Delta f(t-\ell T)-\frac{m}{M}\right)}\right].
\end{multline}
Comparing Eq.~\eqref{Eq:ODDM_s} and Eq.~\eqref{Eq:OTFS_s_alternative}, the ${T}/{M}$-interval stagger in ODDM shifts the shaping pulse $g(t-\frac{mT}{M}-\ell T)$ for different $m$, offering more flexibility in pulse shaping design for ISI reduction.

The time domain transmit signal in ODDM can be rewritten as \cite{ODDM_1}
\begin{align}
	\label{Eq:ODDM_s_alternative}
	s(t) & = \sum_{m=0}^{M-1}\sum_{n=0}^{N-1}u\left(t-\frac{mT}{M}\right)x_{m,n}e^{j2\pi \frac{n}{NT}(t-\frac{mT}{M})},\nonumber\\
	& -Q\frac{T}{M}\leq t\leq NT+(Q-1)\frac{T}{M}, 
\end{align}
where $u(t) = \sum_{\ell=0}^{N-1}g\left(t-\ell T\right)$ is the transmit pulse. This expression shows that the duration of the transmit signal for ODDM is longer than that of OTFS. This alternative representation is more aligned with the pulse-shaped OFDM (PS-OFDM) form, which simplifies the understanding and analysis of the signal, making it more accessible in comparison with the original Eq.~\eqref{Eq:ODDM_s}. When sampling at a period of $T_s$, the discrete signal vector of $N_t$ samples is represented by
\begin{equation}\label{Eq:ODDM_s_Matrix}
	\mathbf{s} = \mathbf{U}\mathbf{x} \in \mathbb{C}^{N_t\times 1},
\end{equation}
where $\mathbf{U}\in \mathbb{C}^{N_t\times MN}$ is the matrix of samples from $u\left(t-\frac{mT}{M}\right)e^{\frac{j2\pi n}{NT}(t-\frac{mT}{M})}$. When using the Nyquist sampling period $T_s = \frac{T}{M}$, the number of samples in the time domain equals $N_t = MN + 2Q - 1$ for the transmit signal without CP. The time domain signal is then transmitted after adding the CP.

The ODDM receiver removes the CP and generates a signal at the $n$-th subcarrier of the $m$-th ODDM symbol as
\begin{equation}\label{Eq:ODDM_r}
	y_{m,n} = \int r(t)u\left(t-\frac{mT}{M}\right)e^{-j2\pi \frac{n}{NT}(t-\frac{mT}{M})}dt.
\end{equation}
Let $\mathbf{r}\in \mathbb{C}^{N_t\times 1}$ be the sampled received signal vector in time domain after CP removal. Similar to Eq.~\eqref{Eq:ODDM_s_Matrix}, the matrix form of  Eq.~\eqref{Eq:ODDM_r} can be represented in the DD domain as
\begin{equation}\label{Eq:ODDM_s_Matrix2}
	\mathbf{y} = \mathbf{U}^H\mathbf{r}.
\end{equation}

\subsection{Summary}
{In this section, we have introduced and analyzed various modulation schemes that share characteristics similar to OTFS. We demonstrated that both OSDM and V-OFDM exhibit mathematical expressions identical to OTFS under specific conditions such as the use of rectangular transceiver pulses and Nyquist sampling rates. Additionally, we examined OTSM and ODDM, which, although not identical to OTFS, follow the same core principle of mapping data symbols into specific transform domains to improve robustness against Doppler effects in high-mobility environments. We compare the key attributes of these modulation schemes in Table~\ref{table:limitataions}.}

\begin{table*}[t]
	\centering
    \renewcommand{\arraystretch}{1.3}
	\setlength{\tabcolsep}{4pt}
	\caption{{Comparison of Key Attributes Across Different Modulation Schemes}}
	\label{table:limitataions}
	\begin{tabular}{|l|ccccccc|}
		\hline
		& \multicolumn{7}{c|}{\textbf{Main Attributes}}                                                                                                                                    \\ \hline
		\textbf{Modulation Scheme}                                        & \textbf{Spectral Efficiency} & \textbf{Complexity} & \textbf{PAPR} & \textbf{Doppler Robustness} & \textbf{Multipath Robustness} & \textbf{Anti-jamming} & \textbf{OoBE Control} \\ \hline
		CP-OFDM                                                           & Moderate                     & Low                 & High          & Sensitive                   & Robust                        & Weak                  & Bad                   \\ \hline
		FBMC                                                              & High                         & High                & High          & Moderate                    & Robust                        & Weak                & Good                  \\ \hline
		\begin{tabular}[c]{@{}l@{}}OTFS-REC\\ (V-OFDM, OSDM)\end{tabular} & High                         & Moderate                & Moderate      & Robust                      & Robust                        & Moderate                & Bad                   \\ \hline
		OTFS-SRRC                                                         & High                         & Moderate                & Moderate      & Robust                      & Robust                        & Moderate                & Good                  \\ \hline
		ODDM                                                              & High                         & High                & Moderate      & Robust                      & Robust                        & Moderate                & Good                  \\ \hline
		OTSM                                                              & High                         & Moderate            & Moderate      & Robust                      & Robust                        & Moderate                & -                     \\ \hline
		OCDM                                                              & High                         & Moderate                & High          & Sensitive                   & Robust                        & Strong                & Good                  \\ \hline
		AFDM                                                              & High                         & Moderate                & High          & Robust                      & Robust                        & Strong                & Good                  \\ \hline
	\end{tabular}
\end{table*}

\section{Numerical Comparisons}
\label{SEC:PSD}

As the first set of numerical comparisons, we provide the power spectral density (PSD) and the bit error rate (BER) comparisons for OTFS and ODDM, since ODDM appears the most complex modulation scheme among all the aforementioned schemes. Both modulations multiplex data symbols in the DD domain but use different methods to transform the source data symbols into transmit signals. The PSD results would demonstrate their characteristics such as bandwidth and OoBE power, whereas the BER comparison demonstrates their error performance under channel noise and robustness against fast-fading channels due to high mobility.

\subsection{Power Spectrum Density and Bandwidth}
We compare the PSD of the baseband transmit signal for several comparable modulations introduced in Section~\ref{SEC:alternatives}. Specifically, we will show the PSD for OTFS with rectangular pulses (OTFS-REC), square-root raised cosine pulses (OTFS-SRRC), and ODDM. Given that OSDM, Vector OFDM, and OTFS with rectangular pulses all share the same expression of the sampled transmit signal, we refrain from repetition and would only show the PSD of OTFS.

To emulate continuous time, we over-sample the transmit signal with the oversampling rate $\gamma$, i.e., the sample period of the signal in our simulation is $T_{os}=\frac{T}{\gamma M}$. For modulations that have continuous time expressions of transmit signal, such as OTFS-REC and ODDM, we sample their continuous time expressions with the sample period of $T_{os}$ in our simulation. For OTFS-SRRC, we first generate the discrete transmit signal vector $\mathbf{s}$ with a rectangular pulse in Eq.~\eqref{Eq:OTFS5}, i.e., matrix $\mathbf{G}=\mathbf{I}_M$. We then upsample $\mathbf{s}$ by the oversampling factor $\gamma$ to match the sample period with other modulations, followed by a SRRC linear filter to mitigate OoBE signal power.
Numerically, we apply Welch's method with a Hamming window of the same length as the transmit signal to estimate the PSD of the transmit signal for each modulation.

As a specific setup, we use $\gamma=8$ and set the sizes of the transmit symbol to $(M,N)=(512,64)$. The subcarrier spacing $\Delta f$ is $15$ kHz with the symbol period of $T=1/\Delta f$. We randomly generate quadrature phase shift keying (QPSK) data symbols for transmission in all modulations. Rectangular pulse has a unit amplitude and duration of $T$. {For OTFS with SRRC pulses, we set the roll-off factor $\beta=0$}. For ODDM, we use the suggested SRRC pulses in \cite{ODDM_1} with $Q=8$ and roll-off factor $\beta=0$ to reproduce the PSD result.

\begin{figure}[h!]
    \centering
    \includegraphics[width = 0.45\textwidth]{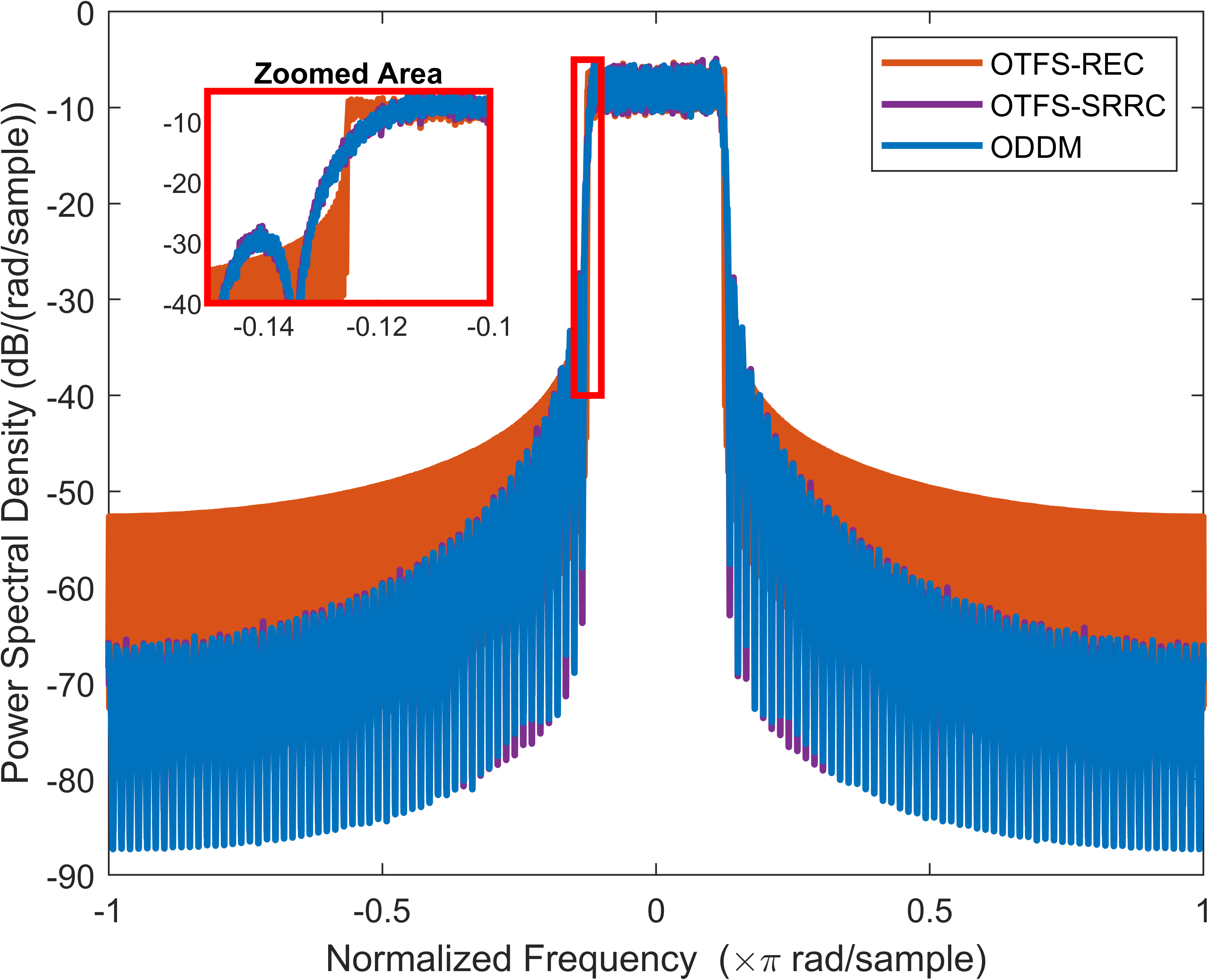}\\
    \caption{PSD for selected modulations.}
    \label{fig:PSD}
\end{figure}

The PSD result is shown in Fig.~\ref{fig:PSD}. Notably, the OTFS with SRRC pulses has a similar PSD as that of ODDM. Conversely, the OTFS with rectangular pulses demonstrates the largest bandwidth of all modulations, posing challenges in suppressing out-of-band power emissions.

\subsection{Receiver Bit Error Rate Comparison}

\begin{figure}
    \centering
    \includegraphics[width = 0.45\textwidth]{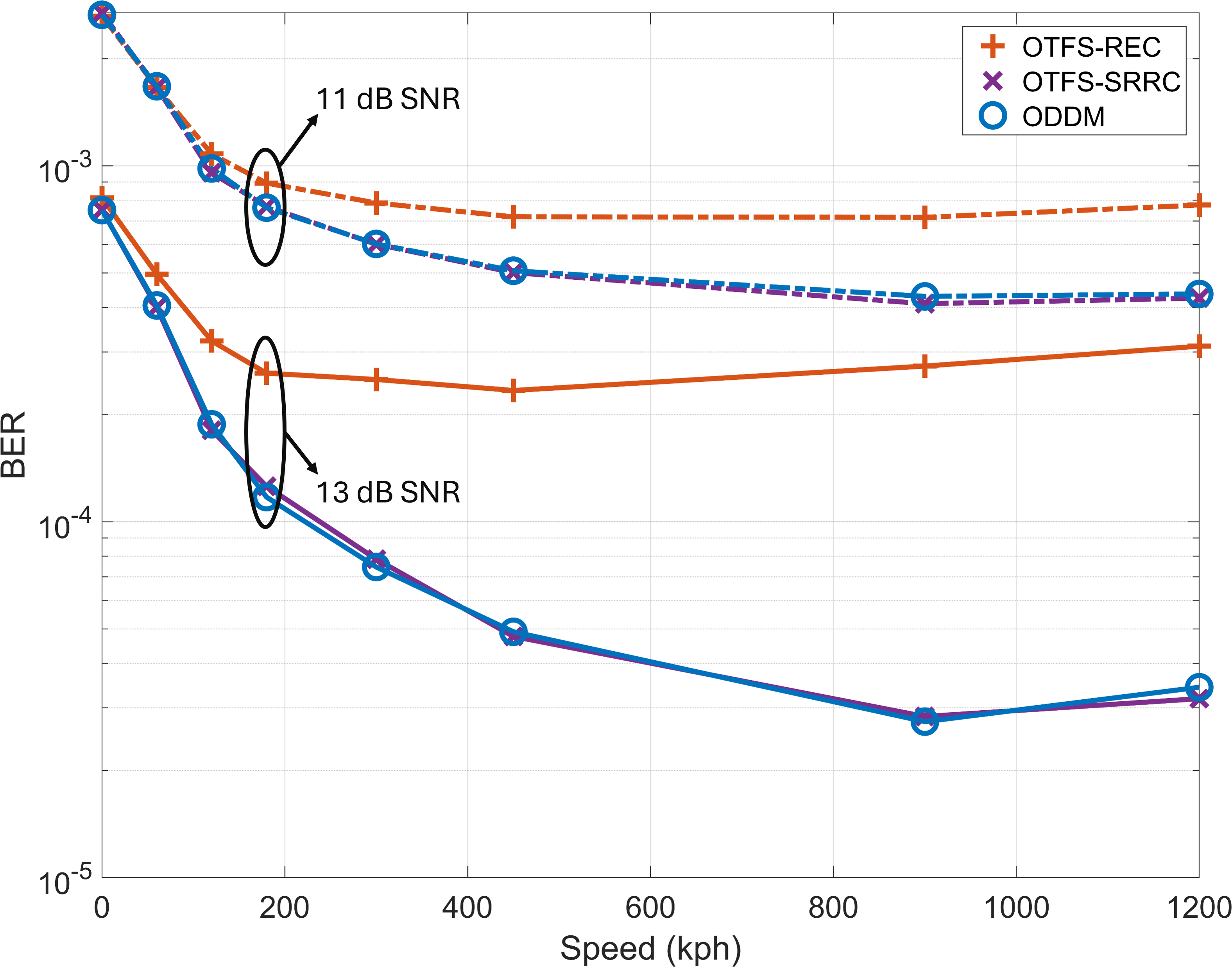}
    \caption{{BER performance of selected modulation schemes under varying user velocities, considering three individual paths and a channel filter bandwidth of $M\Delta f$.}}
    \label{fig:BERSpeed1}
\end{figure}

\begin{figure}
    \centering
    \includegraphics[width = 0.45\textwidth]{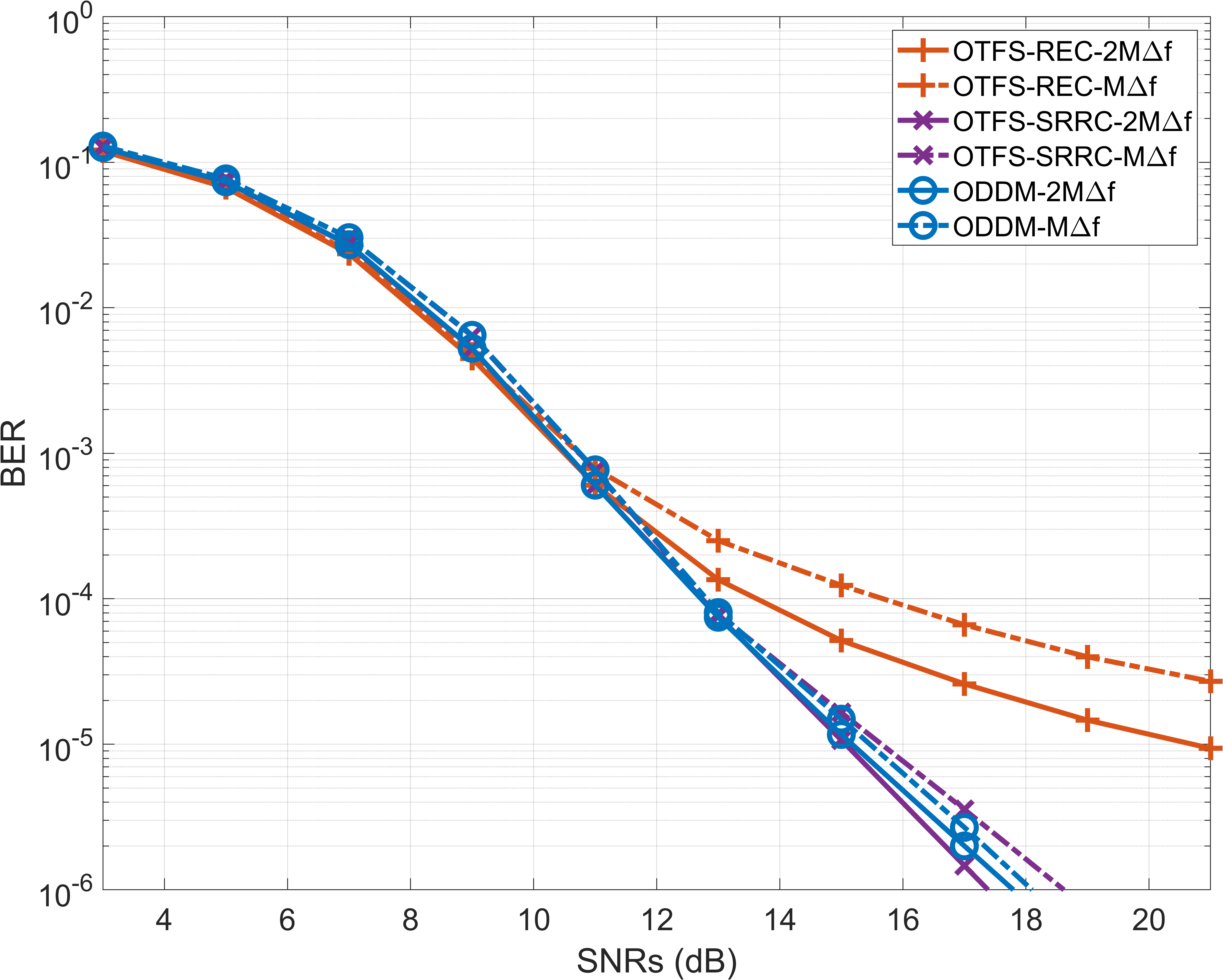}
    \caption{{BER performance of selected modulation schemes at a user velocity of 300 km/h, evaluated for different channel filter bandwidths.}}
    \label{fig:BERBW1}
\end{figure}

To better appreciate how different modulations respond to the effects of Doppler shifts and multipath delays introduced by wireless channels, we evaluate the BER for each modulation under common channel scenarios. In the experiment, we set the parameters for DD domain to $M=64$ and $N=16$. We let subcarrier spacing $\Delta f$ be $15$ kHz and the symbol period be $T=1/\Delta f$. The SRRC pulse for OTFS-SRRC and ODDM has a span of 16 periods, with a roll-off factor of 0.5. Each modulation includes a CP in time domain signal $\mathbf{s}$ to mitigate the inter-block/frame interference. Additionally, we add a band-limiting finite impulse response (FIR) transmit filter and its corresponding matched filter at the receiver sides to control signal bandwidth and reject out-of-band signal/noise power.

In wireless simulation, we model a doubly-selective fading channel discussed in Section~\ref{SUBSEC:doubleSelectChannel}. We configure the channel parameters to include three individual paths, each featuring a random complex gain with a power of $1/3$, an off-the-grid delay $\tau_{i}\in [0, \tau_{max}]$, and a random Doppler $\nu_{i}=\nu_{max}\cos{\theta_{i}}$, where $\theta_{i} \in [-\pi, \pi]$ denotes a random angle-of-arrival (AoA). We set the maximum delay $\tau_{max}$ to $4\frac{T}{M}$, and maximum Doppler shift $\nu_{max}$ according to the user mobile velocity.

After passing the matched filter and CP removal at the receiver we use an orthogonal approximated message passing decoder (OAMP) \cite{Detector_9} discussed later in Section~\ref{subsubsec:NonMemoryMessagePassing} for decoding and BER calculation. We set the bandwidth of the FIR filter to $M\Delta f$. 

We average the BER over 500 Monte Carlo (MC) simulation runs. The average BER results for all three modulations with different user velocities are given in Fig.~\ref{fig:BERSpeed1}. The BER performance of OTFS-REC is worse than either OTFS-SRRC or ODDM across different signal-to-noise ratios (SNRs). On the other hand, OTFS-SRRC and ODDM share similar BER performance, as expected. Additionally, the BER performance of all modulations improves initially with mobile velocity, until it reaches a saturation point before a moderate deterioration. By comparing the effect of channel bandwidth on all modulations in Fig.~\ref{fig:BERBW1}, we observe that the OTFS-REC is more sensitive to narrower channel bandwidth. Such an effect is also expected since our previous PSD comparison found higher OoBE for OTFS-REC and suggested more loss of signal energy at the receiver end under narrow channel bandwidth, leading to worse BER.

\subsection{Summary}
{In this section so far, we have presented the numerical comparisons of OTFS-REC, OTFS-SRRC, and ODDM to evaluate their PSD and BER performance. The results indicate that OTFS-SRRC achieves PSD and BER performances similar to ODDM, reflecting comparable bandwidth usage and effective OoBE control. In contrast, OTFS-REC exhibits higher OoBE power and poorer BER performance than both OTFS-SRRC and ODDM, underscoring its sensitivity to narrower channel bandwidth transmissions.}

\section{Effects of Different Channel Models}
\label{SEC:channelMtx}

\begin{table*}[ht!]
\centering
\renewcommand{\arraystretch}{1.5}
\caption{Input-Output Relationship Expression in Doubly-selective Fading Channels \\(Note that OCDM and AFDM will be discussed in details in Appendix.)}
\begin{tabular}{|c|c|}
\hline
\textbf{Modulation Scheme} & \textbf{Input-Output Expression} \\
\hline
OFDM & $\mathbf{y} = (\mathbf{I}_N \otimes \mathbf{F}_M)\mathbf{H}_{ch}(\mathbf{I}_N \otimes \mathbf{F}_M^H)\mathbf{x} + (\mathbf{I}_N \otimes \mathbf{F}_M)\mathbf{n}$\\
\hline
FBMC & $\mathbf{y} = \mathbf{P}^H \mathbf{H}_{ch}^{(1)} \mathbf{P}\mathbf{x} + \mathbf{P}^H \mathbf{n}$\\
\hline
\makecell{OTFS-REC \\ (V-OFDM, OSDM)} & $\mathbf{y} = (\mathbf{F}_N \otimes \mathbf{I}_M)\mathbf{H}_{ch}(\mathbf{F}_N^H \otimes \mathbf{I}_M)\mathbf{x} + (\mathbf{F}_N \otimes \mathbf{I}_M)\mathbf{n}$ \\
\hline
ODDM &  $\mathbf{y} = \mathbf{U}^H\mathbf{H}_{ch}^{(2)}\mathbf{U}\mathbf{x} + \mathbf{U}^H \mathbf{n}$\\
\hline
OTSM & $\mathbf{y} =(\mathbf{I}_M \otimes \mathbf{W}_N) \mathbf{\Pi}^T\mathbf{H}_{ch}(\mathbf{W}_N \otimes \mathbf{I}_M)\mathbf{\Pi}\mathbf{x} + (\mathbf{I}_M \otimes \mathbf{W}_N)\mathbf{\Pi}^T\mathbf{n}$ \\
\hline
OCDM & $\mathbf{y} = {\bf{\Phi}}\mathbf{H}_{ch}{\bf{\Phi}}^H\mathbf{x} + {\bf{\Phi}}\mathbf{n}$\\
\hline
AFDM & $\mathbf{y} = {\bf{\Lambda }}_{{c_2}}{\bf{F}}_M{\bf{\Lambda }}_{{c_1}}\mathbf{H}_{ch}{\bf{\Lambda }}_{{c_1}}^H{\bf{F}}_M^H{\bf{\Lambda }}_{{c_2}}^H\mathbf{x} + {\bf{\Lambda }}_{{c_2}}{\bf{F}}_M{\bf{\Lambda }}_{{c_1}}\mathbf{n}$\\
\hline
\end{tabular}
\label{Table:DoubleSelective}
\end{table*}

In this section, we derive an effective receive signal model for the examined modulation schemes under different fading channel models. The analysis provides a unified signal model $\mathbf{y}=\mathbf{H}\mathbf{x}$ for efficient detector designs in OTFS related demodulations in Section~\ref{SEC:detector}.

\subsection{Doubly-selective fading channel model}\label{SUBSEC:doubleSelectChannel}
Consider the doubly-selective fading channel model which exhibits both multipath delays and Doppler effects. Once the transmit signal passes through the FIR filter, the resulting filtered time domain signal $\mathbf{s}$ enters the time-varying multipath fading channel, whose baseband model can be written as
\begin{equation}
\label{Eq:doubleFading1}
    h(\tau,\nu) = \Sigma_{i=1}^L h_i \delta(\tau - \tau_i)\delta(\nu - \nu_i),
\end{equation}
where $L$ is the number of propagation paths, $\delta(\cdot)$ is the Dirac delta function, $h_i$, $\tau_i$ and $\nu_i$ denote the complex channel gain, time delay, and Doppler frequency shift for signals on the $i$-th propagation path, respectively. This model lies in the DD domain.

Let $P_{\text{rc}}(t)$ be the linear system impulse response that comprises a band-limited transmit filter and its matched filter at the receiver. A more general expression of the sampled baseband channel response in time domain is given by
\begin{align}
    \label{Eq:doubleFading2}
    h[c,p] & = \Sigma_{i=1}^L h_i e^{j2\pi \nu_i (cT_s - pT_s)}P_{\text{rc}}(pT_s - \tau_i), \\
    c & = 0, \cdots, MN-1; p = 0, \cdots, P-1,
\end{align}
where $P$ is the maximal channel tap determined by the maximum channel delay and the time duration of $P_{\text{rc}}$. $L$ is the number of individual channel paths, and $T_s=\frac{T}{\gamma M}$ is the sampling interval with oversampling rate $\gamma$. 

With the above channel model, we now express the received signal after passing the matched filter and CP removal as
\begin{equation}\label{Eq:generalIO}
    r[c]=\sum_{p=0}^{P-1}h[c,p]s[[c-p]_{MN}] + n[c],
\end{equation}
where $n[\cdot]$ represents the filtered noise after CP removal, $[\cdot]_{MN}$ denotes the modulo-$MN$ operation. 

The model \eqref{Eq:generalIO} can be further compacted into a matrix form:
\begin{equation}\label{Eq:generalH}
    \mathbf{r}=\mathbf{H}_{ch}\cdot\mathbf{s}+\mathbf{n},
\end{equation}
where $\mathbf{s}$, $\mathbf{r}$ and $\mathbf{n}$ are the sampled transmit signal vector before adding CP, sampled received signal vector after CP removal and filtered noise vector after CP removal, respectively. $\mathbf{H}_{ch}$ is the equivalent channel matrix in discrete time domain. Let the CP length be longer than the maximum channel delay span of $P$ taps. Then, we can write $\mathbf{H}_{ch} $ in a circular shift form as Eq.~\eqref{Eq:HMat}.

By incorporating transmitter and receiver of each modulation into signal model Eq.~\eqref{Eq:generalH}, we can directly derive the respective input-output relationship for each OTFS related modulation. By using OTFS as an example, when the transceiver pulses are both rectangular of duration $T$, the input-output (I/O) relationship at sampling interval $T_s=\frac{T}{M}$ can be simplified as
\begin{equation}
    \mathbf{y} = (\mathbf{F}_N \otimes \mathbf{I}_M)\mathbf{H}_{ch}(\mathbf{F}_N^H \otimes \mathbf{I}_M)\mathbf{x},
\end{equation}
where the effective channel matrix $\mathbf{H}=(\mathbf{F}_N \otimes \mathbf{I}_M)\mathbf{H}_{ch}(\mathbf{F}_N^H \otimes \mathbf{I}_M)$. This I/O relationship expression is also the same for both Vector OFDM and OSDM.

Table~\ref{Table:DoubleSelective} summarizes the I/O relationship expression for the various modulations under discussion. Note that for FBMC and ODDM, the inner time domain matrix $\mathbf{H}_{ch}^{(1)}$ and $\mathbf{H}_{ch}^{(2)}$ are different from other modulation schemes because they have a different number of samples in time domain caused by different time duration in their transmit signals. $\mathbf{H}_{ch}^{(1)}$ does not contain the upper right non-zero entries in Eq.~\eqref{Eq:HMat} because FBMC does not add the CP, while $\mathbf{H}_{ch}^{(2)}$ have a similar structure as Eq.~\eqref{Eq:HMat}.

To better illustrate the structure of the effective channel matrices, Table~\ref{Table:ChannelMatrices} graphically illustrates an exemplary set of effective channel matrices. In this example, we let $M=8$ and $N=4$, with the SRRC pulse for OTFS-SRRC and ODDM spanning 6 periods. We set user velocity to 300 km/h with three individual paths, each having an independent off-the-grid delay. From the channel magnitude images of Table~\ref{Table:ChannelMatrices}, we observe that the effective channel matrices for OTFS-SRRC and ODDM exhibit highly similar patterns, thereby further corroborating the similarity of the two modulations beyond the PSD similarity.

\begin{figure*}[t]
	\centering
	\scriptsize
	\setlength{\arraycolsep}{1pt} 
	\renewcommand{\arraystretch}{1} 
	\begin{equation}\label{Eq:HMat}
		{
		\mathbf{H}_{ch} = 
		\begin{bmatrix}
			h[0,0] & 0 & \cdots & \cdots & \cdots & 0 & h[0,P-1] & \cdots & h[0,2] & h[0,1] \\
			h[1,1] & h[1,0] & 0 & \cdots & \cdots & \cdots & 0 & h[1,P-1] & \cdots & h[1,2] \\
			\vdots & \vdots & \ddots & \ddots & \cdots & \cdots & \vdots & \ddots & \vdots & \vdots \\
			h[P-1,P-1] & h[P-1,P-2] & \cdots & h[P-1,0] & 0 & \cdots & \vdots & \cdots & 0 & 0 \\
			0 & h[P,P-1] & \cdots & \cdots & h[P,0] & 0 & \vdots & \cdots & \vdots & 0 \\
			\vdots & \vdots & \cdots & \cdots & \cdots & \ddots & \ddots & \cdots & \vdots & \vdots \\
			\vdots & \vdots & \cdots & \cdots & \cdots & \cdots & \ddots & \ddots & \vdots & \vdots \\
			\vdots & \vdots & \cdots & \cdots & \cdots & \cdots & \vdots & \ddots & \ddots & \vdots \\
			\vdots & \vdots & \cdots & \cdots & \cdots & \cdots & \vdots & \cdots & \ddots & 0 \\
			0 & 0 & \cdots & \cdots & \cdots & h[MN-1,P-1] & h[MN-1,P-2] & \cdots & h[MN-1,1] & h[MN-1,0] 
		\end{bmatrix}
	}
	\end{equation}
	\hrulefill
\end{figure*}


\begin{table*}[t!]
    \centering
    \caption{{Effective Channel Matrices for OTFS-REC, OTFS-SRRC, and ODDM}}
    \begin{tabular}{cccc}
        & Doubly-selective fading channel& Time-selective fading channel& Frequency-selective fading channel\\
        \rotatebox[origin=c]{90}{OTFS-REC} & 
        \raisebox{-.5\height}{\includegraphics[width=0.29\textwidth]{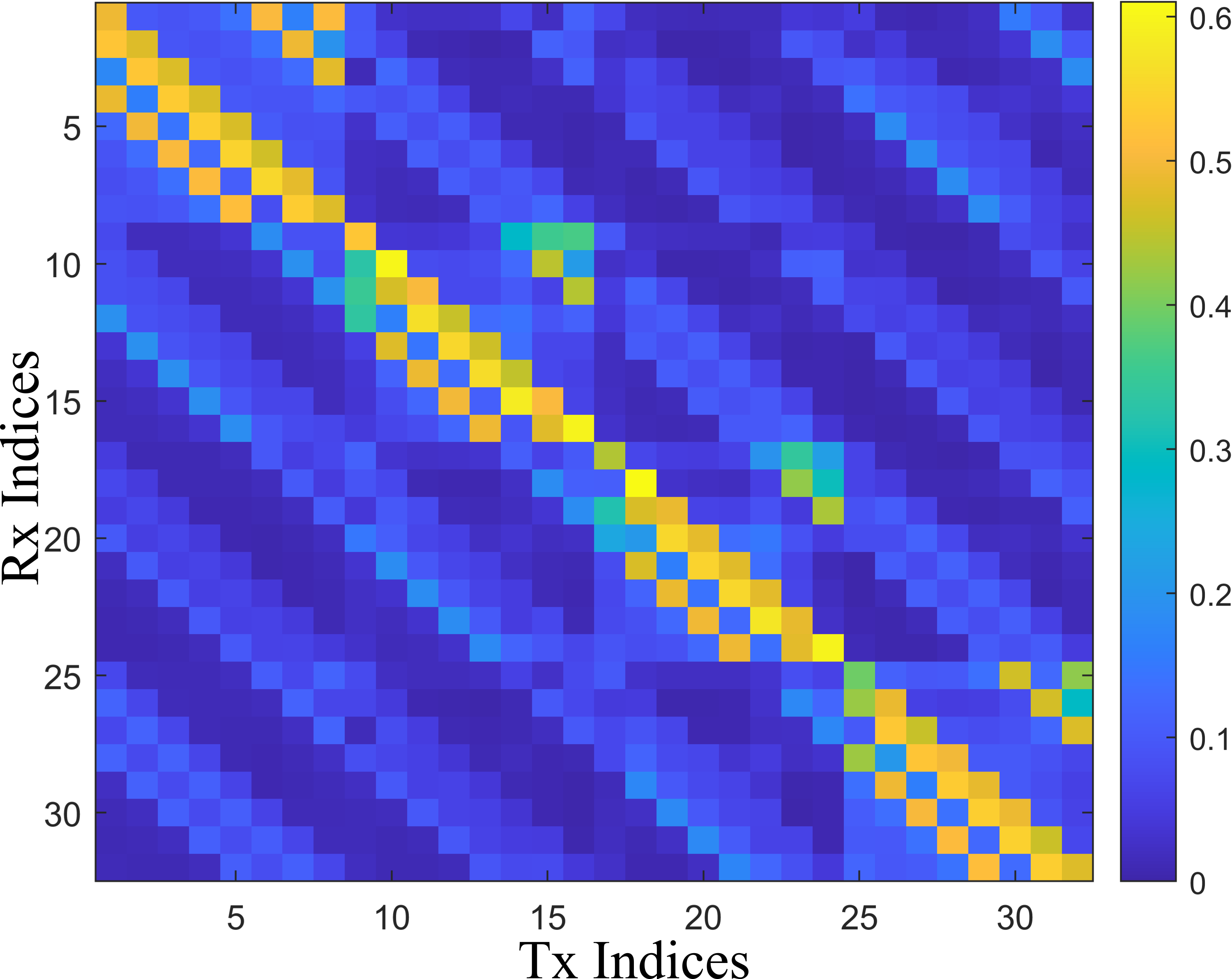}} & 
        \raisebox{-.5\height}{\includegraphics[width=0.29\textwidth]{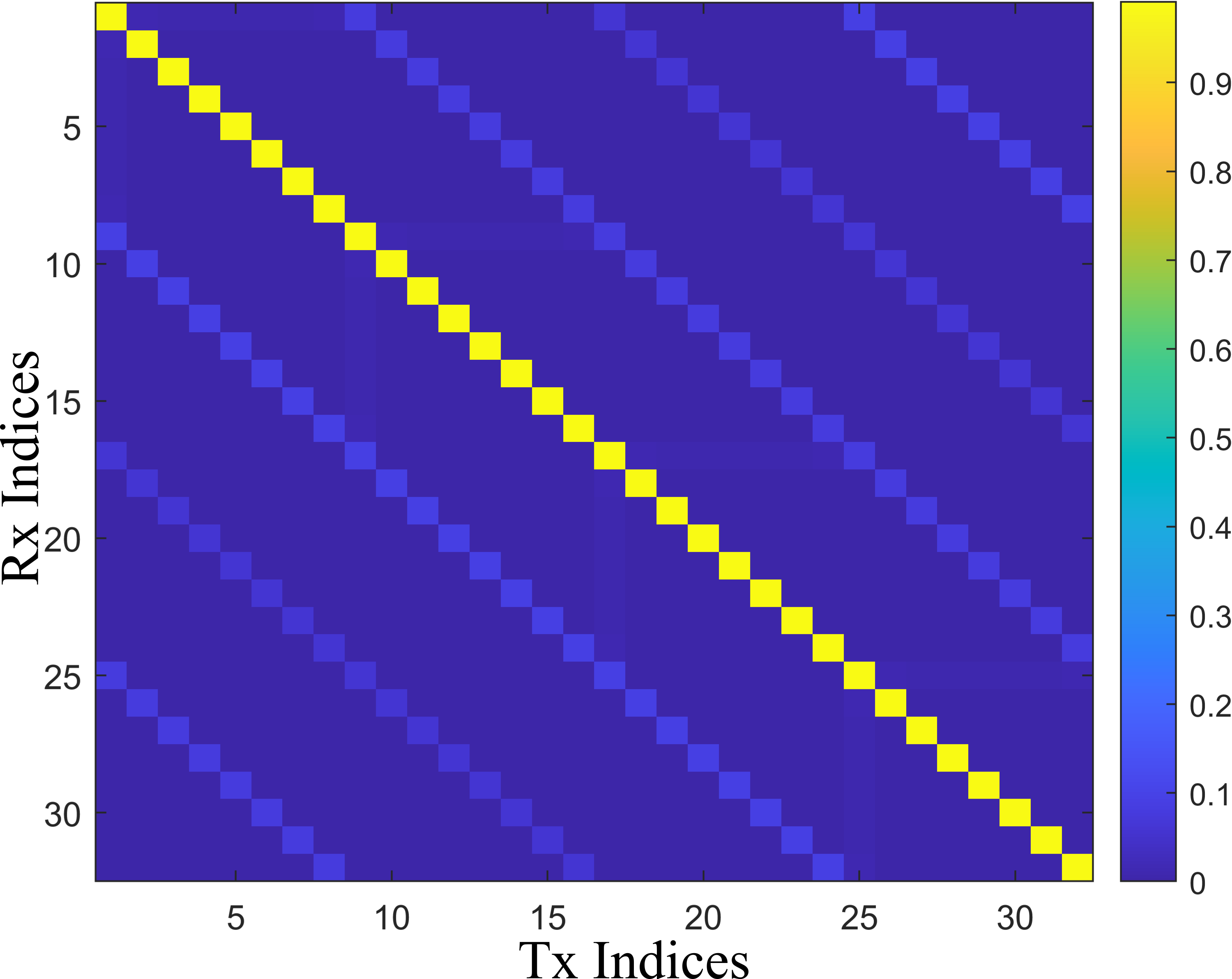}} & 
        \raisebox{-.5\height}{\includegraphics[width=0.29\textwidth]{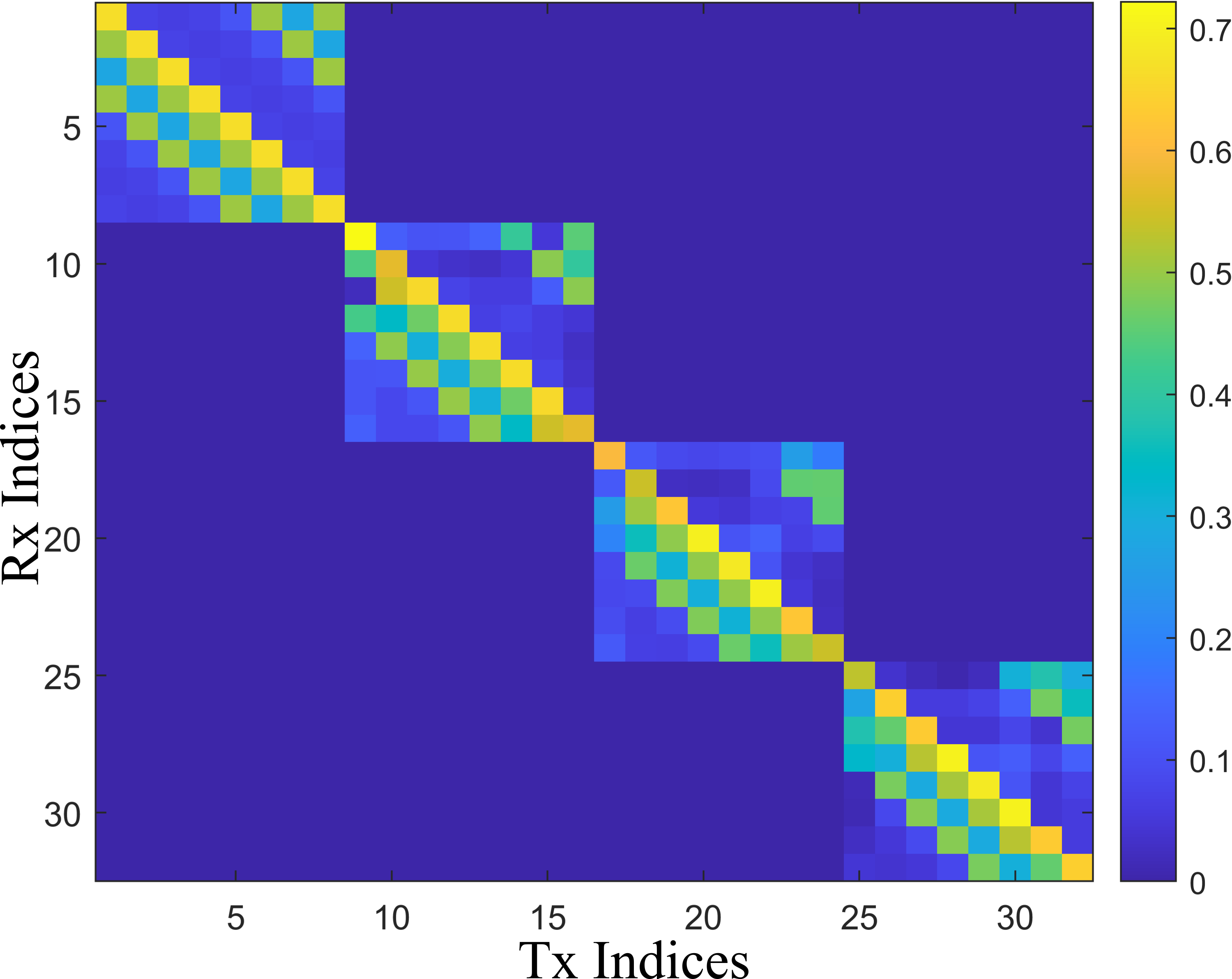}} \\
        [2.2cm]
        
        \rotatebox[origin=c]{90}{OTFS-SRRC} & 
        \raisebox{-.5\height}{\includegraphics[width=0.29\textwidth]{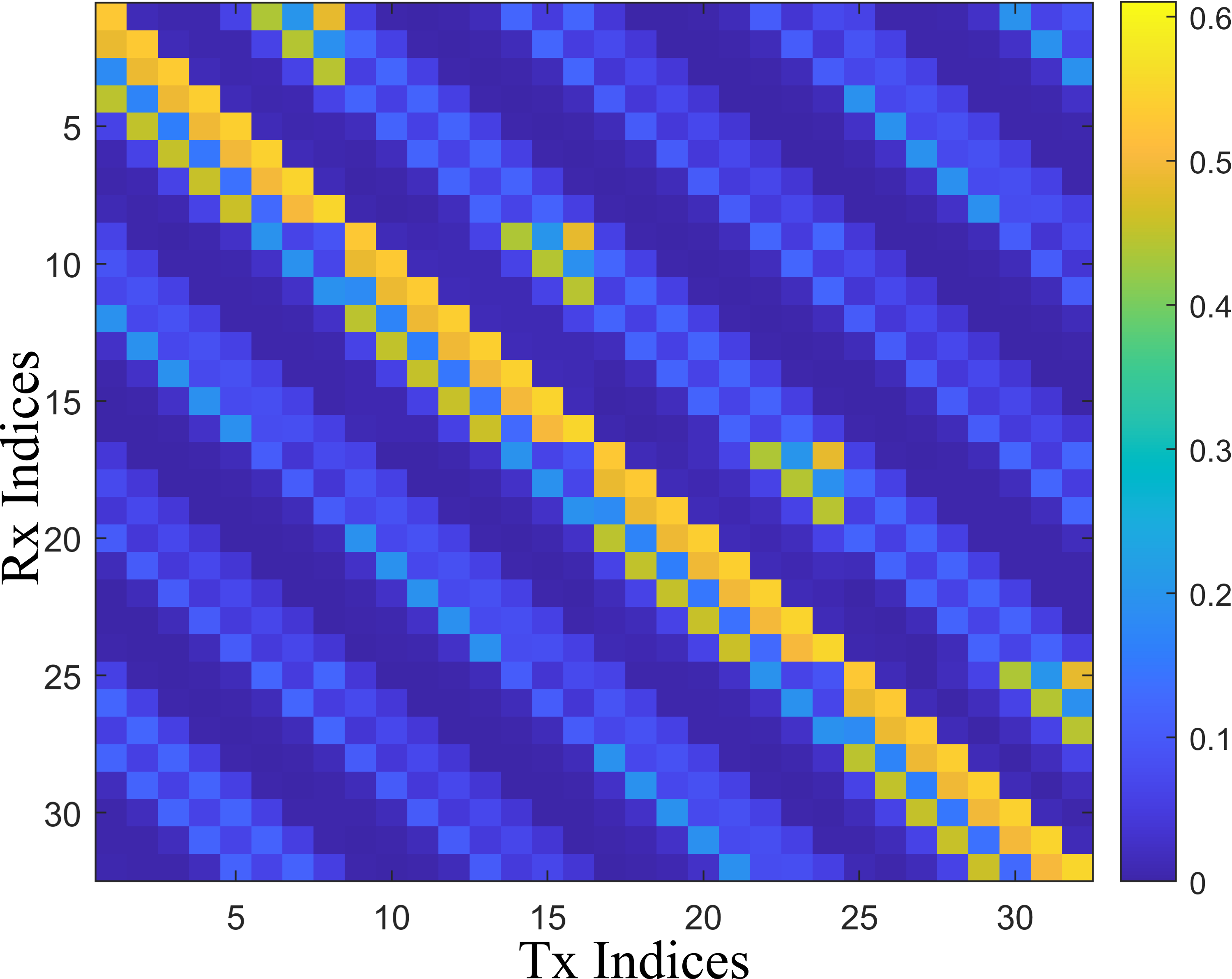}} & 
        \raisebox{-.5\height}{\includegraphics[width=0.29\textwidth]{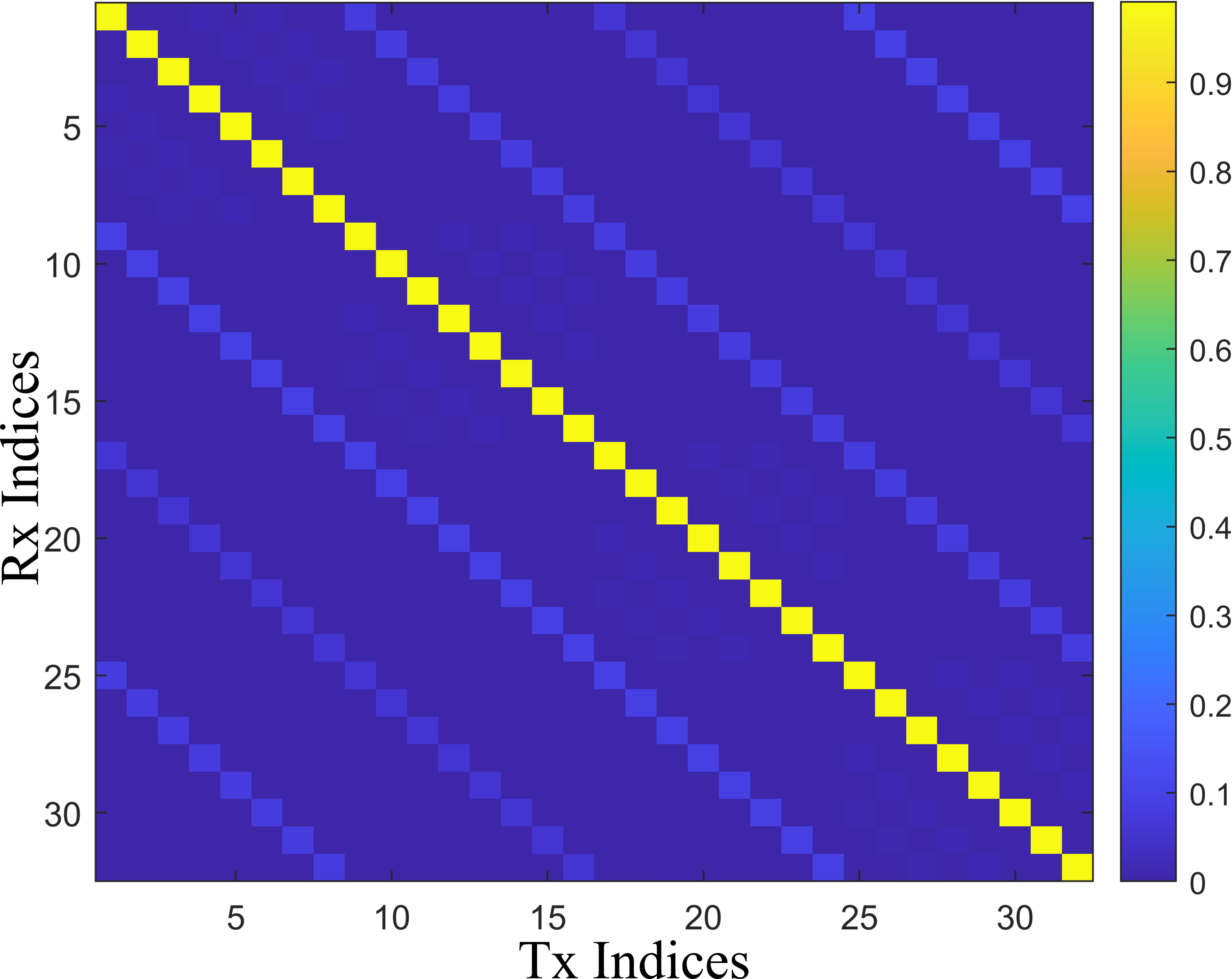}} & 
        \raisebox{-.5\height}{\includegraphics[width=0.29\textwidth]{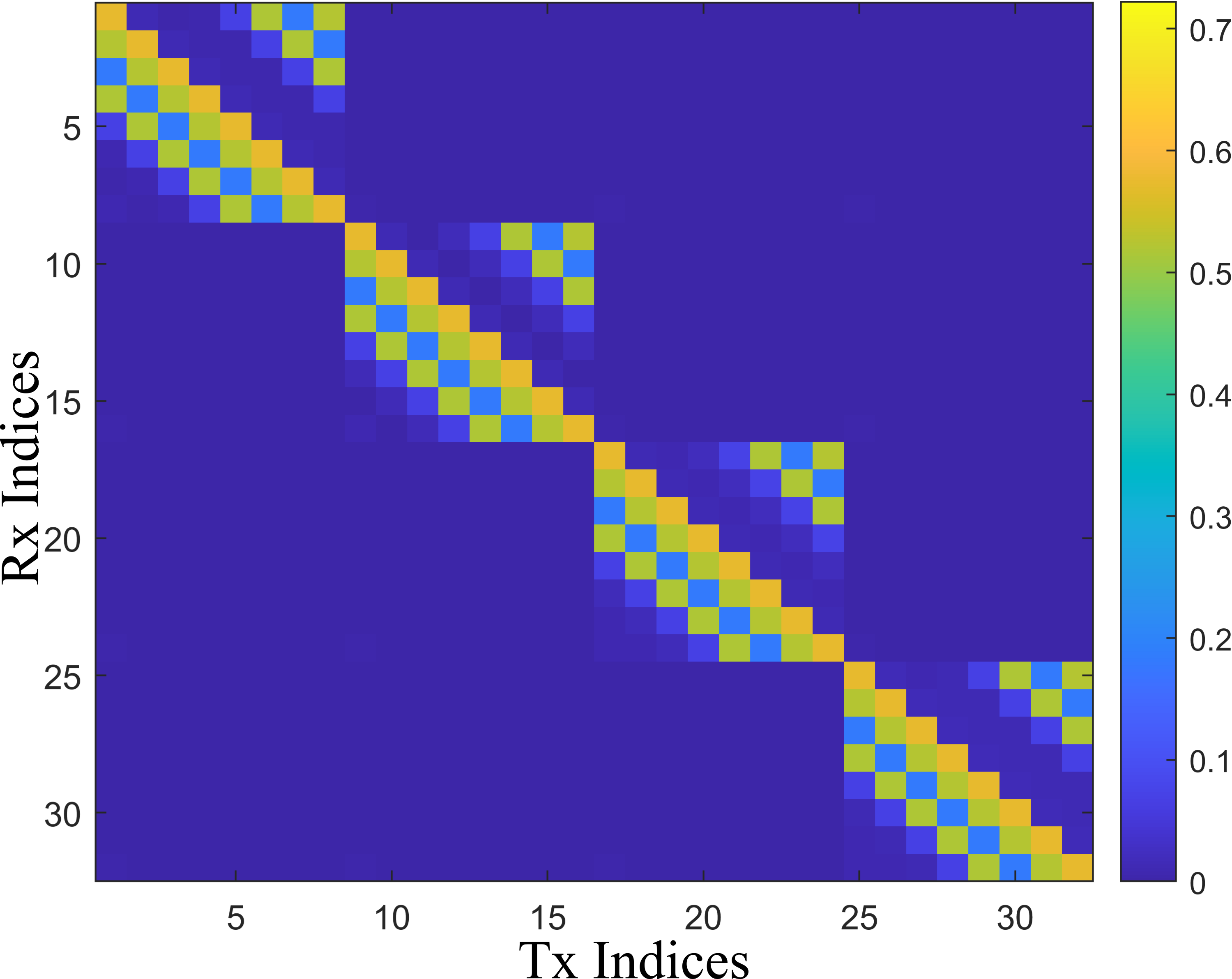}} \\
        [2.2cm]
        
        \rotatebox[origin=c]{90}{ODDM} & 
        \raisebox{-.5\height}{\includegraphics[width=0.29\textwidth]{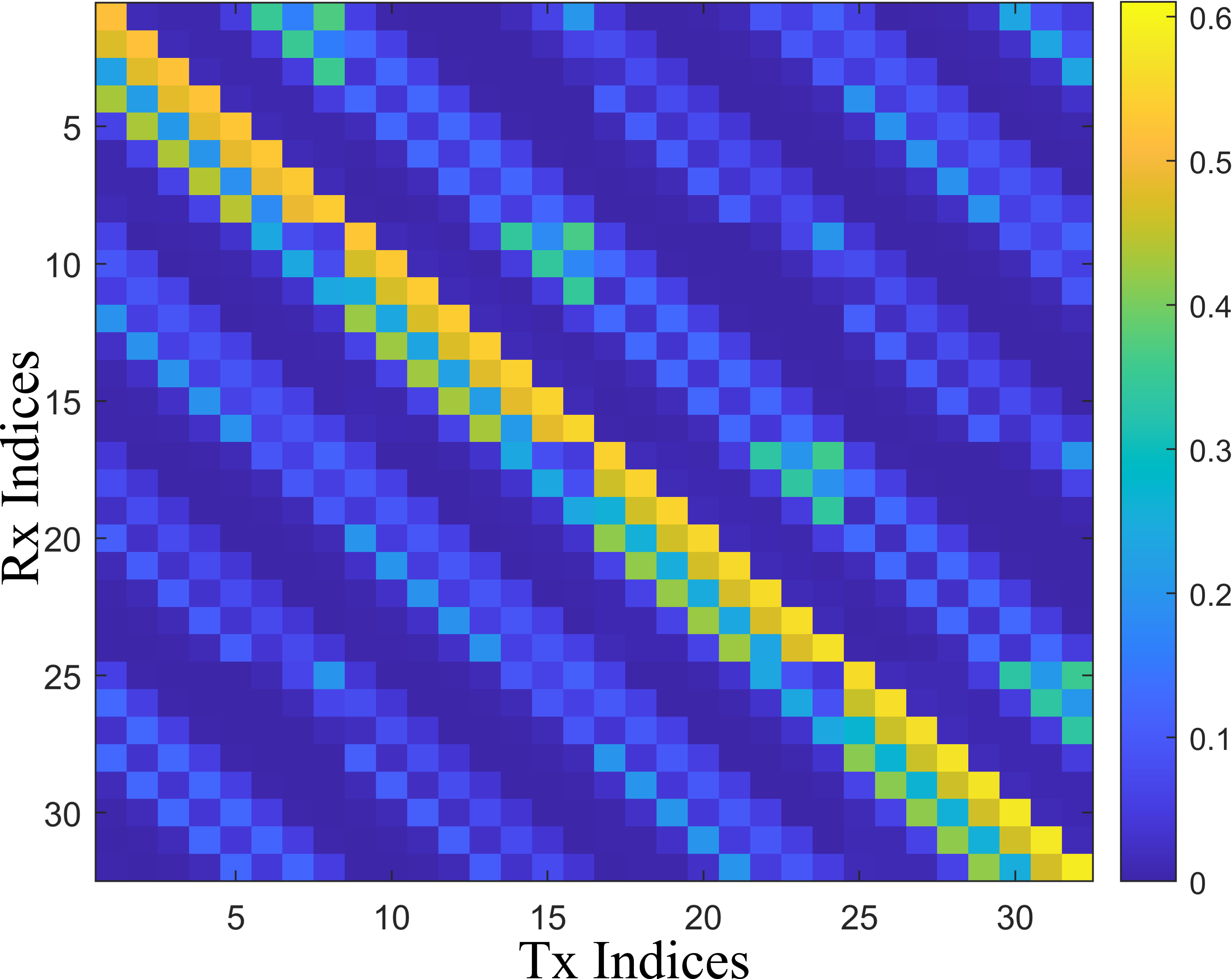}} & 
        \raisebox{-.5\height}{\includegraphics[width=0.29\textwidth]{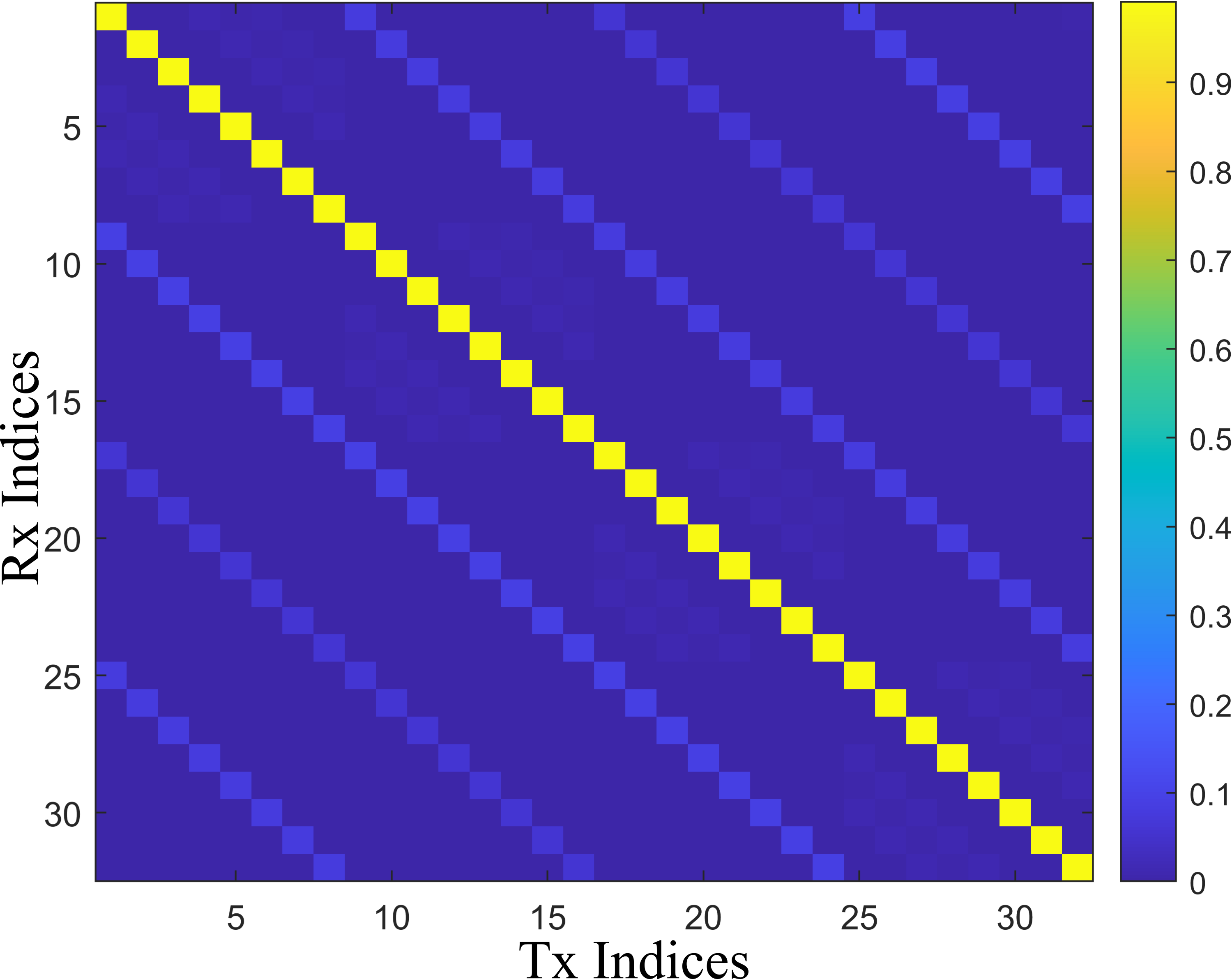}} & 
        \raisebox{-.5\height}{\includegraphics[width=0.29\textwidth]{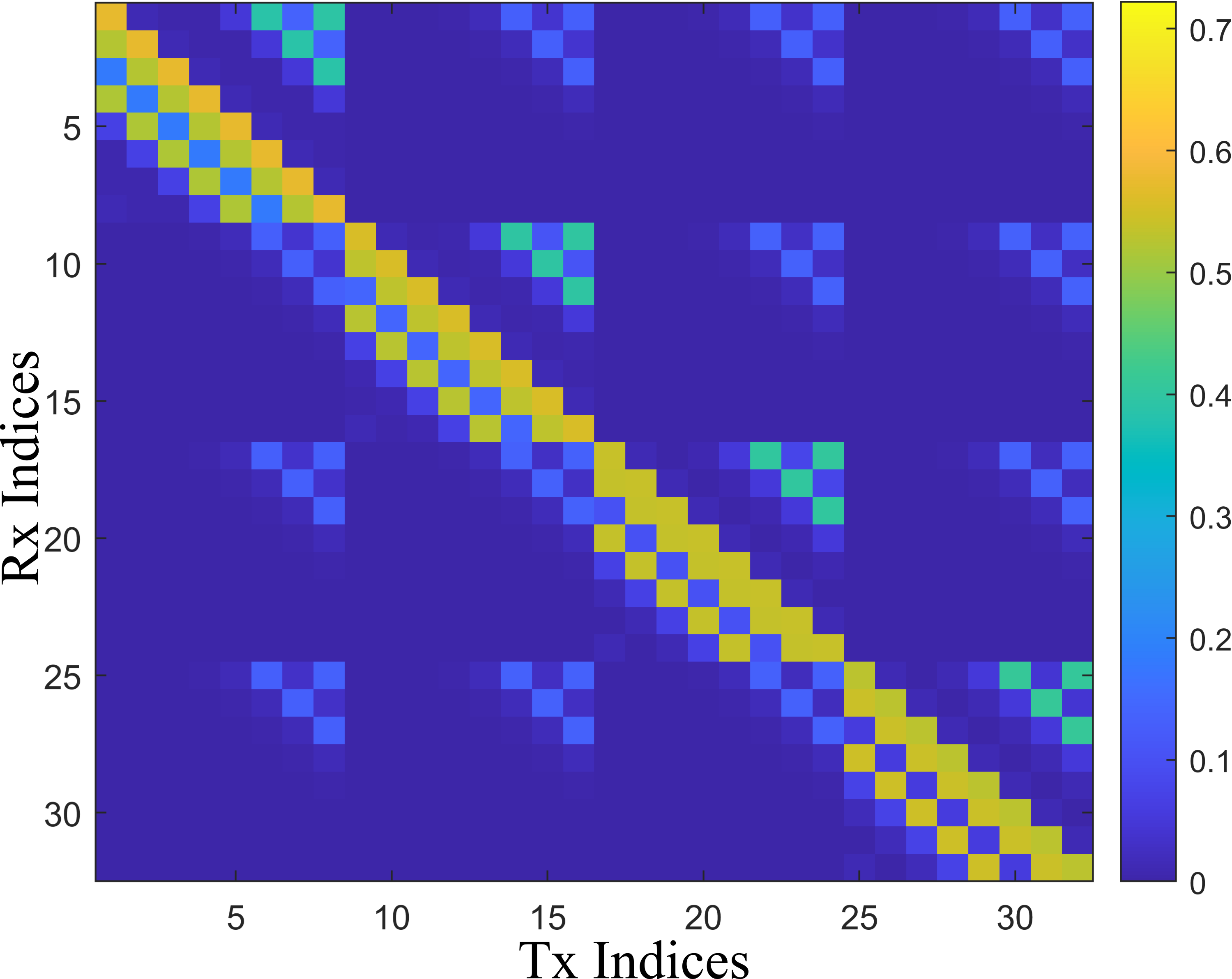}} \\
    \end{tabular}
    \label{Table:ChannelMatrices}
\end{table*}
    
\subsection{Time-selective fading channel model}
In some high-mobility outdoor environments such as UAV and satellite communications, wireless channels may exhibit strong Doppler shift but little or no multipath. Such channels simplify into time-selective fading channels and can be modeled as a special case of Eq.~\eqref{Eq:doubleFading1} and Eq.~\ref{Eq:doubleFading2} with $L=1$. For these time-selective fading channels, the off-grid channel Doppler components spread along the Doppler dimension, leading to inter-symbol-interference (ISI) among nearby symbols in DD domain. Because the effective channel matrix represents the relationship between the column-wise vectorization of the DD grid symbols, the interference for every $M$ sample can be seen from the three figures in the ``Time-selective fading channel'' column of Table~\ref{Table:ChannelMatrices}. Clearly, OTFS-REC, OTFS-SRRC, and ODDM modulations all exhibit a similar pattern for the effective channel matrix of the exemplary time-selective fading channels.
    
\subsection{Frequency-selective fading channel model}

Some low-mobility wireless channels exhibit only multipath effect without Doppler shift, such as indoor wireless local area networks (WLANs) and industrial IoT 4.0. These channels are often modeled as frequency-selective fading channels, which is a special case for Eq.~\eqref{Eq:doubleFading1} and Eq.~\eqref{Eq:doubleFading2} with $\nu_i=0, \; \forall i$. For frequency-selective fading channels, the $N$-point IDFT and DFT pairs in OTFS transform the linear time-invariant channel response into a block-circular form, as shown in the example of Table~\ref{Table:ChannelMatrices}. In this frequency-selective fading channel example, there are three resolvable paths, each having an independent off-the-grid delay and no Doppler shift. These multipath delays lead to ISI in DD domain. As shown in Table~\ref{Table:ChannelMatrices}, OTFS can resolve the effect of these paths in DD domain. Computationally demanding detection algorithms such as message passing (MP) can utilize such independence to effectively mitigate the negative impact of such ISI on BER performance. Compared to OTFS, the effective channel matrix for ODDM includes some additional non-zero elements in certain positions. These elements arise from the finite length of the SRRC pulse truncation used in ODDM. In practical communication systems, we may select larger $M$ and $N$, and allow for a longer SRRC pulse. As a result, the longer pulse tends to lower the amplitude and the (interfering) effect of these additional elements on the performance of receiver data detection.

\subsection{Summary}
{The present section derived a unified effective linear input-output relationship model for the aforementioned modulation schemes under a doubly-selective fading channel. It provides a foundation for efficient detector designs discussed in Section~\ref{SEC:detector}. We presented the expression for the effective channel matrix in doubly selective channels and explored its structure through visualizations. Additionally, we provided the visualizations of the effective channel matrices under time-selective and frequency-selective channel models, highlighting the unique characteristics of each fading scenario. These results offer a cohesive framework for analyzing and optimizing modulation schemes across various wireless environments.}

\section{Effective Receivers for OTFS Related Signals}
\label{SEC:detector}
Based on the discussions in Section~\ref{SEC:channelMtx}, the equivalent transmission of the various OTFS-related modulation schemes under discussion involve non-diagonal effective baseband channel matrices. Thus, OTFS-related modulations tend to suffer from inherent inter-symbol interference (ISI), especially in the doubly-selective fading channels. For this reason, effective receivers must rely on detection algorithms to combat the inherent ISI. The optimal maximum a posteriori (MAP) detector or maximum likelihood (ML) detector tends to suffer from high computational complexity which grows exponentially with the signal dimension. Therefore, significant research efforts have focused on designing simple and effective signal detectors to achieve desirable receiver performance for OTFS related signals. 

In this section, we review some existing OTFS receivers for practical OTFS related signal detection. Without loss of generality, we use the following linear model to represent the discrete I/O relationship of the various modulations under discussion. In particular, let ${\bf{y}} \in {\mathbb{C}^{\mathcal{M} \times 1}}$ be the received signal vector. Let ${\bf{H}} \in {\mathbb{C}^{\mathcal{M} \times \mathcal{N}}}$ be the equivalent effective baseband channel matrix of the transmission.\footnote{It is worth mentioning that the number of inputs $\mathcal{N}$ and the number of outputs $\mathcal{M}$ are typically constant quantities for a given system.} The simple linear I/O relationship is captured by
\begin{equation}\label{linear_model}
    {\bf{y}} = {\bf{Hx}} + {\bm{\omega }},
\end{equation}
where ${\bm{\omega }} \in {\mathbb{C}^{\mathcal{M} \times 1}}$ is the additive random channel noise vector with zero mean and covariance matrix $E\{ {\bm{\omega }}{\bm{\omega }}^H\} =\sigma_\omega^2 \mathbf{I} $. Depending on the specific application under consideration, ${\bf{x}} \in {\mathbb{A}^{\mathcal{N} \times 1}}$ denotes a vector of $\mathcal{N}$ data symbols belonging to a set ${\mathbb{A}^\mathcal{N}}$. Typically for QAM, the constellation $\mathbb{A} = \left\{ {{a_1},{a_2}, \cdots ,{a_Q}} \right\}$ has size $Q$.

The basic task of the detection is to estimate the input vector ${\bf{x}}$ based on the received signal vector ${\bf{y}}$ and the equivalent effective channel matrix ${\bf{H}}$. Bearing in mind that the system model of (\ref{linear_model}) may be established in Time-Frequency domain (e.g., OFDM, FBMC, Vector OFDM and OSDM), Delay-Doppler domain (e.g., OTFS, ODDM), Delay-Sequency domain (e.g., OTSM) or in the chirp domain (e.g., OCDM, AFDM), and may be applied to doubly-selective fading channels, time-selective fading channels, and frequency-selective fading channels, respectively.

\subsection{Linear Detection Receivers}
\begin{figure}
  \centering
  \includegraphics[width=0.35\textwidth]{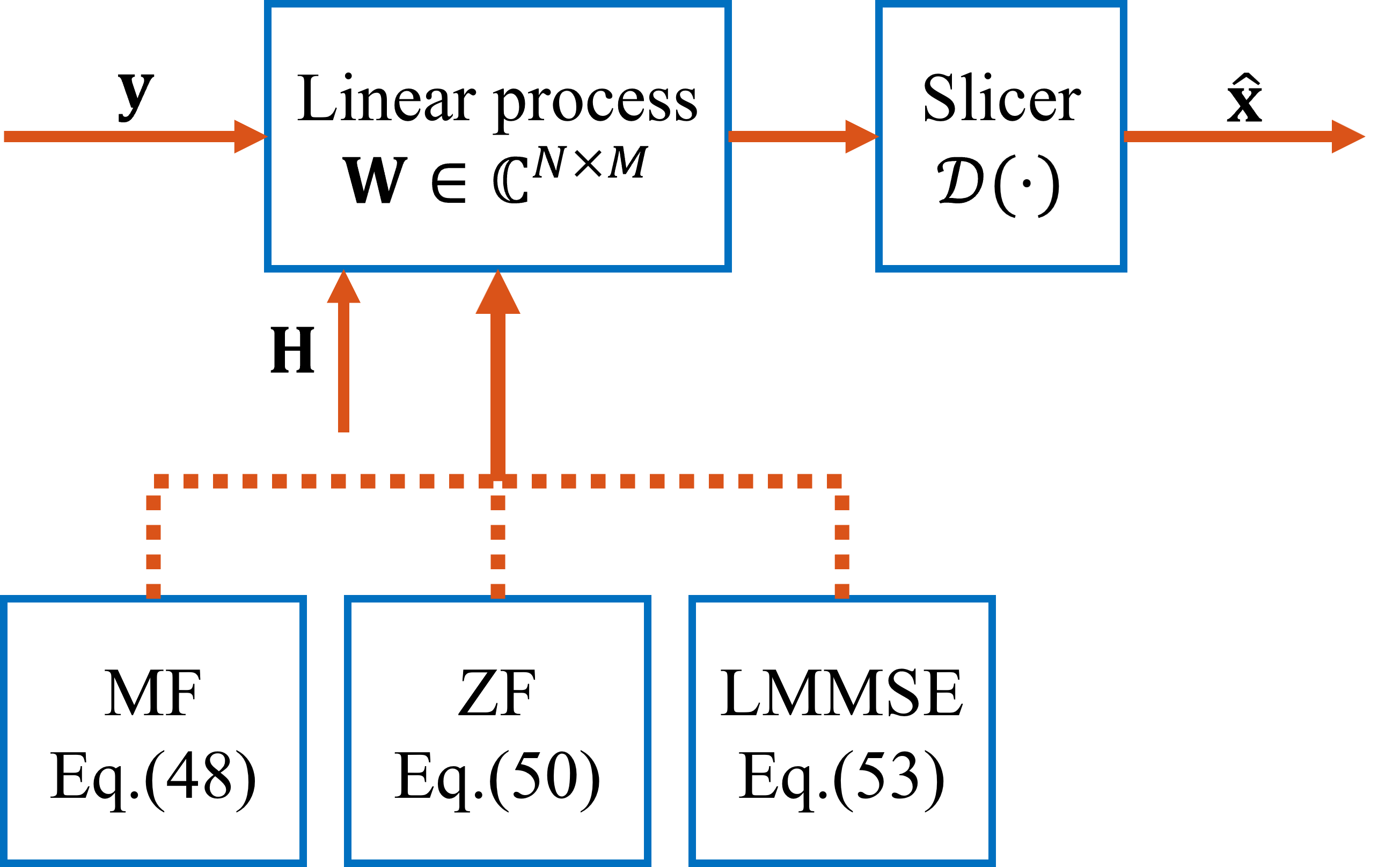}
  \caption{{Block diagram of the linear detector.}}\label{Detector_LE}
\end{figure}

In digital communications in general, linear detection receivers are known for their appealingly low complexity at the expense of certain performance degradation in comparison to the optimum MAP or ML detectors. Fig. \ref{Detector_LE} depicts a standard block diagram of a linear detector for reception. The decision statistics of linear detectors are derived from linear transformation or filtering of the channel output signal ${\bf{y}}$ followed by a memoryless nonlinear decision device $\mathcal{D}\left(  \cdot  \right)$ such as a single symbol slicer which quantizes each input symbol to its nearest neighbor in the constellation alphabet set $\mathbb{A}$. Simply, the linear receiver output is
\begin{equation}
    {\bf{\hat x}} = \mathcal{D}\left( {{\bf{Wy}}} \right),
\end{equation}
where ${\bf{W}} \in {\mathbb{C}^{\mathcal{N} \times \mathcal{M}}}$ denotes the linear transformation chosen according to various optimization criteria. Specifically, the most commonly used linear detection receivers select ${\bf{W}}$ based on criteria such as matched filter (MF) detector to maximize the sampled SNR, zero-forcing (ZF) detector to minimize interference, and the minimum mean square error (MMSE) detector for minimizing mean square error (MSE). We discuss these linear receivers below for OTFS related modulations based on their unified signal model. 

\subsubsection{MF Detector} The MF detector, also named maximum ratio combining (MRC), is well-known as the optimal linear filter developed for maximizing the received SNR of each stream. It treats interference from other sub-streams as purely noise. Hence, its linear transformation matrix is simply
\begin{equation}
	{{\bf{W}}_\text{MF}} = {{\bf{H}}^H},
\end{equation}
and the receiver generates output symbols as 
\begin{equation}
    {{{\bf{\hat x}}}_\text{MF}} = \mathcal{D}\left( {{{\bf{H}}^H}{\bf{y}}} \right).
\end{equation}
In addition to its transformation simplicity, it is also well known that the performance of MF detector degrades significantly for overloaded systems or ill-conditioned channel matrix.

\subsubsection{ZF Detector} 

The principle of ZF detection applies the inverse of channel matrix ${\bf{H}}$ as the linear transformation to fully remove (i.e. to zero out) the interference caused by the non-ideal channel matrix. The linear transformation matrix is the Moore-Penrose pseudo-inverse
\begin{equation}
	{{\bf{W}}_\text{ZF}} = {\left( {{{\bf{H}}^H}{\bf{H}}} \right)^{ - 1}}{{\bf{H}}^H},
\end{equation}
with which the final receiver output symbols are
\begin{equation}\label{ZF_Detector}
    {{{\bf{\hat x}}}_\text{ZF}} = \mathcal{D}\left( {{{\left( {{{\bf{H}}^H}{\bf{H}}} \right)}^{ - 1}}{{\bf{H}}^H}{\bf{y}}} \right).
\end{equation}
Zero-forcing ensures that the co-channel interference amongst the multiple input symbols is completely eliminated. However, the zero-forcing transformation may lead to the enhancement of the transformed noise power
$ \|\left( {{{\bf{H}}^H}{\bf{H}}} \right)^{ - 1}{{\bf{H}}^H} \boldsymbol{\omega}\|^2$ (i.e., noise enhancement) given ill-conditioned channel matrices. Thus, zero-forcing detection is beneficial against well-conditioned channels and under low channel noise.

\subsubsection{Linear MMSE (LMMSE) Detector} 
The LMMSE detection takes into consideration of both co-channel symbol interference and channel noise. To minimize the sum MSE in data symbols resulting from the co-channel symbol interference and noise, the LMMSE receiver transformation matrix can be found via
\begin{equation}
    {{\bf{W}}_\text{LMMSE}} = \mathop {\arg \min }\limits_{{{\bf{W}}_\text{}} \in {\mathbb{C}^{\mathcal{N} \times \mathcal{M}}}} \mathbb{E}\left( {{{\left\| {{\bf{x}} - {{\bf{W}}_\text{}}\cdot {\bf{y}}} \right\|}^2}} \right),
\end{equation}
which minimizes the mean-square error between the original transmit data symbols ${\bf{x}}$ and the transformed symbol array after the linear transformation matrix ${\bf{W}}$. According to \cite{Detector_1}, the closed-form solution is given by
\begin{equation}\label{LMMSE_W}
    {{\bf{W}}_\text{LMMSE}} = {\left( {{{\bf{H}}^H}{\bf{H}} + {\sigma_\omega ^2}{\bf{I}}} \right)^{ - 1}}{{\bf{H}}^H},
\end{equation}
where ${\sigma_\omega ^2}$ is the noise power and signal elements in $\mathbf{x}$ have unit signal power. The LMMSE detector generates output symbols via
\begin{equation}\label{LMMSE_Detector}
    {{{\bf{\hat x}}}_\text{LMMSE}} = \mathcal{D}\left( {{{\bf{W}}_\text{LMMSE}}
    \cdot {\bf{y}}} \right).
\end{equation}
Unlike the ZF detector in (\ref{ZF_Detector}), the LMMSE detector in (\ref{LMMSE_Detector}) achieves a better balance between interference cancellation and noise suppression by jointly minimizing the total MSE caused by both co-channel interference and noise. Hence, the LMMSE detector often achieves better performance than the ZF detector under strong channel noise.

From (\ref{LMMSE_W}), we note that the LMMSE detector also requires a matrix inverse, leading to high computational complexity for high dimensional symbol size $\mathcal{N}$. In response, the authors of \cite{Detector_C1} and \cite{Detector_C2} proposed a low-complexity LMMSE detector by exploiting the sparsity and/or quasi-banded structure of OTFS channel matrix with little or no performance loss. Another lower complexity alternative is the proposal of successive interference cancellation (SIC) receiver based on typical channel matrix decomposition \cite{Detector_C5,Detector_C6}. To further enhance LMMSE receiver performance, an iterative least squares minimum residual (LSMR) detector has shown promise in both OTFS \cite{Detector_C3} and MIMO-OTFS \cite{Detector_C4} systems.

\subsection{High Performance Non-Linear Detection Receivers}

\subsubsection{Decision Feedback Detector} 
\begin{figure}
  \centering
  \includegraphics[width=0.48\textwidth]{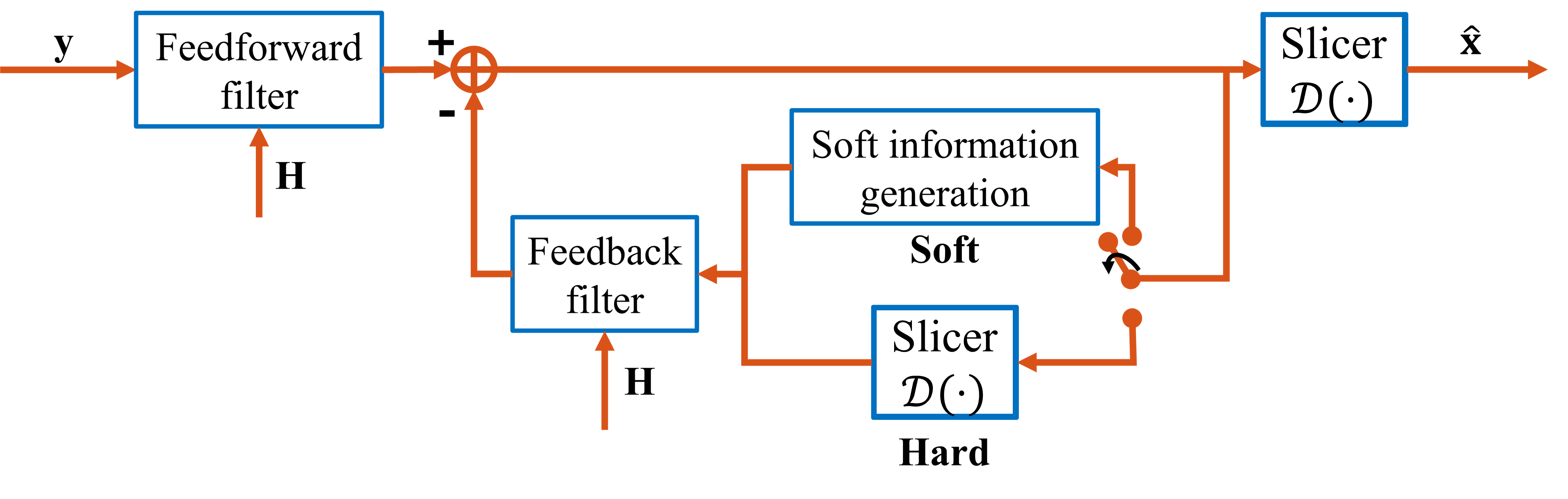}
  \caption{Block diagram of the decision feedback equalization detector.}\label{Detector_DFE}
\end{figure}

Decision feedback equalization (DFE) detector structure is a well-known and popular solution owing to its ability for performance improvement over linear detectors and the relatively low complexity. Generally speaking, the DFE applies the MMSE criteria \cite{Detector_2} in its feedforward processor design and also includes a feedback filter as shown in Fig. \ref{Detector_DFE}. In hard decision DFE,  tentative decisions on past symbols are used for interference removal from current data decisions. In \cite{Detector_3}, the authors propose an iterative hard DFE based on the MRC for zero-padded (ZP)-OTFS systems. Hard decision often suffers from error propagation, i.e., the feedback of incorrect decisions may generate more errors in subsequent symbol decisions in a decision error burst. To reduce error propagation due to erroneous tentative hard decisions, soft DFE \cite{Detector_4,Detector_5,Detector_6} generates soft symbol decisions in accordance with the symbols' associated posterior probabilities to replace the hard finite symbol decision. To further alleviate error propagation and to enhance system performance, a bidirectional structure \cite{Detector_7,10384469} can further take advantage of the different decision errors and noise distributions at the forward and backward DFEs for OTFS system outputs. Such a design can efficiently cancel the causal and non-causal interference and exploit the bidirectional diversity to achieve expected performance. 

\begin{figure}
  \centering
  \includegraphics[width=0.25\textwidth]{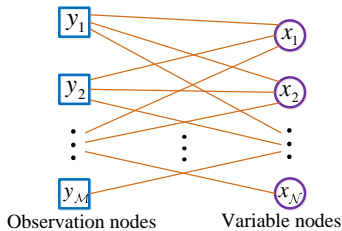}
  \caption{Factor graph example.}\label{Detector_facgra}
\end{figure}
\subsubsection{Non-Memory-Based Message Passing Detector}\label{subsubsec:NonMemoryMessagePassing}
\begin{figure}
  \centering
  \includegraphics[width=0.45\textwidth]{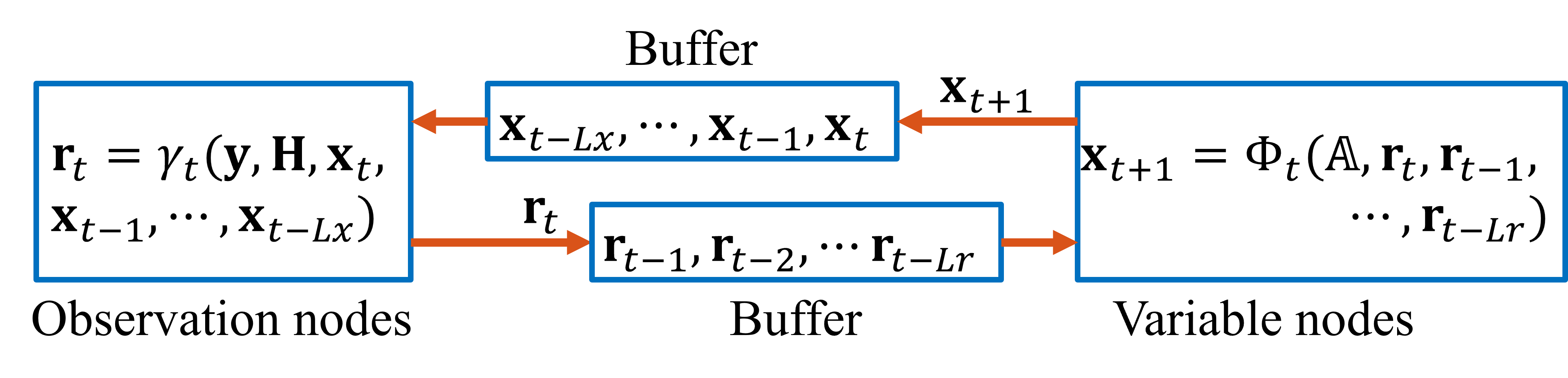}
  \caption{Block diagram of the memory/non-memory message passing detector.}\label{Detector_NMMP}
\end{figure}

The message passing algorithm, also well known as belief propagation (BP) algorithm, may achieve near maximum likelihood (ML) performance but with much lower complexity. Generally speaking, the equivalent effective channel matrix in (\ref{linear_model}) can be illustrated as a certain graphical model as shown in Fig.~\ref{Detector_facgra}. To estimate channel input is equivalent to drawing inference by approximately computing the marginal posterior distribution in the corresponding factor graph. However, exact high-dimensional Bayesian inference is intractable, thereby leading to approximate inference methods. Specifically, the posterior probability of each transmitted symbol is approximated by passing messages that marginalize over other symbols on the factor graph. This message passing process is repeated until convergence. Many existing studies mainly focus on the non-memory-based message passing detectors which have the basic structure of Fig.~\ref{Detector_NMMP} with the memory depth $L_x=0$ and $L_r=0$. 

Orthogonal approximate message passing (OAMP), also known as vector approximate message passing (VAMP) \cite{Detector_8,Detector_9,Detector_10} is a typical non-memory-based message passing detector involving two local processors, i.e., a so-called linear estimator (LE) operating on the observation nodes and a non-linear estimator (NLE) on the variable nodes. The successful performance improvement of OAMP/VAMP is achieved by imposing certain orthogonality constraints between the input and output estimation errors of each processor. However, the high computational cost from either matrix inverse or singular-value decomposition (SVD) within OAMP/VAMP detection limits their application to large-scale systems.

As shown earlier, the equivalent effective channel matrix demonstrates potential sparsity for most OTFS-related modulations. For this reason, efficient low complexity Gaussian message passing (GMP) detector \cite{Detector_11,Detector_12} may take advantage of such sparsity. Specifically, the GMP leverages a sparse factor graph and applies Gaussian approximation of the interference terms to further reduce complexity. To mitigate the potential positive reinforcement in belief propagation caused by the self-correlated signal exchange in iterative message passing, GMP detector tries to discard {\em a priori} message component in its subsequent estimation. The performance of GMP may degrade because its estimate exhibits poorer accuracy than the {\em a posteriori} probability (APP) estimate. Alternatively, expectation propagation (EP) is a powerful Bayesian inference technique \cite{Detector_13,Detector_14} that approximates the true distribution as Gaussian and obtains the posterior probability through iterations on the factor graph. EP presents a good trade-off between the APP and GMP algorithms. EP partially utilizes the {\em a priori} message to improve its estimate while avoiding the positive reinforcement problem caused by the self-correlated signal exchange during the iterations. Therefore, EP could achieve better performance than the GMP. For more detailed discussions, one may refer to \cite{Detector_9,Detector_15}. Like most BP solutions, GMP and EP detectors may suffer from performance loss if the underlying factor graph contains too many short girths (i.e., girth-4). Against the effects of short girths, solutions such as damping could improve the performance. However, optimized damping design remains elusive. 

In summary, existing non-memory-based message passing detectors often suffer from performance loss because of short girths in the factor graph (e.g., GMP and EP) or incur high computational complexity (e.g., OAMP/VAMP). Hence, the design of low-complexity replica Bayes-optimal message passing algorithms for general channel matrices remains a very interesting research direction. 

\subsubsection{Memory-Based Message Passing Detector}
The conventional approximate message passing (AMP) algorithm \cite{Detector_16,Detector_17}, derived from belief propagation (BP) with Gaussian approximation and the first order Taylor approximation is a memory-based detector. In AMP, a so-called ``memory Onsager'' term assists the matched filter in approaching a locally optimal LMMSE. Although AMP only adopts a low complexity matched filter, its performance is guaranteed by state evolution for zero-mean independent and identically distributed (IID) Gaussian channel matrices. However, AMP may perform poorly or even diverge for channel matrices with highly correlated entries. An alternative known as unitary approximate message passing (UAMP) \cite{Detector_18,Detector_19} utilizes a unitary transformation to overcome the issues with correlated channel matrices. Another efficient UAMP receiver aided by message feedback interference cancellation \cite{10384469} utilizes the latest feedback messages from variable nodes for more reliable interference cancellation and performance improvement. Note, however, that the complexity of UAMP algorithm may be as high as that of the OAMP/VAMP detector owing to its use of SVD operation.

More recently, long-memory message passing, where all the preceding messages are adopted for current estimation, was proposed in \cite{Detector_20}. The basic structure of the memory-based message passing is shown in Fig. \ref{Detector_NMMP}. Specifically, Takeuchi proposed a convolutional AMP (CAMP) \cite{Detector_21} that replaces the Onsager term of the AMP with a convolution of all preceding messages. CAMP has lower complexity than AMP but its convergence is relatively slow. CAMP exhibits convergence issues, particularly against ill-conditioned channel matrices. Although empirical damping may improve the convergence of CAMP, such damping in CAMP could weaken the orthogonality and the asymptotic Gaussianity of estimation errors, leading to significant performance loss.

To tackle the various shortcomings of AMP, UAMP, and CAMP, the framework of Memory AMP (MAMP) proposed in \cite{Detector_20} applies the orthogonality principle by utilizing a low-complexity long-memory matched filter for interference suppression. MAMP's use of long memory requires stricter orthogonality to induce the asymptotic IID Gaussianity for estimation errors. Specifically, the traditional principle of orthogonality between the current input estimation error and current output estimation error in non-memory-based message passing is generalized to require that the current output estimation error be orthogonal to all preceding input estimation errors. For simplicity, the MAMP detector \cite{Detector_22} only involves a low-complexity matched filter in each iteration, and applies finite terms of matrix Taylor series to approximate matrix inverse. Based on the expanded orthogonality principle and a closed-form damping solution, MAMP detector delivers superior performance over GMP and EP detectors based on heuristic damping, and achieves similar performance to that of OAMP/VAMP at lower complexity.

\begin{figure*}
    \centering
    \subfloat[]{\includegraphics[width=.32\textwidth]{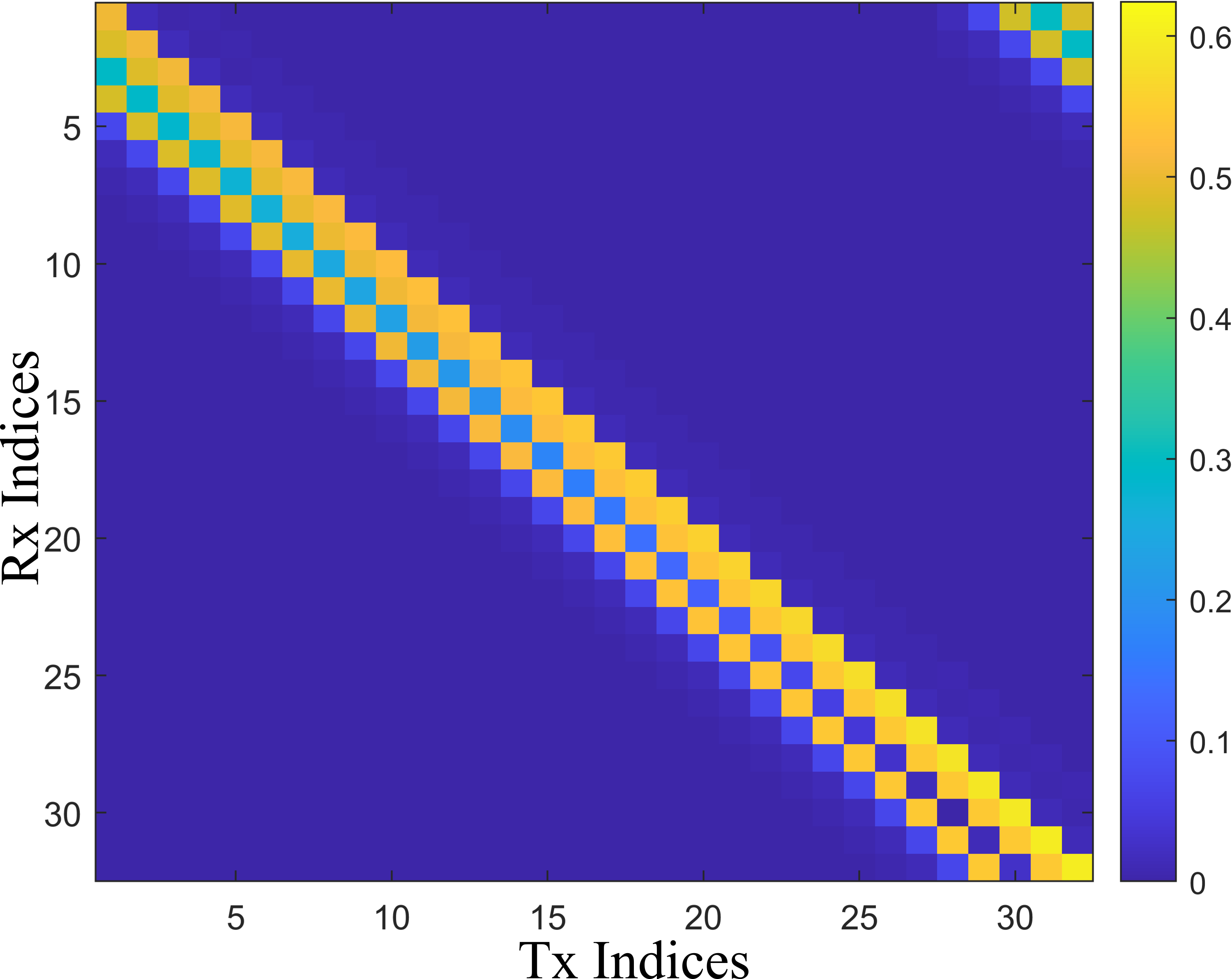}}
    \hfill
    \subfloat[]{\includegraphics[width=.32\textwidth]{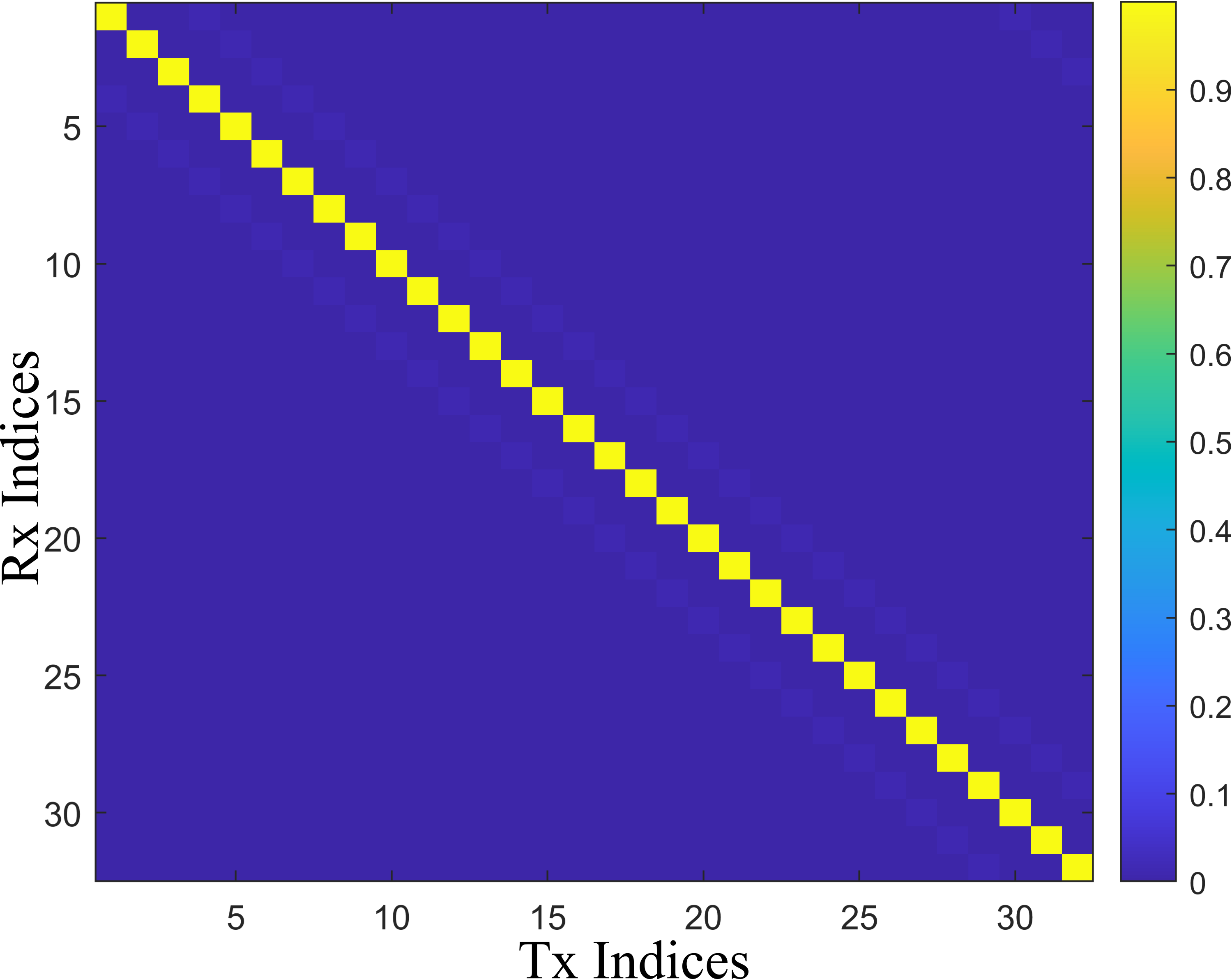}}
    \hfill
    \subfloat[]{\includegraphics[width=.32\textwidth]{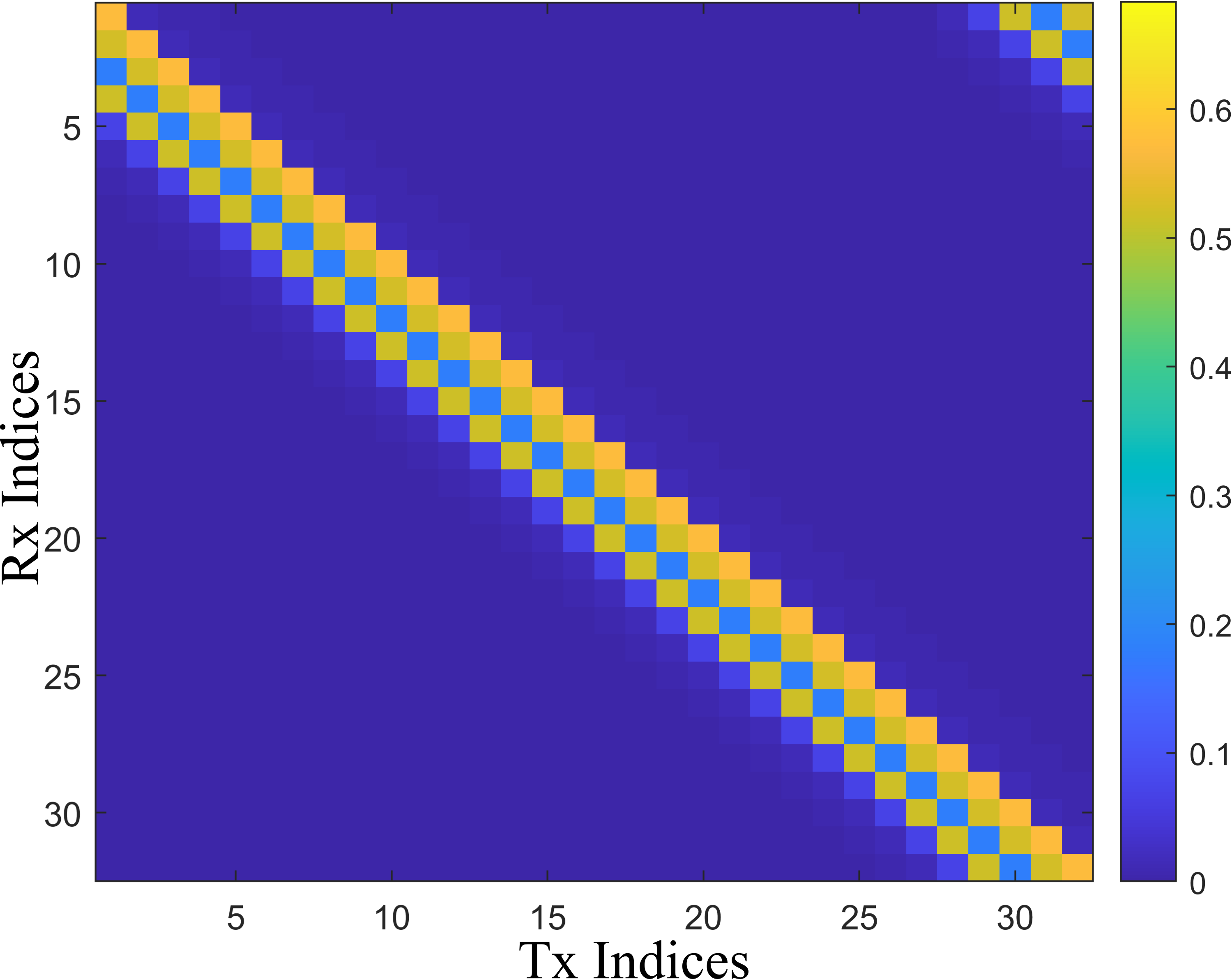}}
    \caption{Example of time domain channel matrix. (a) Doubly-selective fading; (b) Time-selective fading; (c) Frequency-selective fading.}
    \label{fig:TimeChannelMatrices}
\end{figure*}

\subsubsection{Cross-Domain Iterative Detector}
\begin{figure*}
  \centering
  \includegraphics[width=0.82\textwidth]{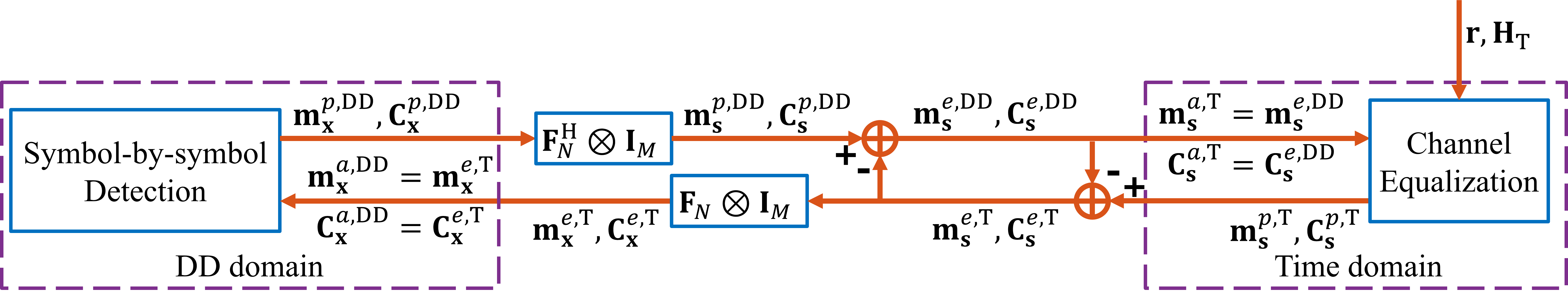}
  \caption{Block diagram of the cross domain iterative detector.}\label{Detector_CDMP}
\end{figure*}
As discussed above, a majority of existing works on OTFS detection take advantage of channel sparsity in DD domain to reduce receiver complexity. Nevertheless, some DD domain channels can still be dense due to the fractional (off-grid) Doppler shifts given a limited resolution of the Doppler dimension, particularly for insufficiently long OTFS frame duration. For moderate to short OTFS frames, conventional detection in DD domain may become quite complex since the corresponding channel matrix is not sufficiently sparse in the DD domain. To tackle this problem, it is helpful to notice that the effective channel matrix in time domain remains sparse and shows banded support even in fractional Doppler scenarios, as our Fig.~\ref{fig:TimeChannelMatrices} has shown. Such observation inspired a novel cross domain iterative detection algorithm for OTFS modulation in \cite{Detector_23}. The proposed cross-domain iterative algorithm can exploit not only the time domain channel sparsity but also the DD domain symbol constellation constraints to enhance the error performance of OTFS systems with low complexity. As illustrated in Fig. \ref{Detector_CDMP}, its basic detection considers both the time domain and delay-Doppler (DD) domain such that the algorithm updates extrinsic information iteratively between these two domains under a unitary transformation. It is worth mentioning that the concept of cross-domain iterative detection can directly extend to other OTFS-related demodulation of ODDM, OTSM, OCDM, and AFDM by exploiting time domain channel sparsity for low complexity and utilizing their corresponding symbol domain constellation constraints for performance enhancement. Additionally, a new cross-domain MAMP (CD-MAMP) detector proposed in \cite{10522098} further reduces the complexity by maintaining a receiver performance similar to that of the original OAMP. Further investigation in designing low complexity cross-domain iterative detection remains a promising research direction for OTFS-related modulation schemes.

\subsection{Deep Learning Inspired Detection Receivers}
Most conventional approaches discussed above rely heavily on sufficiently accurate signal models. They tend to be more sensitive to modeling errors and/or error propagation in message passing due to decision errors. In addition, these receivers require explicit knowledge of channel state information (CSI) in the form of an effective channel matrix acquired via channel estimation with the support of pilot overhead. With the recent success of various state-of-the-art deep learning architectures in natural image processing, media processing, and computer vision, the design of detection algorithms based on deep learning is attracting rapidly growing attention from both academia and industry.

Deep-learning inspired detection (DLID) possesses powerful non-linear representation and processing abilities to generate improved detection results. DLID is less sensitive to modeling errors and decision inaccuracy such as imperfect CSI and symbol errors. Specifically, existing DLID receivers can be divided into two categories: data-driven methods and model-driven methods. Data-driven methods \cite{Detector_24,Detector_25,Detector_26,Detector_27} often adopt some known neural network architecture such as convolutional neural network (CNN) to replace traditional detection functionalities, whereas model-driven approaches \cite{Detector_28,Detector_29,Detector_30,Detector_31,10516684,Detector_32,Detector_33} exploit the conventional detection principles to design a customized neural network to achieve more accurate and/or more robust performance.

Consider examples of data-driven DLID receivers. A recent neural network OTFS detector proposed in \cite{Detector_24} does not require explicit CSI, providing robust equalization performance, even in low SNR regimes. Another proposal is a two-dimensional (2D) CNN based detector by \cite{Detector_25}. The authors of \cite{Detector_26} presented a data-driven intelligent receiver from the perspective of signal processing for OTFS systems. Deep-learning systems can also tackle the effect of hardware impairments in OTFS, as shown in \cite{Detector_27}. We note, however, that the data-driven deep learning methods remain poorly interpretable and show over-reliance on empirical results without strong analytical support. In addition, the number of trainable deep-learning neural network parameters is generally too large for small wireless devices. Therefore, an important and challenging open question is the investigation to design lightweight detection models with strong convergence assurance.

Model driven DLID receiver leverages existing signal models as its basic foundation.  For example, the proposed OTFS signal detections in \cite{Detector_28,Detector_29} exhibit low training cost and good interpretability by unfolding the conventional detection algorithm into a deep learning solution. Specifically, each iteration of the detection algorithm is unfolded into a layer-wise network. The trainable parameters are then added to this network to accelerate the convergence of the algorithm, as well as improve the detection performance. Graph neural network (GNN) \cite{Detector_30} and Bayesian neural network \cite{Detector_31} can also take advantage of deep learning versatility and Markov random field to efficiently extract hidden features of receiver data to achieve more accurate detection. In \cite{10516684}, an AMP and a GNN module collaborate to enhance OTFS detection accuracy by iteratively exchanging their intermediate estimation results. Another autoencoder (AE)-based enhanced OTFS (AEE-OTFS) modulation is proposed in \cite{Detector_32} to improve the communication reliability. Recognizing the limitations of deep learning receivers under model uncertainty in time varying wireless environment, a novel detection framework inspired by contrastive learning \cite{Detector_33} provides the potential for achieving high detection accuracy, strong robustness, and fast convergence for ODDM signal reception. 

We note that despite many promising results, the applications of DLID in receiver designs for many related modulation schemes such as OTFS, ODDM, and chirp-based AFDM are still in an early stage. The sensitivity of DLID to environmental variation and training dataset are only two of many open challenges. In the presence of many challenges and opportunities, significant research efforts are still needed to tailor neural network architectures specifically for the unique characteristics of OTFS-related modulation schemes. 

\subsection{Some Head-to-Head Detection Comparisons}
\begin{figure}
  \centering
  \includegraphics[width=0.48\textwidth]{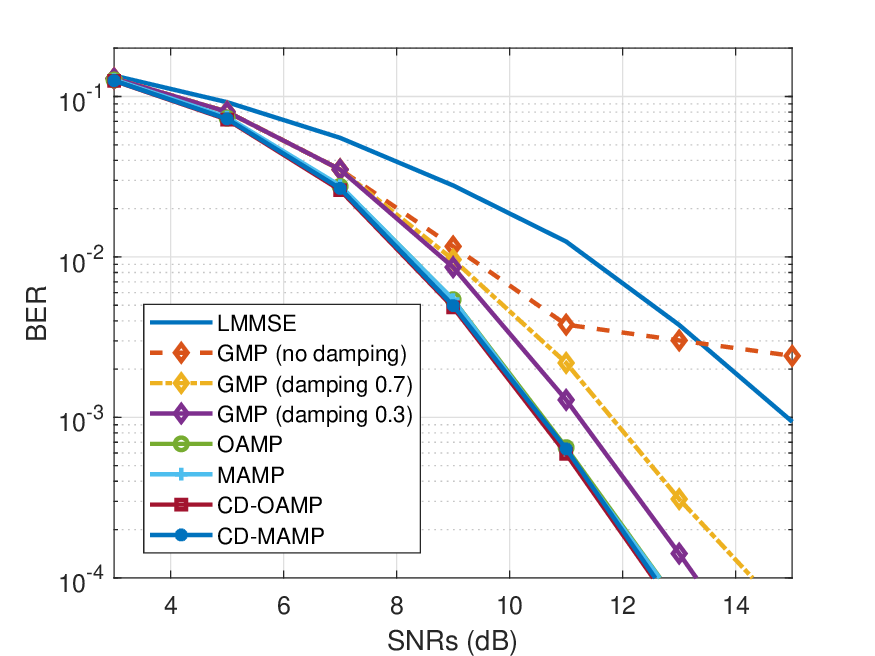}
  \caption{BER performance of different detectors in OTFS-SRRC system.}\label{Fig:BERDetector}
\end{figure}
Now we provide a sneak peek at the performance comparison of several popular detectors for OTFS-related modulations. Our focus is the BER evaluation of various detection principles implemented for an OTFS-SRRC receiver. In this numerical experiment, we generate random QPSK data symbols of size $M=64$ and $N=16$. The subcarrier spacing $\Delta f$ remains at $15$ kHz and the symbol period equals $T=1/\Delta f$. We use an SRRC pulse with a roll-off factor of 0.5 and a time duration of $16T$ at both transmitter and receiver for pulse shaping and spectrum constrain. In our tests, we configure channel parameters to have three distinct paths, each featuring a random complex gain of power $1/3$, an off-the-grid delay $\tau_{i}\in [0, \tau_{max}]$, and a random Doppler $\nu_{i}=\nu_{max}\cos{\theta_{i}}$, where $\theta_{i} \in [-\pi, \pi]$ denotes a random angle-of-arrival (AoA). We set the maximum delay $\tau_{max}$ to $4\frac{T}{M}$, and set the maximum Doppler shift $\nu_{max}$ according to mobile velocity of 300 km/h.

We select the LMMSE detector in our linear receiver. For non-memory-based message passing, we test GMP \cite{Detector_11,Detector_12} and OAMP \cite{Detector_8,Detector_9,Detector_10} detectors as receivers, respectively. For memory-based message passing, we choose the Memory AMP (MAMP) \cite{Detector_22} detector for receiver comparison. In the case of cross-domain iterative detection, we integrate the foundational cross-domain framework from \cite{Detector_23} with OAMP and MAMP, leading to an OAMP-based cross-domain iterative receiver (CD-OAMP) and an MAMP-based cross-domain iterative receiver (CD-MAMP). 

Fig.~\ref{Fig:BERDetector} presents the BER performance of the aforementioned receivers. It is clear that the linear receiver based on LMMSE delivers the weakest receiver performance. In comparison, our tests of non-linear detectors provide similar BER performance superior to LMMSE. Among the nonlinear receivers, GMP detection exhibits worse performance and is sensitive to the damping parameters, which may even experience an error floor if inappropriate damping values are chosen. Unfortunately, there is no efficient damping solution for GMP detectors currently. All other non-linear detectors, regardless of their complexity, generate similar detection accuracy at different levels of SNR. TABLE~\ref{tab1} further presents a complexity comparison of the considered detectors, where $\mathcal{T}$ denotes the iteration number, $\mathcal{L}$ is the average number of non-zero entries in each row of the effective channel matrix ${\bf{H}}$, and $P$ is the maximal channel tap ($P < \mathcal{L} \ll \mathcal{N}$). It is obvious that CD-MAMP can yield practical implementation advantages with low complexity and desired BER performance.
\begin{table}
  \begin{center}
    \caption{Complexity comparison between different detectors.}
    \renewcommand{\arraystretch}{1.3}
    \label{tab1}
    \begin{tabular}{c|c} 
    \hline
      Detector & Computational Complexity\\
      \hline
      \thead{LMMSE} & $\mathcal{O}\left( {{\mathcal{N}^3}} \right)$\\
      \hline
      GMP & $\mathcal{O}\left( \mathcal{MLT} \right)$\\
      \hline
       OAMP & $\mathcal{O}\left( {{\mathcal{N}^3}\mathcal{T}} \right)$\\
      \hline
       MAMP & $\mathcal{O}\left( \mathcal{MLT} \right)$\\
      \hline
      CD-OAMP & $\mathcal{O}\left( {{\mathcal{N}^3}\mathcal{T} + 2\mathcal{NT}\log \mathcal{N}} \right)$\\
      \hline
      \thead{CD-MAMP} & \thead{$\mathcal{O}\left( {\mathcal{M}P\mathcal{T} + 2\mathcal{NT}\log \mathcal{N}} \right)$}\\
      \hline
    \end{tabular}
  \end{center}
  \vspace{-1.5em}
\end{table}

\section{Opportunities and Challenges}

In this section, we outline the opportunities and challenges for future OTFS research, which is also applicable to other related modulation schemes. We stress issues associated with modulations in high mobility scenarios and identify areas for future research and development to enhance receiver performance in such environments. Because of the similarity of the various OTFS related modulations, we do not find it necessary to enumerate all the variants in this section. Instead, we use OTFS as an example modulation scheme in our presentation for opportunities and challenges.

\label{SEC:Opportunities}
\subsection{Channel Estimation}\label{subsec:CE}
One of the fundamental challenges in implementing OTFS is effective channel estimation, which is critical for achieving reliable system performance. OTFS places transmit data symbols in the DD domain over which the channel characteristics, particularly channel delay and Doppler shifts, are more sparse in comparison with the time-frequency domain. Despite the DD domain channel sparsity, channel estimation in the DD domain \cite{meng2023flag} lends to some distinct difficulties previously unseen in other traditional modulations. 

One particular obstacle to channel estimation arises from interference from the data-bearing symbols to the pilot symbols that channel estimation relies on \cite{9539066,8727425,ranasinghe2024low}. Such OTFS inter-symbol-interference is exacerbated by the interference spread from multipath delays and Doppler shifts off the DD resource grid \cite{9440710,9456029}. Thus, effective channel estimation in OTFS must jointly involve pilot design, algorithm development, and a deeper understanding of underlying channel models to optimize performance and reliability.

Existing pilot designs for channel estimation belong to the categories of embedded pilots, superimposed pilots, interleaved pilots, and training sequences. As an early exploration for OTFS channel estimation, embedded pilot design uses discrete impulse pilots plus null guard symbols around each pilot to separate the pilot on the received signal from the channel effect of data symbols, as proposed in \cite{CE_1}. These guard symbols mitigate the cross-interference from data symbols to the received pilots in the DD domain. Though experiments in \cite{CE_1} demonstrate good performance over a practical Extended Vehicular A model \cite{EVA1} with off-grid channel Doppler shifts, embedded pilot design requires large pilot power to suppress data symbol interference due to off-grid multipath delays and Doppler shifts. Higher pilot power further increases the peak-to-average power ratio (PAPR) for OTFS transmit signal, thereby leading to worsened energy efficiency. Equally important is the increasing use of guard resources for channels with large multipath delay spread and Doppler shift, which significantly degrades spectral efficiency. The loss of significant channel resources to guard symbols would become more severe in MIMO-OTFS systems in which more antennas require more channel estimation pilots. 

One notable channel estimation algorithm proposed for OTFS is sparse Bayesian learning (SBL) \cite{CE_SBL1,10436557,10574336}. Instead of estimating the effective channel response in DD domain by considering off-grid delays and Doppler shifts, SBL estimates the original and more sparse DD domain channel response. By taking advantage of the sparsity of the DD domain channel responses, the proposed SBL channel estimation framework can achieve better channel estimation performance by using a smaller guard source area under the same pilot design as \cite{CE_1}. Additional SBL-inspired channel estimation proposals \cite{CE_SBL2,CE_SBL3,CE_SBL4,10370744,9590508} have further reduced the complexity and extended the framework to MIMO systems and satellite communications.

\subsection{Synchronization}\label{subsec:Sync}
{
Synchronization often poses a significant challenge in the design of practical communication systems, particularly under time-varying channel conditions. Errors such as timing offset (TO) and carrier frequency offset (CFO) can severely degrade receiver performance in OTFS systems. For example, integer TO and CFO lead to cyclic shifts in the delay and Doppler dimensions, causing substantial interference in the received OTFS signal \cite{9303350}. Moreover, uplink timing synchronization of users is crucial for ensuring efficient operation in multi-user OTFS (MU-OTFS) systems \cite{9241423}.}

{
Existing OTFS synchronization methods can be categorized into two groups. The first group uses a preamble frame in addition to the OTFS frame, employing training sequences specifically designed for synchronization. Alok et al. \cite{9241423} proposed an OTFS-based random access preamble waveform for user terminals and developed a TO estimation method at the base station for uplink MU-OTFS systems. The estimated TO values are then fed back to users for uplink timing correction. Mohammed et al. \cite{9429220} focused on the downlink direction by designing a detection preamble along with two OTFS-based pilot and secondary cell identity blocks for each downlink frame. Furthermore, Char-Dir et al. \cite{10614751} designed a rectangularly-pulsed, comb-type preamble waveform for initial time synchronization in OTFS systems.}

{
The second group, on the other hand, relies on pilots and sequences embedded within the DD grid to achieve synchronization. Suvra et al. \cite{9303350} examined the impact of synchronization errors on channel matrix and proposed a channel estimation method alongside a SIC receiver to mitigate the interference caused by the residual TO and CFO. Mohsen et al. \cite{9916227} proposed TO and CFO estimation techniques based on a 2D correlation function for OTFS systems, utilizing the same pilot design in \cite{CE_1}. Mohammed et al. \cite{10437461} employed a Zadoff-Chu sequence as the primary synchronization signal embedded in the DD grid for downlink synchronization scenario. Additionally, Mohsen et al. \cite{10278633} proposed pilots that combine a constant amplitude pilot sequence with a CP to reduce the PAPR of the transmitted signal. This concept also extended to uplink MU-OTFS systems in \cite{bayat2024synchronizationmultiuseruplinkotfs}. Songyan et al. \cite{10314509} tackled CFO estimation in OTFS-based LEO satellite communication systems by embedding maximum length sequences (MLS) along the delay dimension to estimate the large CFO. Meanwhile, Jiazheng et al. \cite{10742117} utilized a DFT processed MLS along the Doppler dimension as the pilot signal and proposed a joint time synchronization and channel estimation algorithm.
}

\subsection{MIMO and Multi-user OTFS Systems}

The integration of OTFS with MIMO and multiuser systems brings forth new design complexities and challenges. The complexity of MIMO-OTFS systems lies in the need to effectively coordinate the data and pilot transmissions on multiple transmitting antennas. Additionally, the design of efficient channel estimation and data detection algorithms for MIMO-OTFS systems poses multi-fold difficulties in view of the intricate signal interactions across time, frequency, and spatial domains. As discussed in Section~\ref{subsec:CE}, MIMO increases the burden for OTFS channel estimation as the number of unknown channels swells. Similarly, the multi-user will also bring similar challenges for OTFS channel estimation. To improve the coverage, capacity, and reliability of mobile wireless communications, distributed antenna system \cite{6180090} is more efficient and promising. Large-scale application of distributed antenna in OTFS systems \cite{9906092}, requiring enhanced collaboration among multiple distributed antennas and advanced transceiver design. 

MU-OTFS system design is significantly more challenging due to the greater difficulty in managing inter-user interference than the corresponding and traditional orthogonal frequency division multiple access (OFDMA). In both uplink and downlink directions, a diverse range of user Doppler shifts on or off grids lead to highly complex and often dynamic inter-user interference scenarios. Such complication only worsens with an increasing number of users. To support massive multiple access and explosive transmission needs, the integration of NOMA \cite{8786203,Detector_8,10022044,9411900} and rate splitting multiple access (RSMA) \cite{10462183,10804646} with OTFS is also an interesting research direction for high mobility communication systems.

In addition, massive machine-type communication (MTC) \cite{9537931} is a crucial enabler for IoT applications, playing a key role in the development of smart cities, industrial 4.0, and connected vehicles. A key characteristic of MTC is that a vast number of devices sporadically transmit short data packets. The traditional grant-based scheme used for human-type communication (HTC) involves a complicated handshaking procedure, which can no longer meet the demands of large-scale MTC. Recently, grant-free random access has emerged as a promising technology to accommodate massive connectivity with reduced signaling overhead and transmission latency for 6G MTC. Efforts have already been made in \cite{9849120,9928043,10110015,9496190,10274133} to explore an efficient grant-free random access scheme based on OTFS transmissions for performance enhancement of high-mobility MTCs.

These challenges necessitate future research endeavors to develop new analytic tools, advanced signal processing techniques, and optimization strategies to fully exploit the potential of various MU-OTFS and MIMO-OTFS configurations. 

\subsection{PAPR}
For wireless communication signals, a high PAPR significantly impacts power amplifiers at the transmitter, leading to nonlinear distortion, low energy efficiency, and limited maximum transmission power and coverage range \cite{PAPR1}. In OTFS systems, the maximum PAPR grows linearly with the number of temporal slots $N$ \cite{OTFS_PAPR1}. Consequently, OTFS can have a lower PAPR compared to the conventional OFDM system as $N<M$ in general, where $M$ is the number of subcarriers and controls the PAPR of OFDM. However, the PAPR of OTFS signals still grows as $N$ increases. For a relatively large value of $N$, OTFS signals with rectangular pulses exhibit a similar PAPR to conventional OFDM signals\cite{OTFS_PAPR2}. Another cause of high PAPR comes from the pilot design. As introduced in Section~\ref{subsec:CE}, some channel estimation methods, such as the single embedded pilot-aided channel estimation scheme in \cite{CE_1}, require large pilot power to suppress data symbol interference due to the off-grid multipath delays and Doppler shifts, which further increases the PAPR of OTFS signals.

Efforts have already been made to address the PAPR issue in OTFS. Gao et al. \cite{PAPR2} reduce the PAPR of pilot-embedded OTFS signals by using the iterative clipping and filtering (ICF) scheme. Francis et al. \cite{PAPR3} propose an indexing design of DD bins to reduce the PAPR and increase the diversity order of OTFS. Liu et al. \cite{PAPR4} propose a deep learning based PAPR reduction method with an autoencoder architecture. A unique OTFS frame structure is further proposed in \cite{PAPR5} to reduce the PAPR. Kumar et al. \cite{PAPR6} proposed a hybrid technique that combines selective mapping with partial transmission sequence (SLM + PTS) that can lower the PAPR of the OTFS signal. On the other hand, different patterns of superimposed pilots are also proposed in \cite{PAPR7,PAPR8} for PAPR reduction. Another efficient approach to reduce the PAPR is to use the precoder at the transmitter \cite{PAPR9,PAPR10,PAPR11,yuan2024papr,tao2024daft}.

\subsection{Applications in Wideband Time-Varying channels}
Wideband doubly-spread fading channels are commonly countered in underwater acoustic (UWA) and ultra wideband (UWB) radio communications, where severe Doppler effect results in a frequency-dependent non-uniform shift of signal frequencies across the relatively large spectrum band, leading to a time-scaling in the received signal waveform. Although OTFS can exploit both channel delay and Doppler diversities for performance enhancement in doubly-selective narrowband fading channels, its performance degrades significantly in doubly-spread wideband channels. For example, the wideband OTFS systems suffer from the Doppler squint effect (DSE), where the Doppler shift is frequency-dependent \cite{OTFS_OC6}. Such frequency-dependent Doppler shifts can increase the interference and degrade the channel estimation accuracy for OTFS. One recent waveform design, known as orthogonal delay scale space (ODSS) modulation \cite{9772941}, can tackle the time-scaling effect of wideband time-varying channels. The ODSS pre-processing uses a 2D transformation from the Fourier-Mellin domain to the delay-scale domain and improves the performance when compared to OTFS. In view of its promising potential, further works should study the design of efficient ODSS detectors, both to improve performance robustness and to control receiver complexity. Other open research challenges include the generalization of ODSS multi-antenna systems at the transceivers, analysis of diversity-multiplexing trade-off,  and development of efficient pilot placement for channel estimation.  

\subsection{Integrated Sensing and Communications}
Because OTFS modulation has significantly better performance than OFDM in time-varying channels, it has also emerged as a promising candidate for integrated sensing and communications (ISAC). The OTFS-ISAC system design proposed in \cite{OTFSISAC_1} has considered a relatively simple mono-static radar scenario. This ISAC proposal provides an efficient approximate maximum likelihood (ML) algorithm for range and velocity estimation, integrated with a message passing detector for data symbol recovery. Another proposal of \cite{OTFSISAC_2,9724198} considered a more general multi-antenna system with a novel beamforming design. In \cite{9667103}, the authors propose a low-complexity OTFS sensing method for ISAC in Industrial IoT applications. An efficient delay-Doppler-angle estimation algorithm for MIMO-OTFS radar sensing is proposed in \cite{10463758} to mitigate ISI and inter-carrier interference (ICI) effects. Overall, the integration of sensing and communication functionalities \cite{10478980RIS} in OTFS systems is poised to become an integral part of future wireless networks, particularly in high-mobility scenarios involving vehicular, UAV, and robotics.

\subsection{Jamming and Anti-Jamming Considerations}
Even though OTFS modulation places data symbols on the DD domain, classic jamming signals such as narrowband interference (NBI) and periodic impulse noise (PIN) can still effectively impact OTFS signal quality. Such targeted jamming can drastically worsen the error performance of affected users \cite{OTFS_Jamming1}. In response, a simple yet effective method is to introduce resource hopping \cite{OTFS_Jamming1}, where the system pseudo-randomly changes the resource bins allocated to different users from one OTFS block to another. The study of OTFS jamming and effective counter-measures is still in its infancy. Future research may investigate OTFS vulnerability under a whole host of potential jamming sources and the development of robust anti-jamming strategies. 

\subsection{Other Challenges}
OTFS has shown great potential in suppressing the interference in doubly-selective fading channels, making it a promising candidate for next-generation wireless communication systems, particularly in scenarios with high mobility. However, its application in specific mobile communication scenarios presents unique challenges. For example, real-time vehicle-to-everything (V2X) systems demand low latency and high reliability, which requires robust and efficient channel estimation and decoder algorithms to mitigate the effects of mobility on OTFS performance \cite{OTFS_OC1}. Air-to-ground (A2G) platforms such as UAVs experience high path-loss and severe Doppler effects due to their high mobility, making wireless communication unstable. They are also equipped with limited battery life, so an energy-efficient communication scheme is needed. In these platforms, OTFS requires sophisticated trajectory planning and energy optimization to minimize the BER and power consumption \cite{OTFS_OC2}. Furthermore, the low earth orbit (LEO) satellite moves at a very high speed relative to the Earth's surface, causing extreme Doppler spread that can severely affect the communication quality. Despite OTFS can effectively overcome such Doppler spread, the severe path-loss caused by the long distance transmission requires a unique design in OTFS-based LEO satellite transmission \cite{OTFS_OC3,OTFS_Survey2}. 

In addition, millimeter-wave (mmWave) and terahertz (THz) communications are attracting widespread attention due to their abundant spectrum resources and higher spectral efficiency. For high-speed mobile communication systems, although OTFS has been applied to mmWave and THz communications \cite{OTFS_OC4,OTFS_OC5}, some additional challenges need to be further considered and addressed in the future. Similar to the UWB radio communications, the large bandwidth offered by mmWave and THz communications \cite{OTFS_OC7} and the high Doppler effects make the DSE more severe for OTFS. Another challenge brought by the mmWave and THz communications is the severe path loss caused by the high spreading loss. As a result, beam alignment and management are crucial for mmWave and THz communcations. Therefore, an efficient beamforming design and tracking scheme are required for OTFS-based mmWave and THz communications \cite{OTFS_OC8,OTFS_OC9,10164149}.

Recently, reconfigurable intelligent surfaces (RIS) received significant attention due to their potential to improve the capacity and coverage of wireless communication systems \cite{OTFS_OC10,OTFS_OC11,10478980RIS}. However, the precise estimation of the CSI is crucial for the performance enhancement of RIS-aided systems. In a high mobility scenario, the challenge of channel estimation is exacerbated due to channel aging and large Doppler effect. Thus, effective channel estimation in RIS-aided OTFS systems must jointly involve pilot design, RIS configuration, and advanced signal processing techniques to optimize performance and reliability. Several studies have been explored to integrate OTFS with RIS for performance enhancement of high-mobility communications. For example, Li et al. \cite{OTFS_OC12} develop a new transmission scheme to activate part of the hybrid RIS (HRIS) and estimate the CSI with low pilot overhead. Xu et al. \cite{OTFS_OC13} leverage a virtual Doppler frequency to optimize the configuration of RIS in DD domain, and propose an efficient DD-domain channel estimation algorithm. Li et al. \cite{OTFS_OC14} further propose a new transmission scheme that can reduce the channel training overhead with desired estimation accuracy, and optimize the configuration of RIS based on the predicted channel parameters.

\section{Conclusions}

This survey aims to provide a unified signal model for OTFS modulation and its many variants under well-known fading channel models. We provide a detailed overview of OTFS-related modulation schemes, including their connections to the conventional OFDM. By presenting a unified signal model under doubly-selective fading channel models, we demonstrate the clear relationship among the many OTFS-related modulation schemes, each viewed as a special case of OTFS through linear transformation and pulse-shaping. Through PSD and BER comparisons, we further verify the underlying relationship and highlight the robustness of these modulations in high-mobility scenarios. We offer a broad introduction of various detection schemes, focusing on their respective fundamental principles, performance, and complexity. To motivate future research endeavors, we discuss the opportunities and challenges that OTFS-related systems must overcome to achieve broad applications. Overall, our comprehensive review dispels the possible of confusion associated with the many OTFS-related proposals and lays the groundwork for continued exploration and development of OTFS-related system designs to address high speed mobile scenarios in future wireless technologies.

\appendix
\label{SEC:chirp}
In this appendix, we introduce another class of OTFS variants in the form of {\bf chirp-based multi-carrier modulations}. These transmission schemes utilize chirp signals to spread the data symbols to the full domain of the assigned time-frequency resources and share similar features as ISFFT in OTFS. They can also be captured by using the same signal model provided in Section~\ref{SEC:channelMtx} and TABLE~\ref{Table:DoubleSelective}.

\subsection{OCDM}
\begin{figure}[h!]
    \centering
    \subfloat[]{\includegraphics[width = 0.45\textwidth]{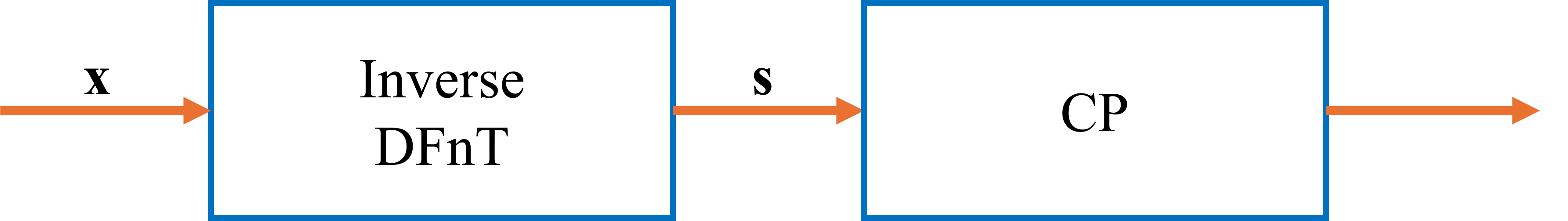}}\\
    \subfloat[]{\includegraphics[width = 0.45\textwidth]{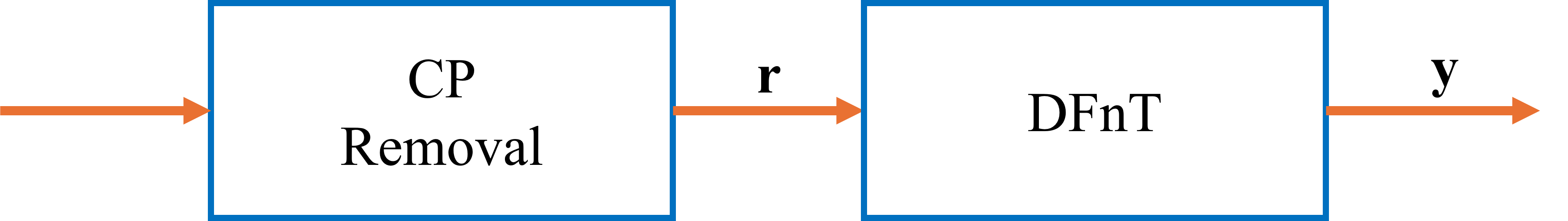}}\\
    \caption{Diagram of OCDM: (a) modulation; (b) demodulation.}
    \label{fig:ocdm}
\end{figure}
OCDM is orthogonally multiplexing a bunch of chirp waveforms overlapped temporally and spectrally. Let ${\bf{x}} \in {\mathbb{A}^{M \times 1}}$ denote the transmitted chirp symbols, the discrete time-domain OCDM signal is then obtained by inverse discrete Fresnel transform (DFnT) as below
\begin{equation}
    {\bf{s}} = {{\bf{\Phi }}^H}{\bf{x}},
\end{equation}
where ${\bf{\Phi }} \in {\mathbb{C}^{M \times M}}$ is the DFnT matrix with the $(m,n)$-th entry defined as 
\begin{equation*}
\begin{aligned}
\Phi \left( {m,n} \right) \!=\! \frac{1}{{\sqrt M }}{e^{ - j\frac{\pi }{4}}} \!\times\!
\begin{cases}
{e^{j\frac{\pi }{M}{{\left( {m - n} \right)}^2}}},&M \equiv 0\;(\bmod \; 2),\\
{e^{j\frac{\pi }{M}{{\left( {m + \frac{1}{2} - n} \right)}^2}}}, &M \equiv 1\;(\bmod \; 2).
\end{cases}
\end{aligned}
\end{equation*}
Note that the DFnT matrix is unitary, and its specific important properties can be found in \cite{OCDM_1,9346006} for details. To avoid inter-block interference, a CP that is at least larger than the maximum channel delay spread is added before ${\bf{s}}$.

After the transmission over the channel and discarding CP, the received time domain signal ${\bf{r}}\in {\mathbb{C}^{M \times 1}}$ is transformed back to the chirp symbols via DFnT as described below
\begin{equation}
    {\bf{y}} = {\bf{\Phi r}}.
\end{equation}
The system diagram of OCDM is shown in Fig.~\ref{fig:ocdm} for details.

\subsection{AFDM}
\begin{figure}[h!]
    \centering
    \subfloat[]{\includegraphics[width = 0.45\textwidth]{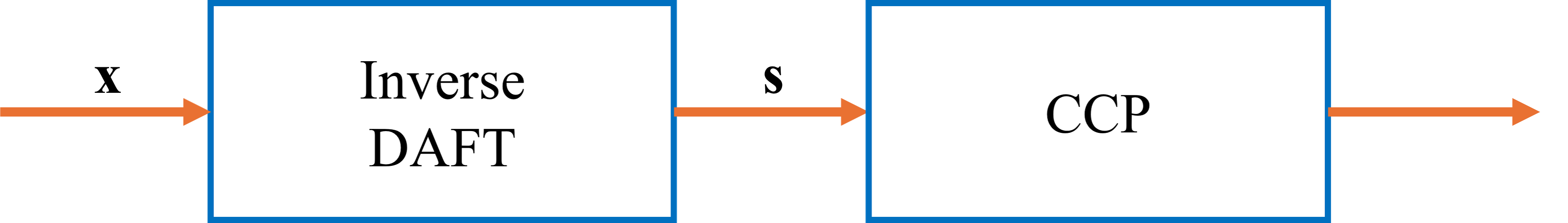}}\\
    \subfloat[]{\includegraphics[width = 0.45\textwidth]{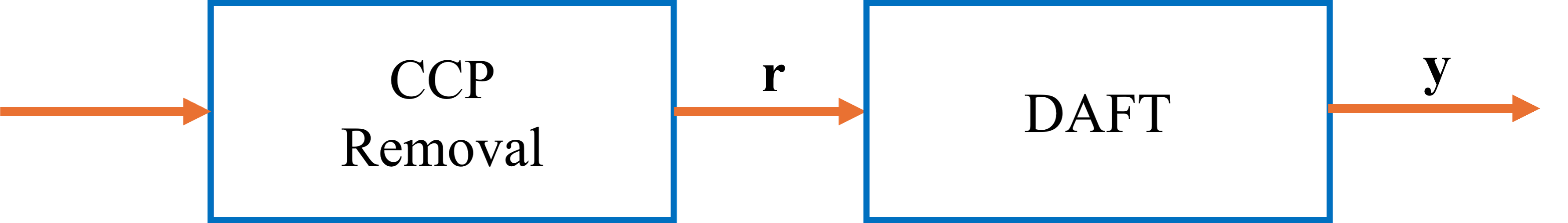}}\\
    \caption{Diagram of AFDM: (a) modulation; (b) demodulation.}
    \label{fig:afdm}
\end{figure}
AFDM is a new chirp-based multicarrier waveform tailored for high-mobility communications. The key idea of AFDM is to modulate the transmitted symbols over the discrete affine Fourier (DAF) domain in such a way that all the paths are separated from each other and each symbol experiences full diversity by adapting the underlying AFDM parameters. As shown in Fig.~\ref{fig:afdm}, let ${\bf{x}} \in {\mathbb{A}^{M \times 1}}$ denote the transmitted symbols in the DAF domain. Then, the inverse discrete affine Fourier transform (DAFT)\cite{AFDM_1,tao2024daft} is applied to convert ${\bf{x}}$ into a time domain signal ${\bf{s}}$,
\begin{equation}\label{AFDM_T}
    s(t) = \frac{1}{{\sqrt M }}\sum\limits_{m = 0}^{M - 1} {{x_m}{e^{j2\pi \left( {\frac{{{c_1}{t^2}}}{{T_s^2}} + {c_2}{m^2} + \frac{{mt}}{{M{T_s}}}} \right)}}},
\end{equation}
where ${{c_1}}$ and ${{c_2}}$ are the AFDM parameters, which can be adjusted according to the channel delay and Doppler distributions for full diversity \cite{AFDM_1,tao2024affine}. By sampling $s(t)$ with the interval ${T_s} = \frac{1}{{M\Delta f}}$, (\ref{AFDM_T}) can be rewritten in a matrix form as
\begin{equation}
    {\bf{s}} = {\bf{\Lambda }}_{{c_1}}^H{\bf{F}}_M^H{\bf{\Lambda }}_{{c_2}}^H{\bf{x}},
\end{equation}
where ${{\bf{\Lambda }}_{{c_i}}} \!=\! \text{diag}\left\{ {{e^{ - j2\pi {c_i}{m^2}}},m \!=\! 0,1, \cdots ,M - 1;i \!=\! 1,2} \right\}$. Then, a chirp-periodic prefix (CPP) \cite{AFDM_1} is added before $\bf{s}$ instead of an OFDM CP. The length of CPP ${L_\text{CPP}}$ is larger than or equal to the maximum channel delay spread. The specific CPP is given by
\begin{equation}
    {s_n} = {s_{M + n}}{e^{ - j2\pi {c_1}\left( {{M^2} + 2Mn} \right)}},n =  - {L_\text{CPP}}, \cdots , - 1.
\end{equation}
Note that CPP is simplified to CP if $2M{c_1}$ is an integer value and $M$ is even.

After the transmission over the channel and discarding CPP, the resulted time domain signal ${\bf{r}}\in {\mathbb{C}^{M \times 1}}$ is transformed back to the DAF domain via DAFT as described below
\begin{equation}
    {\bf{y}} = {{\bf{\Lambda }}_{{c_2}}}{\bf{F}}_M{{\bf{\Lambda }}_{{c_1}}}{\bf{r}}.
\end{equation}

It is worth mentioning that DFT and DFnT are two special cases of DAFT when ${{c_1}}$ and ${{c_2}}$ are chosen by zero or $\frac{1}{{2N}}$.
\bibliographystyle{IEEEtran}
\footnotesize
\bibliography{ref_OTFS_new}

\begin{IEEEbiography}[{\includegraphics[width=1in,height=1.25in,clip,keepaspectratio]{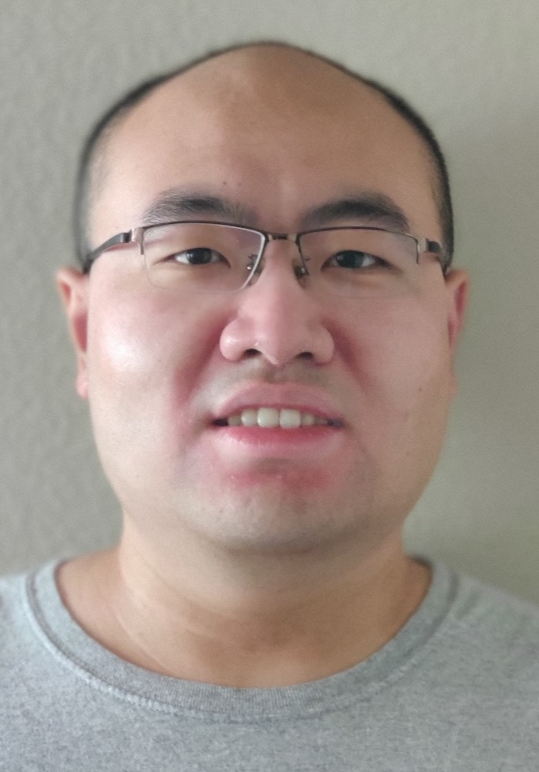}}]
	{Qinwen Deng}~(Member, IEEE)~received his Ph.D. degree in Electrical and Computer Engineering from the University of California at Davis, Davis, CA, USA, in 2024. He is set to join Nanyang Technological University as a postdoctoral researcher. His current research interests include orthogonal time frequency space (OTFS) modulation, channel estimation, decoder design, machine learning, hypergraph signal processing, and multilayer graph signal processing.
\end{IEEEbiography}

\begin{IEEEbiography}[{\includegraphics[width=1in,height=1.25in,clip,keepaspectratio]{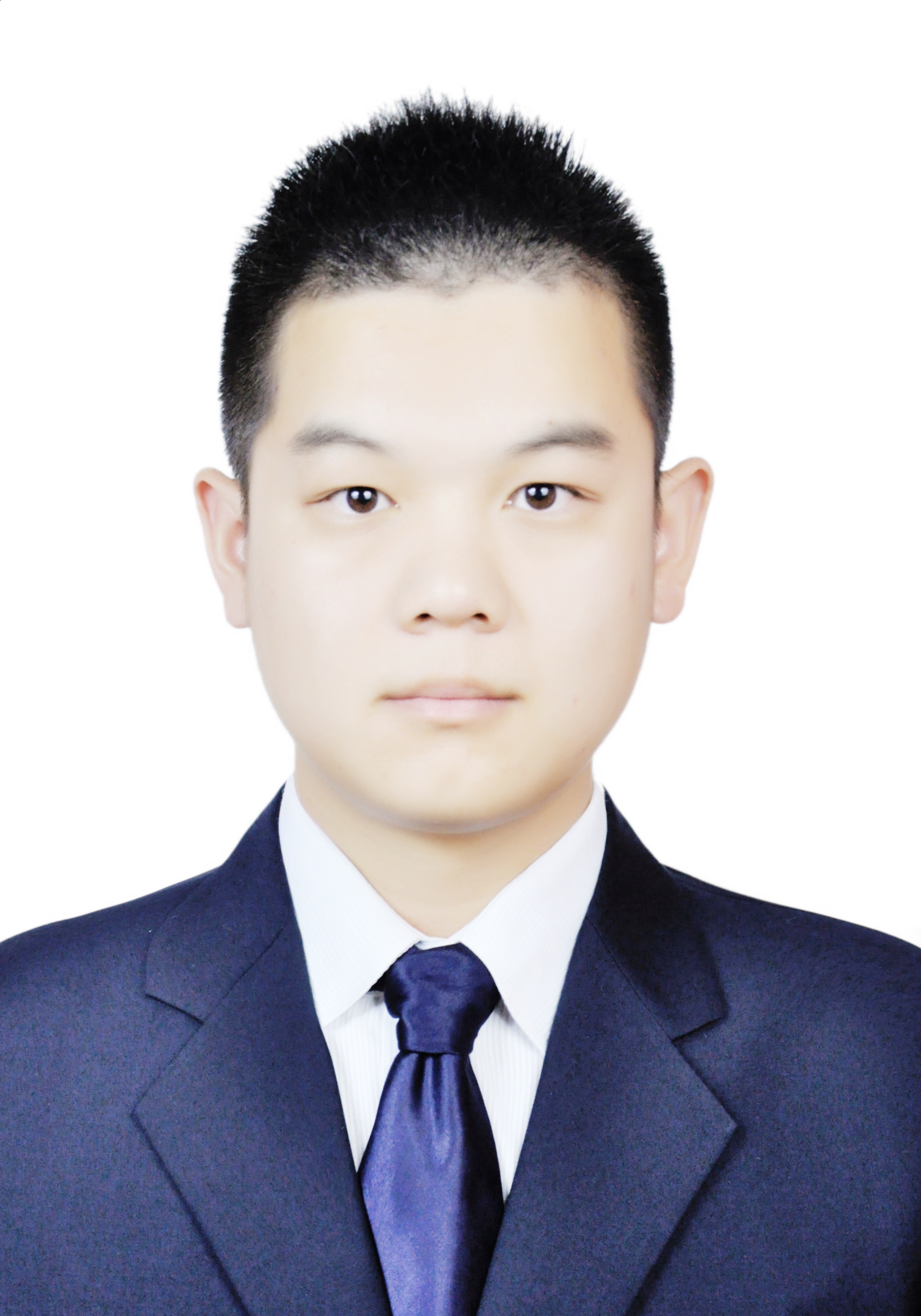}}]
	{Yao GE}~(Member, IEEE)~received the Ph.D. degree in Electronic Engineering from The Chinese University of Hong Kong (CUHK), Shatin, Hong Kong, in 2021, and the M.Eng. degree (research) in Communication and Information System and the B.Eng. degree in Electronics and Information Engineering from Northwestern Polytechnical University (NPU), Xi’an, China, in 2016 and 2013, respectively. He is currently a Research Fellow with Continental-NTU Corporate Lab, Nanyang Technological University (NTU), Singapore. From October 2015 to March 2016, he was a Visiting Scholar with the Department of Electrical and Computer Systems Engineering, Monash University, Melbourne, VIC, Australia. From April 2016 to August 2016, he was a Visiting Scholar with the Department of Computer, Electrical and Mathematical Science and Engineering, King Abdullah University of Science and Technology (KAUST), Thuwal, Saudi Arabia. From May 2019 to December 2019, he was a Visiting Scholar with the Department of Electrical and Computer Engineering, University of California at Davis (UC Davis), Davis, CA, USA. His current research interests include wireless communications and system design, the Internet of Things, cognitive radio networks, automotive vehicle signal processing and communications, integrated sensing and communication, wireless network security, statistical signal processing, optimization, and game theory. He received the Best Paper Award from the International Conference on Wireless Communications and Signal Processing (WCSP) 2022. He is a Founding Member of the IEEE ComSoc Special Interest Group (SIG) on OTFS and has served as the Co-Chair for the IEEE/CIC ICCC 2022 Workshop on OTFS. He is also a Youth Editor of the Journal of Information and Intelligence.
\end{IEEEbiography}

\begin{IEEEbiography}[{\includegraphics[width=1in,height=1.25in,clip,keepaspectratio]{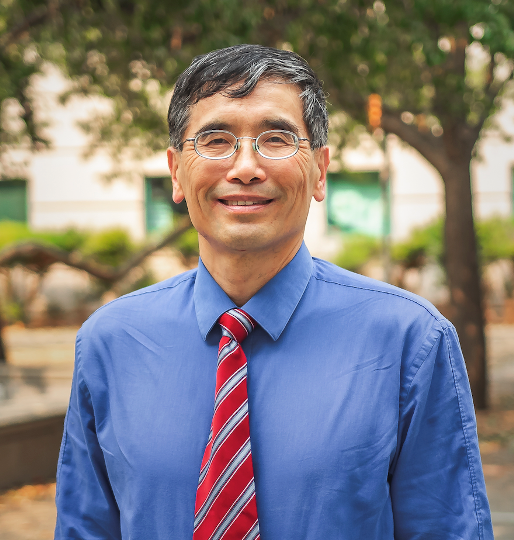}}]
	{Zhi Ding}~(S'88-M'90-SM'95-F'03)~is with the Department of Electrical and Computer Engineering at the University of California, Davis, where he holds the position of distinguished professor. He received his Ph.D. degree in Electrical Engineering from Cornell University in 1990. From 1990 to 2000, he was a faculty member of Auburn University and later, University of Iowa. Prof. Ding joined the College of Engineering at UC Davis in 2000. His major research interests and expertise cover the areas of wireless networking, communications, signal processing, multimedia, and learning. Prof. Ding supervised over 30 PhD dissertations since joining UC Davis. His research team of enthusiastic researchers works very closely with industry to solve practical problems and contributes to technological advances. His team has collaborated with researchers around the world and welcomes self-motivated young talents as new members. 
	
	Prof. Ding is a Fellow of IEEE and has served as the Chief Information Officer and Chief Marketing Officer of the IEEE Communications Society. He was associate editor for IEEE Trans. Signal Process. from 1994-1997, 2001-2004, and associate editor of IEEE Signal Processing Letters 2002-2005. He was a member of technical committee on Statistical Signal and Array Processing and member of technical committee on Signal Processing for Communications (1994-2003).  Dr. Ding was the General Chair of the 2016 IEEE International Conference on Acoustics, Speech, and Signal Processing and the Technical Program Chair of the 2006 IEEE Globecom. He was also an IEEE \textit{Distinguished Lecturer} (Circuits and Systems Society, 2004-06, Communications Society, 2008-09). He served on as IEEE Transactions on Wireless Communications Steering Committee Member (2007-2009) and its Chair (2009-2010). Dr. Ding is a coauthor of the textbook: \textit{Modern Digital and Analog Communication Systems,} 5th edition, Oxford University Press, 2019. Prof. Ding received the IEEE Communication Society’s WTC Award in 2012 and the IEEE Communication Society’s Education Award in 2020. 
\end{IEEEbiography}
\vfill\pagebreak

\end{document}